\documentclass[twocolumn]{revtex4-1}
\usepackage[utf8]{inputenc}
\usepackage{amsmath}
\usepackage{bm}
\usepackage{amsfonts}
\usepackage{amssymb}
\usepackage[dvipsnames]{xcolor}
\usepackage{graphicx}
\usepackage{physics}
\usepackage{textcomp}
\usepackage{dsfont}
\usepackage{lipsum}
\usepackage{bbold}
\usepackage{url}
\usepackage{hyperref}
\usepackage{ulem}
\graphicspath{ {Figs-manuscript/} }

\begin{document}

\title{Two-qubit sweet spots for capacitively coupled exchange-only spin qubits}
\author{MengKe Feng} \thanks{Present address: School of Electrical Engineering and TeleComm.,
University of New South Wales, Sydney, NSW 2052, Australia.}
\affiliation{Division of Physics and Applied Physics, School of Physical and Mathematical Sciences, Nanyang Technological University, 21 Nanyang Link, Singapore 637371, Singapore.}
\author{Lin Htoo Zaw} \thanks{Present address: Centre for Quantum Technologies, National University of Singapore, 3 Science Drive 2, Singapore 117543, Singapore.}
\affiliation{Division of Physics and Applied Physics, School of Physical and Mathematical Sciences, Nanyang Technological University, 21 Nanyang Link, Singapore 637371, Singapore.}
\author{Teck Seng Koh} \thanks{Corresponding author:~\href{mailto:kohteckseng@ntu.edu.sg}{kohteckseng@ntu.edu.sg} }
\affiliation{Division of Physics and Applied Physics, School of Physical and Mathematical Sciences, Nanyang Technological University, 21 Nanyang Link, Singapore 637371, Singapore.}

\begin{abstract}
The implementation of high fidelity two-qubit gates  is a bottleneck in the progress towards universal quantum computation in semiconductor quantum dot qubits.  We study capacitive coupling between two triple quantum dot spin qubits encoded in the $S=1/2, S_z=-1/2$ decoherence-free subspace -- the exchange-only (EO) spin qubits. We report exact gate sequences for CPHASE and CNOT gates, and demonstrate theoretically, the existence of multiple two-qubit sweet spots (2QSS) in the parameter space of capacitively coupled EO qubits. Gate operations have the advantage of being all-electrical, but charge noise that couple to electrical parameters of the qubits cause decoherence. Assuming noise with a $1/f$ spectrum, two-qubit gate fidelities and times are calculated, which provide useful information on the noise threshold necessary for fault-tolerance. We study two-qubit gates at single and multiple parameter 2QSS. In particular, for two existing EO implementations -- the resonant exchange (RX) and the always-on exchange-only (AEON) qubits -- we compare two-qubit gate fidelities and times at positions in parameter space where the 2QSS are simultaneously single-qubit sweet spots (1QSS) for the RX and AEON. These results provide a potential route to the realization of high fidelity quantum computation. 
\end{abstract}

\maketitle


\section{Introduction}

Semiconductor quantum dots are one of the leading platforms for building a quantum computer.
They present promises of scalability, coherence and integration with existing microelectronics technologies~\cite{Loss1998, Zwanenburg2013}. High fidelity gate operations have been demonstrated in
single quantum dot (QD)~\cite{Yoneda2018, Zajac2018}, double QD~\cite{Foletti2009, Bluhm2011a, Kim2014, Kim2015, Thorgrimsson2017, Cerfontaine2020} and triple quantum dot (TQD)~\cite{Schroer2007, Laird2010, Gaudreau2012} architectures.
In particular, qubits encoded in the decoherence-free subspace of three electron spins~\cite{DiVincenzo2000, Shi2012, Koh2012, Russ2017}
have the advantage of fast, all-electrical control.  Single qubit gates are based on the exchange interaction, hence its namesake, the exchange-only (EO) qubit~\cite{DiVincenzo2000}.
The total spin of three electrons
comprise a $S=3/2$ quadruplet and two $S=1/2$ degenerate doublets, whose degeneracy can be lifted by an external magnetic field. Logical qubit states are encoded in the
total spin $S=1/2, S_z=-1/2$  doublet, which provides immunity against collective decoherence.

The implementation of high fidelity two-qubit gates is a bottleneck in the progress toward universal, fault-tolerant quantum computation.
Two-qubit entangling gates can be based on  exchange~\cite{DiVincenzo2000} or capacitive~\cite{Srinivasa2015, Calderon-Vargas2015, Neyens2019a, MacQuarrie2020} coupling.
Exchange is fast but short-ranged, giving rise to hybrid approaches like spin-shuttling~\cite{Fujita2017, Feng2018, Mills2019, Yoneda2020, Ginzel2020} and circuit QED~\cite{Mi2017, Stockklauser2017}.
Because exchange arises from spin-conserving tunneling, two-qubit exchange gates have potential for leakage.
On the other hand,  capacitive coupling, which arises from electrostatic Coulomb interaction, allows for a  longer range of interaction, has less stringent QD addressability requirements, and alleviates the problem of leakage.

We study two capacitively coupled EO qubits, and report exact gate sequences for CPHASE and CNOT gates. The non-local gate is implemented in a single time interval, in contrast to exchange gating which requires several steps~\cite{DiVincenzo2000, Fong2011, Shi2012}. A major progress in Refs.~\cite{Doherty2013, Shim2016} was to propose a single pulse exchange gate, in the negligible transverse coupling limit.  Single and two-qubit operations are all-electrical, but charge noise, ubiquitous in the solid state environment~\cite{Paladino2014}, couple to electrical parameters of the qubits and cause decoherence.

Sweet spots~\cite{Fei2015, Frees2019} are flat points in the energy landscape which provide protection from parameter fluctuations due to noise. For two existing implementations of the EO qubit -- always-on exchange-only (AEON)~\cite{Shim2016} and resonant exchange (RX)~\cite{Medford2013, Taylor2013, Doherty2013, Russ2015, Wardrop2016} qubits -- single-qubit sweet spots (1QSS) have been studied. In Ref.~\onlinecite{Shim2016}, two-qubit sweet spots (2QSS) for exchange coupling was reported.
However, sweet spots  for capacitive coupling have not been studied, a knowledge gap which we address. 
We show theoretically, in the weak noise, perturbative limit, that 2QSS exist for single-qubit parameters, $\varepsilon_{\mathrm{m}}, t_{\mathrm{l}}$ and $t_{\mathrm{r}}$, in both control and target qubits. 
This enables two-qubit gates to be operated at positions in parameter space which are either (1) a single parameter 2QSS, or (2) simultaneously two parameter 2QSS in $\varepsilon_{\mathrm{m}}$ and $t_{\mathrm{l}}$ (or $t_{\mathrm{r}}$), or (3) simultaneously $\varepsilon_{\mathrm{m}}$ 2QSS and  $\varepsilon$ 1QSS. 
For (3), this requires tuning to a particular tunnel coupling ratio, so that the $\varepsilon_{\mathrm{m}}$ 2QSS overlaps with the 1QSS of RX and AEON.  Finally, we discuss existence conditions for 2QSS, optimal choices of working points, and address further fidelity optimization.


\section{Results}


\subsection{Two-qubit Hamiltonian}

\begin{figure*}
\centering
\includegraphics[width=0.45\textwidth]{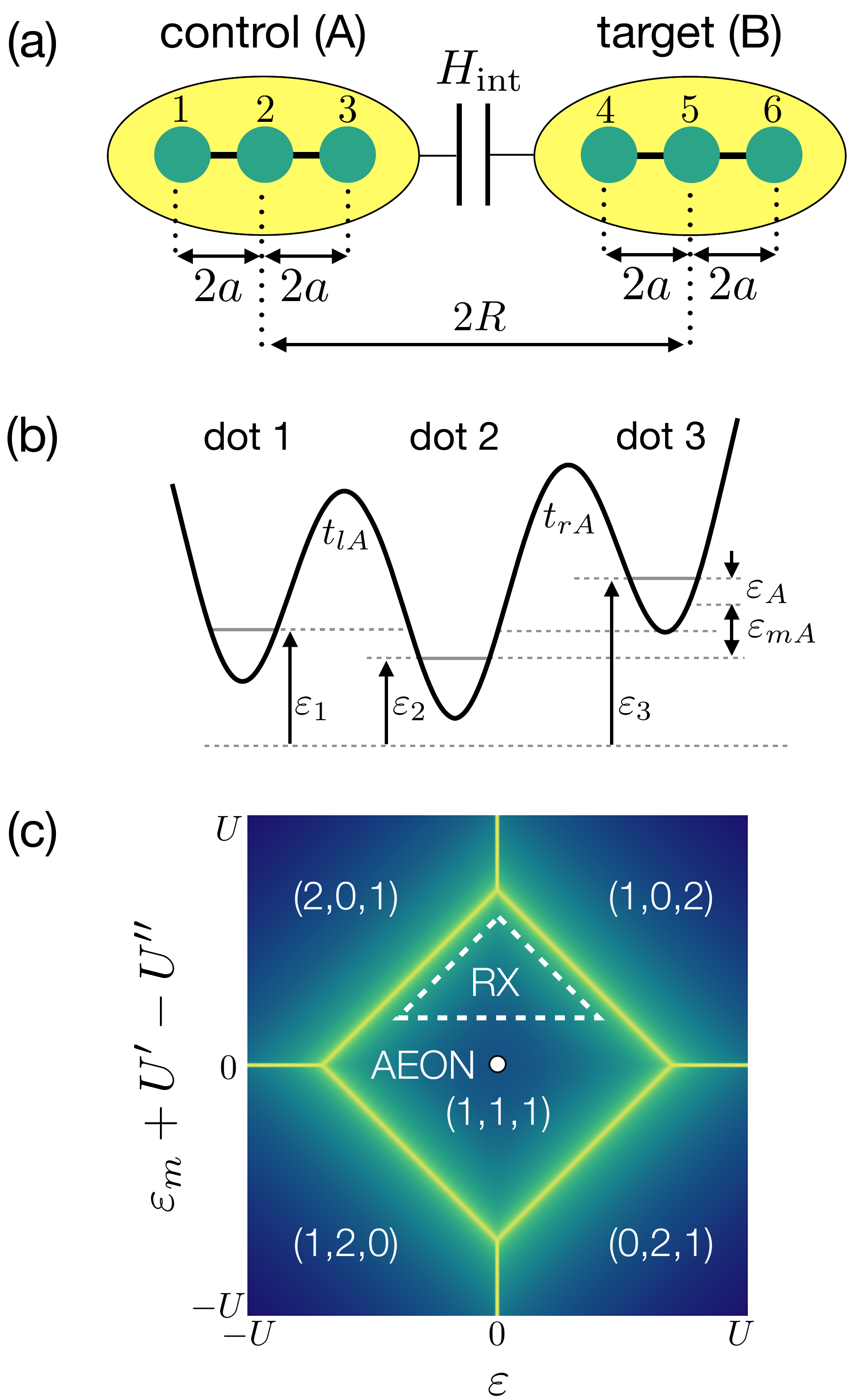}
\caption{\linespread{1}\selectfont{}
Exchange-only (EO)  qubit parameters. 
(a) Two EO qubits  in a linear quantum dot (QD) array, with qubit A (B) in QDs $1 - 3$ ($4 - 6$).
We take 
$a = 50~\mathrm{nm}$, $R = 160~\mathrm{nm}$, with QD Bohr radius, 
$a_\mathrm{B} = 25~\mathrm{nm}$
in our calculations. Qubit A (B) is the control (target) qubit in the CPHASE and CNOT gates studied. The capacitive coupling between qubits is given by $\hat{H}_\mathrm{int}$ (Eq.~\eqref{eq:Hint}). 
(b) Schematic of parameters in qubit A. Left/right tunnel couplings are given by $t_\mathrm{lA/rA}$. Detuning parameters $\varepsilon_i$ control the relative single-particle energies between QDs, represented by arrows from a reference energy to the ground orbital energy of each QD. Outer and middle detunings, $\varepsilon_{\mathrm{A}} \equiv \left(\varepsilon_1 - \varepsilon_3 \right)/2$ and $\varepsilon_{\mathrm{mA}} \equiv \varepsilon_2- \left(\varepsilon_1 + \varepsilon_3 \right)/2$, with the tunnel couplings are sufficient to describe single qubit dynamics. For all our results, we take a reflection symmetry in the parameters of the two qubits: $\varepsilon_{\mathrm{B}} = -\varepsilon_{\mathrm{A}}, \varepsilon_{\mathrm{mB}} = \varepsilon_{\mathrm{mA}}$,  $t_{\mathrm{lB}} = t_{\mathrm{rA}}$ and  $t_{\mathrm{rB}} = t_{\mathrm{lA}}$.
(c) Detuning space of an EO qubit and charge occupation numbers. The AEON single-qubit double ($\varepsilon$ and $\varepsilon_{\mathrm{m}}$) sweet spot is at the center of the (1,1,1) region at $(\varepsilon, \varepsilon_{\mathrm{m}}) = (0, U^{\prime \prime} - U^\prime)$ (white circle). The RX single-qubit working region is indicated by the dashed triangle. In our study, the RX working point is taken to be at  $(\varepsilon, \varepsilon_{\mathrm{m}}) = (0,  U^\prime)$ (i.e. at its single-qubit $\varepsilon$ sweet spot  for equal tunnel couplings).
}\label{fig:tqd}
\end{figure*}

We study capacitively coupled EO qubits in a linear array of two TQDs~(Fig.~\ref{fig:tqd}(a)). The left/right TQD (qubit A/B) is the control/target qubit in CPHASE and CNOT gates. QDs are numbered as shown in Fig.~\ref{fig:tqd}(a). Within each TQD, there are four independently tunable parameters --  left/right tunnel couplings $t_{\mathrm{l/r}}$, and detunings for outer  and middle QDs, $\varepsilon$ and $\varepsilon_{\mathrm{m}}$. They are defined for qubits A and B:  $\varepsilon_{\mathrm{A}} \equiv \left(\varepsilon_1 - \varepsilon_3 \right)/2$, $\varepsilon_{\mathrm{B}} \equiv \left(\varepsilon_4 - \varepsilon_6 \right)/2$, $\varepsilon_{\mathrm{mA}} \equiv \varepsilon_2- \left(\varepsilon_1 + \varepsilon_3 \right)/2$, and $\varepsilon_{\mathrm{mB}} \equiv \varepsilon_5 - \left(\varepsilon_4 + \varepsilon_6 \right)/2$, and  schematically represented for qubit A in Fig.~\ref{fig:tqd}(b).

Each EO qubit is fully described by the Hubbard Hamiltonian~\cite{DasSarma2011},
   \begin{align}
\hat{\mathcal{H}} = \hat{H}_\mathrm{t} + \hat{H}_\mathrm{\varepsilon} + \hat{H}_\mathrm{U}.
 \end{align}   
Here, $\hat{H}_\mathrm{t} = \sum_{<i,j>}\sum_{\sigma} t_{ij} \hat{c}^\dag_{i\sigma}\hat{c}_{j\sigma}$ is nearest-neighbor tunneling, with spin index $\sigma = \{ \uparrow, \downarrow \}$ and QD indices $i,j$ run from dots 1 to 3 (4 to 6) for qubit A (B). Tunnel couplings $t_{12}\equiv t_{\mathrm{lA}}$, $t_{23}\equiv t_{\mathrm{rA}}$, $t_{45}\equiv t_{\mathrm{lB}}$, and $t_{56}\equiv t_{\mathrm{rB}}$.  There is no tunnel coupling between the TQDs, i.e. $t_{34}=0$. The detuning term is $\hat{H}_\mathrm{\varepsilon} = \sum_{i,\sigma} \varepsilon_i \hat{n}_{i\sigma}$, where the number operator $\hat{n}_{i\sigma}\equiv \hat{c}^\dag_{i\sigma}\hat{c}_{i\sigma}$. The Coulomb energy term is $\hat{H}_\mathrm{U} = \sum_{i} U_i \hat{n}_{i \uparrow} \hat{n}_{i \downarrow} +  \frac{1}{2}\sum_{i}\sum_{j\neq i} U_{ij}   \hat{n}_{i }\hat{n}_{j }$, where the first term is intra-dot, the second term is inter-dot Coulomb energies, and $\hat{n}_i = \hat{n}_{i\uparrow}+\hat{n}_{i \downarrow}$. We consider inter-dot Coulomb energies between all neighbours.

In the perturbative limit $t_{\mathrm{l/r}} \ll U_{ij}  < U_i$, each EO qubit operates in the (1,1,1) charge state, where numbers represent  electron occupation in left, middle and right dots. This is shown by the central region in detuning space for one qubit in Fig.~\ref{fig:tqd}(c). However, the encoded logical ``1'' state of the qubit comprises small charge admixtures of doubly occupied charge states (1,0,2), (2,0,1), (0,2,1) and (1,2,0), with a singly occupied (1,1,1) state. This results in an effective exchange coupling with the encoded logical ``0'' state.

The basis states for qubit A are the singly-occupied encoded qubit states 
$|1_{\mathrm{A}}\rangle \equiv \left( |\! \downarrow_1 \downarrow_2 \uparrow_3\rangle - |\! \uparrow_1 \downarrow_2 \downarrow_3\rangle \right)/ \sqrt{2}$,
and
$|0_{\mathrm{A}}\rangle \equiv \left(2|\! \downarrow_1 \uparrow_2 \downarrow_3\rangle -  |\! \downarrow_1 \downarrow_2 \uparrow_3\rangle - |\! \uparrow_1 \downarrow_2 \downarrow_3\rangle \right)/ \sqrt{6}$, 
and four states comprising a singly- and a doubly-occupied QD residing in the same spin space, 
$ | \mathrm{A}_{1}\rangle\equiv |\! \downarrow_1 \uparrow_3 \downarrow_3\rangle$,  $|\mathrm{A}_{2}\rangle\equiv |\! \uparrow_1 \downarrow_1 \downarrow_3\rangle$, $|\mathrm{A}_{3} \rangle \equiv |\! \downarrow_1 \uparrow_2 \downarrow_2\rangle$, and $|\mathrm{A}_{4}\rangle \equiv |\! \uparrow_2 \downarrow_2 \downarrow_3\rangle$. 
Subscripts label QD numbers. Basis states  $\{|0_{\mathrm{B}}\rangle, |1_{\mathrm{B}}\rangle, | \mathrm{B}_{i=1\dots 4}\rangle\}$ for qubit B can be written by mapping the QD indices $(1,2,3) \rightarrow (4,5,6)$.
With the six-dimensional basis of each qubit, the Schrieffer-Wolff transformation~\cite{Schrieffer1966, Gros1987, MacDonald1988} gives effective single-qubit Hamiltonians,
   \begin{align}
\hat{H}_{\mathrm{eff,A}} =& -\frac{\hbar \omega_\mathrm{A}}{2} \hat{\sigma}_z - \frac{\hbar g_\mathrm{A}}{2} \hat{\sigma}_x, \label{eq:SWA}\\
\hat{H}_{\mathrm{eff,B}} =& -\frac{\hbar \omega_\mathrm{B}}{2} \hat{\sigma}_z - \frac{\hbar g_\mathrm{B}}{2} \hat{\sigma}_x,\label{eq:SWB}
 \end{align}   
where $\hbar\omega \equiv 2 \left( t_{\mathrm{l}}^2 b_{-} + t_{\mathrm{r}}^2 b_{+} \right)$, $\hbar g \equiv 2\sqrt{3} \left(t_{\mathrm{l}}^2 b_{-} - t_{\mathrm{r}}^2 b_{+}\right)$ and
$b_{\pm} \equiv 1/\left(U- U^{\prime \prime} \pm \varepsilon + \varepsilon_{\mathrm{m}}  \right)+ 1/\left(U-2U^\prime + U^{\prime \prime} \mp \varepsilon - \varepsilon_{\mathrm{m}} \right) $. (Qubit subscripts A, B have been omitted for brevity here.)
In the derivation, we assumed identical QDs: $U_i \equiv U $ for all $i$, $U_{12}=U_{23}=U_{45}=U_{56}\equiv U^\prime$ and $U_{13}=U_{46} \equiv U^{\prime \prime}$. (For numerical values, see Methods.) Encoded states, dressed by charge admixtures, and denoted with a prime, are 
   \begin{align}
|{1}^\prime_\mathrm{A}\rangle =& \frac{1}{\sqrt{N_\mathrm{A}}}\left(|1_{\mathrm{A}}\rangle + \sum_{i=1}^4 \alpha_i |\mathrm{A}_{i}\rangle \right),\\
|{1}^\prime_\mathrm{B}\rangle =& \frac{1}{\sqrt{N_\mathrm{B}}}\left(|1_{\mathrm{B}}\rangle +  \sum_{i=1}^4 \beta_i |\mathrm{B}_{i}\rangle \right),
 \end{align}   
where $\sqrt{N_{\mathrm{A/B}}}$ are  normalization constants. Admixtures $\alpha_i$, $\beta_i$ depend on  electrical parameters $t_{\mathrm{l/r}}$, $\varepsilon$ and $\varepsilon_{\mathrm{m}}$, which are detailed in Supplementary Method 1. 

Two-qubit capacitive coupling arises from the inter-dot Coulomb interaction between the  TQDs, $\hat{H}_\mathrm{int} = \sum_{i=1}^{3}\sum_{j=4}^{6} \mathcal{V}_{ij} \hat{n}_{i }\hat{n}_{j } $. 
We denote  inter-TQD Coulomb terms by $\mathcal{V}_{ij}$ to distinguish them from intra-TQD terms $U_{ij}$. This distinction is for notational clarity; the physical basis -- electrostatic interaction -- is the same.
In the computational basis 
$\{ | 0_{\mathrm{A}} {0}_{\mathrm{B}}\rangle, |{0}_A 1_{\mathrm{B}}^\prime\rangle, |1_{\mathrm{A}}^\prime {0}_{\mathrm{B}}\rangle, |1_{\mathrm{A}}^\prime 1_{\mathrm{B}}^\prime \rangle\}$,   
the two-qubit capacitive coupling is diagonal, given by
   \begin{align}\label{eq:Hint}
\hat{H}_\mathrm{int}  = \hbar \times \text{diag}\{V_1, V_2, V_3, V_4 \},
 \end{align}   
which comprises a global phase, single-qubit energy shifts and a $\hat{\sigma}_z \otimes \hat{\sigma}_z$ term equivalent to the Ising (ZZ) model (see Supplementary Note 1). Tri-quadratic confinement potentials (see Methods) provide analytical expressions for $V_i$ terms (see Supplementary Methods 2, 3). The analytical expressions are crucial for the calculation of 2QSS and gate fidelities, explained in the rest of this paper.
Finally, the full two-qubit Hamiltonian in the computational basis is
   \begin{align}\label{eq:Heff}
\hat{H} = \hat{H}_{\mathrm{eff,A}} \otimes \hat{\mathbb{1}}_\mathrm{B} +  \hat{\mathbb{1}}_\mathrm{A} \otimes \hat{H}_{\mathrm{eff,B}} + \hat{H}_\mathrm{int}.
 \end{align}   
Because  capacitive coupling arises from the overlap of qubit wavefunctions, by increasing the barrier between QD 3 and 4, the exponential ``tail'' of the wavefunctions can be arbitrarily reduced, thereby turning off the interaction. Thus, energy shifts from the presence of the other qubit (equivalent to a redefinition of detunings) does not affect 1QSS and single-qubit gates. This redefinition shifts the entire charge boundary diagram, and hence 1QSS (Fig.~\ref{fig:tqd}c) by an approximately constant amount, $\Delta \varepsilon = 0.54$~meV, $\Delta \varepsilon_{\mathrm{m}} = 0.21$~meV, for parameters used in our study.

The $\hbar V_i$ terms in Eq.~\eqref{eq:Hint} contain all pairwise inter-dot Coulomb energies between the two TQDs, weighted by the charge admixtures. The latter is key in modeling the effect of charge noise, which has been measured to be $1/f$-like over a wide range of frequencies~\cite{Paladino2014, Yoneda2018, Chan2018, Struck2020a}.


\subsection{Charge noise}

We introduce noise by simulating random fluctuations in tunneling and detuning~\cite{Russ2016, Huang2018a}. These  fluctuations perturb charge admixtures,  leading to noisy two-qubit interaction. At this point, we write qubit parameters in vectorized form $\bm{n} = ( t_{\mathrm{lA}}, t_{\mathrm{rA}}, \varepsilon_{\mathrm{A}}, \varepsilon_{\mathrm{mA}},  t_{\mathrm{lB}}, t_{\mathrm{rB}}, \varepsilon_{\mathrm{B}}, \varepsilon_{\mathrm{mB}} )$ for notational simplicity, where the first (last) four components belong to qubit A (B). Noisy  parameters, denoted with a tilde, is the sum of the noiseless parameter with a random time-dependent fluctuation, $\tilde{n}_i(t) = n_i + \delta n_i(t)$. We assume uncorrelated noise and each random time series is generated independently in simulations. To avoid confusion, symbol $t$ with or without numeral subscripts indicates time, while $t_\mathrm{l/r, A/B}$ indicate left/right tunnel couplings for qubits A/B. It should be clear from the context which is being referred to. Random variables are characterized by the time correlation function $C_{n_i} (t_1 - t_2) = \langle \delta {n_i}(t_1) \delta {n_i}(t_2) \rangle$, where angular brackets $\langle . \rangle$ denote average over noise realizations. The corresponding (two-sided) power spectral density is the Fourier transform of the time correlation function, ${S}_{n_i}(\omega) = \int_{-\infty}^{\infty} C_{n_i}(t) \mathrm{e}^{-\mathrm{i} \omega t}dt$.
Adapting the algorithms of Refs.~\onlinecite{Timmer1995, pycolorednoise},
each noisy parameter $\delta n_i (t)$ is simulated with a $1/f$ power spectral density with a low frequency roll-off, 
   \begin{align}\label{eq:S(w)}
{S}_{n_i} (\omega) =
\begin{cases}
\Delta ^2_{n_i} \frac{2\pi}{\omega_\mathrm{l}}, &\text{for } \left |\omega  \right | \le \omega_\mathrm{l} \\
\Delta ^2_{n_i} \frac{2\pi}{\left | \omega  \right | }, & \text{for } \omega_\mathrm{l} \le  \left | \omega  \right |  \le \omega_\mathrm{h}\\
0, & \text{otherwise}
\end{cases}
 \end{align}   
where $\omega_\mathrm{l}$ ($\omega_\mathrm{h}$) is the lower (higher) cutoff frequency. $\Delta_{n_i}^2$ is related to the noise variance $\sigma_{n_i}^2$ by $\Delta_{n_i}^2 = \sigma_{n_i}^2 \tfrac{1}{2}\pqty{1+\ln(\omega_\mathrm{h}/\omega_\mathrm{l})}^{-1}$ (see Supplementary Note 2). Although spectroscopy experiments~\cite{Yoneda2018, Chan2018, Struck2020a},
which are limited by acquisition time, have not observed low frequency saturation in $1/f$ noise, we model the roll-off to avoid unphysical, infinite noise power at $f=0$. (See Supplementary Discussion 1).

Because our objective is to study  fidelities and sweet spots of two-qubit gates under noise, and  1QSS have already been found, we assume noisy two-qubit interactions and ideal (noiseless) single-qubit operations for all calculations.


\subsection{Non-local gate time}

CPHASE and CNOT gates are given by unitary evolution operators
   \begin{align}
\hat{U}_\mathrm{CPHASE} =
\begin{pmatrix}
1 & 0 & 0 & 0\\
0 & 1 & 0 & 0\\
0 & 0 & 1 & 0\\
0 & 0 & 0 & -1
\end{pmatrix},~
\hat{U}_\mathrm{CNOT} =
\begin{pmatrix}
1 & 0 & 0 & 0\\
0 & 1 & 0 & 0\\
0 & 0 & 0 & 1\\
0 & 0 & 1 & 0
\end{pmatrix}.
 \end{align}

\begin{figure*}
\centering
\includegraphics[width=0.65\textwidth]{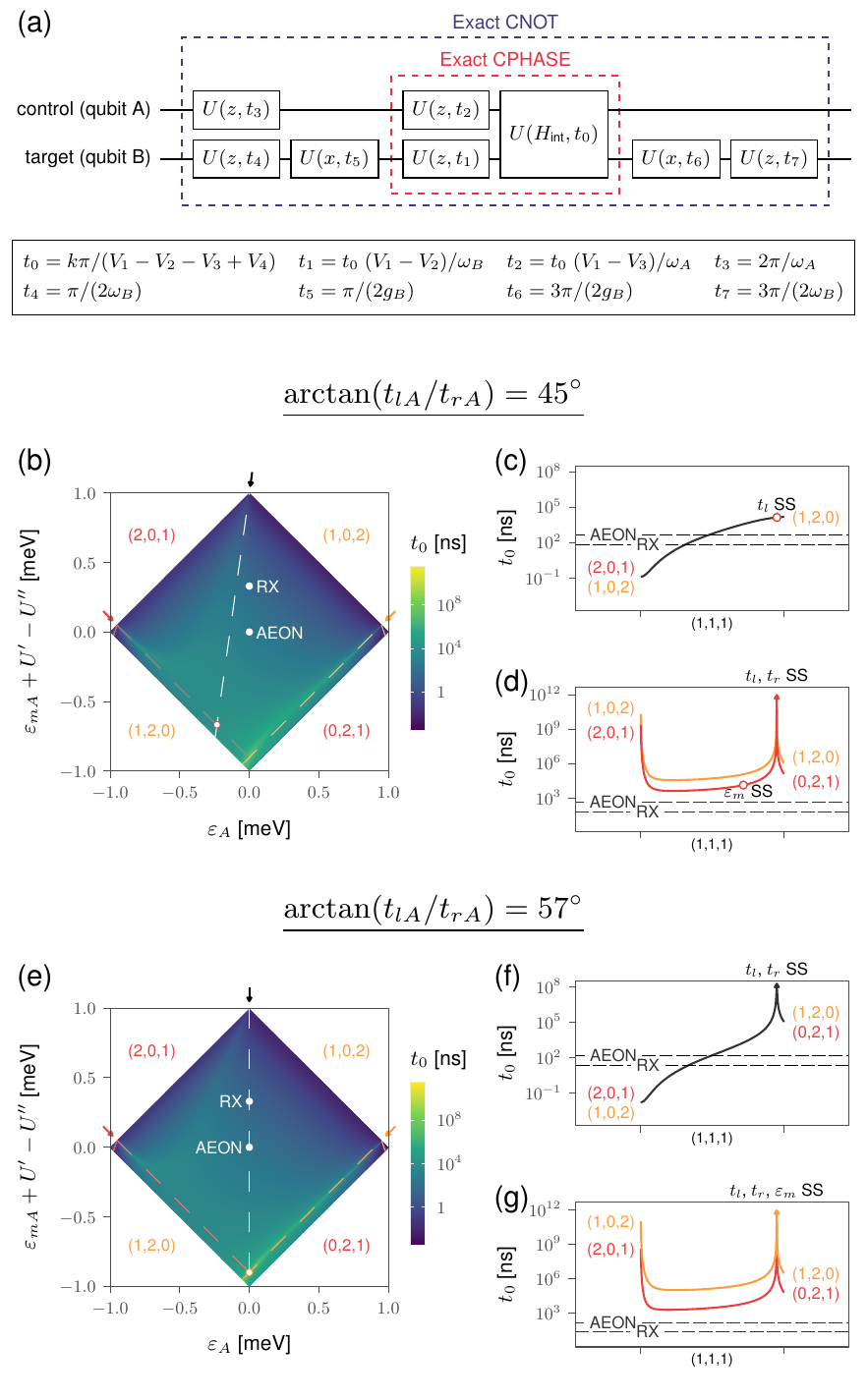}
\caption{\linespread{1}\selectfont{}
Gate sequences and gate times in detuning space.
(a) Exact gate sequences and timings for CPHASE (red dashed box) and CNOT (blue dashed box). Effective single-qubit Hamiltonians (Eqs.~\eqref{eq:SWA},~\eqref{eq:SWB}) generate unitaries for qubit A/B, $U_{\mathrm{A/B}}(x,t) \equiv \exp\left(\mathrm{i} \hat{\sigma}_x g_{\mathrm{A/B}} t/2 \right)$ and $U_{\mathrm{A/B}}(z,t) \equiv \exp\left(\mathrm{i} \hat{\sigma}_z \omega_{\mathrm{A/B}} t/2 \right)$. The non-local unitary is given by Eq.~\eqref{eq:Uint} without noise.  The non-local gate time $t_0$ is identical for CPHASE and CNOT, where $k$ is an odd positive integer (Eq.~\eqref{eq:t0}). (b) Color plot of gate times for equal tunneling $t_{\mathrm{lA}} = t_{\mathrm{rA}}$, with two-qubit sweet spots (2QSS) indicated for $\varepsilon_{\mathrm{mA}}$ (white), $t_{\mathrm{lA}}$ (red), and $t_{\mathrm{rA}}$ (orange) with dashed lines. (c) Linecuts of $\varepsilon_{\mathrm{mA}}$ 2QSS. Fastest gate times are at the top corner and increase down the $\varepsilon_{\mathrm{mA}}$ 2QSS line. Fastest ($k=1$) AEON and RX gate times are 450~ns and 64~ns respectively. At the intersection with $t_{\mathrm{lA}}$ 2QSS, gate time is 13.6~$\mu$s. (d) Linecuts of $t_{\mathrm{lA}}$ and $t_{\mathrm{rA}}$ 2QSS. Gate times go to infinity at the point where the two tunneling 2QSS intersect. Panels (e--g) repeats panels (b--d), but with tunneling  $\arctan(t_{\mathrm{lA}}/t_{\mathrm{rA}})=57^\circ$, chosen such that the $\varepsilon_{\mathrm{m}}$ 2QSS is along $\varepsilon_{\mathrm{A}} = 0$, which is the 1QSS, and which the operating points of  AEON and RX lie on. As before, gate times increase as $\varepsilon_{\mathrm{mA}}$ decreases, and goes to infinity at the $t_{\mathrm{lA}}$, $t_{\mathrm{rA}}$ double 2QSS. 
}\label{fig:gateresults}
\end{figure*}

Using Mahklin invariants~\cite{Makhlin2004a}, we find the non-local interaction time required for these gates to be identical,
\begin{equation}\label{eq:t0}
t_0 =  k\pi/[(V_1 + V_4)- ( V_2 + V_3 )],
\end{equation}
for any odd positive integer $k$ (see Supplementary Note 3). While  non-local gating has been studied in the literature~\cite{Doherty2013, Taylor2013, Shim2016, Pal2015, Luczak2018}, the exact gate sequence (including local gates) for capacitively coupled EO qubits has not been reported, to our knowledge. We show the exact sequence and timings in Fig.~\ref{fig:gateresults}(a). 
Importantly, it is the  energy difference in the denominator, $(V_1 + V_4) - (V_2 + V_3)$, that is important for capacitive gating. This difference is dominated by Coulomb interactions between electrons in doubly-occupied charge admixtures of qubits A and B, which have a centre of mass situated away from the middle dot (unlike the dominant (1,1,1) configuration), giving rise to a net electric dipole moment. The energy difference can be qualitatively understood as a dipole-dipole interaction between the two TQDs. This is what gives rise to the dependence on TQD parameters for the non-local gate time.

Color plots of gate time $t_0$ in detuning space for qubit A are shown, for equal tunneling ratios where $\arctan(t_{\mathrm{lA}}/t_{\mathrm{rA}})=45^\circ$ (Fig.~\ref{fig:gateresults}(b)) and tunneling ratios given by $\arctan(t_{\mathrm{lA}}/t_{\mathrm{rA}})=57^\circ$ (Fig.~\ref{fig:gateresults}(e)). The equivalent color plot for qubit B (not shown) is a mirror reflection about the $\varepsilon = 0$ line for qubit A, because of our choice of  parameters $\varepsilon_{\mathrm{B}} = -\varepsilon_{\mathrm{A}}$ and $t_{\mathrm{lA/B}} = t_{\mathrm{rB/A}}$. 
Gate times are faster near the boundaries of (1,1,1) with (2,0,1) and (1,0,2) charge occupations.  This can be understood as a stronger capacitive interaction arising from a larger mean dipole moment of each TQD as a result of proportionately larger charge admixtures. Comparatively,  sitting near (1,2,0) or (0,2,1) boundaries gives rise to a smaller net TQD dipole moment and thus a significantly longer gate time. In addition, for qubit A, gate times decrease faster as one goes from the  central (1,1,1)  region towards the (1,0,2) boundary, compared to moving towards the (2,0,1) boundary.  This is because of the stronger dipole-dipole interaction arising from proximity of the doubly occupied (1,0,2) state of qubit A with B. The converse is true for qubit B, i.e. gate times decrease faster moving from the  central region towards the (2,0,1) boundary, compared to the (1,0,2) boundary.

The AEON works at the $\varepsilon=0, \varepsilon_{\mathrm{m}} = U^{\prime \prime} - U^\prime = -0.9$~meV double 1QSS~\cite{Shim2016}, and is independent of tunnel coupling.
The RX operates within the upper triangular region (Fig.~\ref{fig:tqd}(c))~\cite{Medford2013, Russ2015}. We take the RX operating point to be at $\varepsilon=0$, (which is a 1QSS for symmetric tunnel coupling), and $\varepsilon_{\mathrm{m}} = -0.57$~meV. 
Asymmetric tunnel couplings shift the RX $\varepsilon$ SS. We assume tunnel couplings are tunable; they can be tuned to the desired ratio, if necessary, during non-local gating and back to equal tunnel coupling (for RX SS) during single-qubit operations.
Consequently, the fastest ($k=1$) non-local gate times for RX and AEON are 64~ns and 450~ns respectively, for numerical parameters used. The choice of $\varepsilon_{\mathrm{m}}$ for RX is slightly arbitrary; it is possible for the RX gate time to be faster with larger $\varepsilon_{\mathrm{m}}$, thereby moving the operating point closer to the doubly occupied regime while remaining in the (1,1,1) configuration. The trade-off is greater susceptibility to charge noise.


\subsection{Two-qubit gate fidelity}

Noisy two-qubit evolution given by 
   \begin{align}\label{eq:Uint}
\tilde{U}_\mathrm{int}(t_0) = \exp\left \{-\frac{\mathrm{i}}{\hbar} \int_0^{t_0} \tilde{H}_\mathrm{int}(t^\prime) ~dt^\prime  \right \},
 \end{align}   
can be decomposed into ideal and noisy evolution,
\begin{widetext}
   \begin{align}
\tilde{U}_\mathrm{int} (t_0) =&  \hat{U}_\mathrm{int}(t_0) \hat{U}_\mathrm{\delta}(t_0) \\
=&
\begin{pmatrix}
\mathrm{e}^{-{\mathrm{i}} V_1 t_0} & 0 & 0 & 0\\
0 & \mathrm{e}^{-{\mathrm{i}} V_2 t_0} & 0 & 0\\
0 & 0 & \mathrm{e}^{-{\mathrm{i}} V_3 t_0} & 0\\
0 & 0 & 0 & \mathrm{e}^{-{\mathrm{i}} V_4 t_0}
\end{pmatrix} 
\begin{pmatrix}
\mathrm{e}^{-{\mathrm{i}} \int_0^{t_0} \delta V_1(t^\prime) dt^\prime} & 0 & 0 & 0\\
0 & \mathrm{e}^{-{\mathrm{i}} \int_0^{t_0} \delta V_2(t^\prime) dt^\prime} & 0 & 0\\
0 & 0 & \mathrm{e}^{-{\mathrm{i}} \int_0^{t_0} \delta V_3(t^\prime) dt^\prime} & 0\\
0 & 0 & 0 & \mathrm{e}^{-{\mathrm{i}} \int_0^{t_0} \delta V_4(t^\prime) dt^\prime}
\end{pmatrix}. \nonumber
 \end{align}   
\end{widetext}

We use the two-qubit gate fidelity~\cite{Wang2008}, $ \mathcal{F}  = \frac{1}{d^2}   | \text{Tr} (U^\dag \tilde{U} ) |^2  $, where $d$ is dimensionality, to evaluate the effect of charge noise. We compute the gate fidelity for exact CPHASE and CNOT gate sequences (Fig.~\ref{fig:gateresults}(a)), averaged over noisy two-qubit interactions. Since  CPHASE and CNOT  gates share the same non-local interaction, the fidelity expression for both are identical. The average gate fidelity~\cite{Green2013} is
   \begin{align}\label{eq:fid}
\langle \mathcal{F}\rangle
=& \frac{1}{16} \left \langle \left | \text{Tr}  \left( \tilde{U}_\mathrm{\delta}(t_0) \right) \right|^2 \right \rangle \nonumber \\
=&
\frac{1}{16}\sum_{i,j}\ev{
  \mathrm{e}^{
    -\mathrm{i} \int_0^{t_0}\dd{t'}
    \pqty{\delta V_i(t')-\delta V_j(t')}
  }
}.
 \end{align}   
This formula is used for all our numerical simulations. We note that this formula is identical to process fidelity~\cite{Nielsen2010}; the averaging over noise realizations done in our numerical simulations and analytical calculations  correspond to the  measurement protocols performed in quantum process tomography experiments (see Supplementary Note 4).

\subsection{Analytical fidelity formula}

Next, we derive an approximate analytical expression for average gate fidelity. Assuming stationary, Gaussian noise  with zero mean, and making use of the series expansion to linear order, $\delta V_i = \sum_{j}\frac{\partial V_i}{\partial n_j}\delta n_j$, we perform a cumulant expansion~\cite{Kubo1962, Kubo1963} of Eq.~\eqref{eq:fid} to obtain an analytical expression for average gate fidelity,
\begin{widetext}
\begin{equation}
	\begin{aligned}
		\ev   {\mathcal{F_\mathrm{an}}} 
		&=
		\frac{1}{4}\pqty{1+\exp\Bqty{
				-\frac{1}{2}\sum_{i=1}^{4}\pqty{\pqty{\bm{F}_\mathrm{A}-C_{11}\bm{1}}^\mathrm{T} \frac{\partial \bm{\xi}}{\partial n_i}}^2\sigma_{n_{i}}^2
				\varsigma( t_0 )
		}} \\
		& \times \pqty{1+\exp\Bqty{
				-\frac{1}{2}\sum_{i=5}^{8}\pqty{\pqty{\bm{F}_\mathrm{B}-C_{11}\bm{1}}^\mathrm{T} \frac{\partial \bm{\nu}}{\partial n_i}}^2\sigma_{n_{i}}^2
				\varsigma( t_0 )
		}}.
	\end{aligned}\label{eq:fidelity-formula}
\end{equation}
\end{widetext}
The derivation is detailed in Supplementary Note 5. Here $\varsigma(t_0) = \int_0^{t_0}\dd{t'}\int_0^{t'}\dd{t''} \overline{C}(t'')$  is the double integral of the noise correlation function, normalized to $\overline{C}(0) = 1$, and $\sigma_{n_{i \mathrm{A/B}}}$ is the noise standard deviation in first/last four $\bm{n}$ components for qubit A/B. The terms $\bm{F}_\mathrm{A}, \bm{F}_\mathrm{B}, \bm{\xi}, \bm{\nu}$  comprise linear combinations of Coulomb integrals and are detailed in Supplementary Method 3. The term $C_{11} \equiv \sum_{i=1}^{3} \sum_{j=1}^{3} \mathcal{V}_{ij}$ comprises all possible pairwise Coulomb energies between electrons of qubits A and B in the $(1,1,1)_{\mathrm{A}}-(1,1,1)_{\mathrm{B}}$ configuration.
The function $\varsigma(t_0)$ is general; it can be obtained from any  noise power spectrum. In this study, $\varsigma(t_0)$ is calculated from the $1/f$ power spectrum in Eq.~\eqref{eq:S(w)}. The exact expression is given by Supplementary Eq.~(101) and its derivation shown in Supplementary Note 2.


\subsection{Two-qubit sweet spots (2QSS)}

Simplification of the noise terms of Eq.~\eqref{eq:fidelity-formula} yields  insights into the existence of 2QSS. 
A 2QSS is defined to be a point in parameter space for which all partial derivatives, $\partial V_i / \partial n_j = 0$, for a particular parameter $n_j$. At these points, each qubit is protected from noise in parameter $n_j$. 
Firstly, interaction terms $V_i$ for $i=2,3,4$, 
contain charge admixtures through the 
$|1^\prime_{\mathrm{A/B}}\rangle$ state; these admixtures depend on TQD parameters which couple to charge noise. This means $\delta V_1=0$ because the $|0_{\mathrm{A}} 0_{\mathrm{B}}\rangle$ state does not contain  admixtures. Secondly, $|0_\mathrm{A} 1_{\mathrm{B}}^\prime \rangle$ and $|1_{\mathrm{A}}^\prime 0_{\mathrm{B}} \rangle$ states contain admixtures in qubits 
B and A respectively, implying that $\delta V_2$ and $\delta V_3$ are only dependent on noise in the respective qubits. 
Explicitly  evaluating the expansions 
$\delta V_2 \approx \sum_j \frac{\partial V_2}{\partial n_j} \delta {n_j}=  \frac{\partial V_2}{\partial \varepsilon_{\mathrm{B}}}\delta \varepsilon_{\mathrm{B}} +  \frac{\partial V_2}{\partial \varepsilon_{\mathrm{mB}}}\delta \varepsilon_{\mathrm{mB}} +  \frac{\partial V_2}{\partial t_{\mathrm{lB}}} \delta t_{\mathrm{lB}} +   \frac{\partial V_2}{\partial t_{\mathrm{rB}}} \delta t_{\mathrm{rB}}$,
and
$\delta V_3 \approx  \sum_j \frac{\partial V_3}{\partial n_j} \delta {n_j}=  \frac{\partial V_3}{\partial \varepsilon_{\mathrm{A}}}\delta \varepsilon_{\mathrm{A}} +  \frac{\partial V_3}{\partial \varepsilon_{\mathrm{mA}}}\delta \varepsilon_{\mathrm{mA}} +  \frac{\partial V_3}{\partial t_{\mathrm{lA}}} \delta t_{\mathrm{lA}} +   \frac{\partial V_3}{\partial t_{\mathrm{rA}}} \delta t_{\mathrm{rA}}$,
we indeed find that noise in 
qubit B (A)
affects only $\delta V_2~(\delta V_3)$. 
In truncating the series expansions to terms linear in fluctuating parameters, we have assumed the weak noise limit. In addition,  we keep only leading order terms within the partial derivatives in the perturbative limit. As a result, $\delta V_4 \approx  \delta V_2 + \delta V_3$. These  approximations allow us to separate the contributions of noise in the two qubits.
Explicit forms of the  derivatives are provided in Supplementary Note 5.

The locus of 2QSS are shown as dashed lines in detuning space of qubit A in Fig.~\ref{fig:gateresults}(b,e). As described in the preceding section, equivalent color plots for qubit B are reflections about the $\varepsilon=0$ line, and the discussion about qubit A here also applies to qubit B.
The $\varepsilon_{\mathrm{mA}}$  2QSS for qubit A is indicated as the white dashed line in Fig.~\ref{fig:gateresults}(b,e). The angle of the $\varepsilon_\mathrm{mA/B}$ 2QSS line depends on tunnel coupling ratio, and equations of the lines are
\begin{widetext}
   \begin{align}
\frac{t_\mathrm{lA}}{t_\mathrm{rA}}
		&= \sqrt{\frac{C_{51}-C_{11}}{C_{21}-C_{11}}}\pqty{\frac{U-2U'+U''-\varepsilon_\mathrm{mA}-\varepsilon_\mathrm{A}}{U-2U'+U''-\varepsilon_\mathrm{mA}+\varepsilon_\mathrm{A}}}^{\frac{3}{2}}, & (\varepsilon_{\mathrm{mA}} ~\text{ 2QSS)} \label{eq:epsmA2QSS}, \\
\frac{t_\mathrm{lB}}{t_\mathrm{rB}}
	&=
	\sqrt{\frac{C_{13}-C_{11}}{C_{14}-C_{11}}}\pqty{
		\frac{U-2U'+U''-\varepsilon_\mathrm{mB}-\varepsilon_\mathrm{B}}{U-2U'+U''-\varepsilon_\mathrm{mB}+\varepsilon_\mathrm{B}}
	}^{\frac{3}{2}}, & (\varepsilon_{\mathrm{mB}} ~\text{ 2QSS)}.\label{eq:epsmB2QSS}
 \end{align}   
\end{widetext}
Here, $C_{11}, C_{51}, C_{21}, C_{13}$ and $C_{14}$ comprise all pairwise Coulomb energies between electrons of the TQDs for $(1,1,1)_{\mathrm{A}}-(1,1,1)_{\mathrm{B}}$, $(0,2,1)_{\mathrm{A}}-(1,1,1)_{\mathrm{B}}$, $(1,0,2)_{\mathrm{A}}-(1,1,1)_{\mathrm{B}}$, $(1,1,1)_{\mathrm{A}}-(2,0,1)_{\mathrm{B}}$, and $(1,1,1)_{\mathrm{A}}-(1,2,0)_{\mathrm{B}}$ configurations respectively. 

Left and right tunneling 2QSS are lines close to and parallel to the (1,2,0) and (0,2,1) to (1,1,1) detuning boundaries respectively. Working at these tunneling 2QSS require non-local gate times of the order $10 - 100 ~\mu\mathrm{s}$ in general, so the protection from tunneling noise has to be weighed against decoherence  from long gate times. Linecuts along tunneling 2QSS are shown in Fig.~\ref{fig:gateresults}(d,g). At the intersection of left and right tunneling 2QSS lines, the conditions imposed by Makhlin invariants for non-local gates are not satisfied and Eq.~\eqref{eq:t0} gives non-physical infinite gate time. It is therefore not possible to take advantage of the double tunneling 2QSS.  Equations of lines corresponding to tunneling 2QSS are    \begin{align}
\varepsilon_{\mathrm{mA}} =& -\varepsilon_{\mathrm{A}} - 2 (U^\prime - U^{\prime \prime}), &(t_{\mathrm{lA}} \text{ 2QSS)} \label{eq:tlA2QSS}, \\
\varepsilon_{\mathrm{mA}} =& \varepsilon_{\mathrm{A}} - 2 (U^\prime - U^{\prime \prime}), &(t_{\mathrm{rA}} \text{ 2QSS)} \label{eq:trA2QSS}, \\
\varepsilon_{\mathrm{mB}} =& \varepsilon_{\mathrm{B}} - 2 (U^\prime - U^{\prime \prime}), &(t_{\mathrm{lB}}\text{ 2QSS)} \label{eq:tlB2QSS}, \\
\varepsilon_{\mathrm{mB}} =& -\varepsilon_{\mathrm{B}} - 2 (U^\prime - U^{\prime \prime}). &(t_{\mathrm{rB}}\text{ 2QSS)}\label{eq:trB2QSS}.
 \end{align}   
The  $\varepsilon_{\mathrm{mA/B}}$ and $t_{\mathrm{lA/B}}$ double 2QSS conditions for the qubits are given by the solution to the relevant pairs of simultaneous equations (Eqs.~\eqref{eq:epsmA2QSS}, \eqref{eq:tlA2QSS} and Eqs.~\eqref{eq:epsmB2QSS}, \eqref{eq:tlB2QSS}). There is no $\varepsilon$ 2QSS in the linear QD geometry. The full derivation of the 2QSS equations is given in Supplementary Note 6.


\subsection{Infidelity results}

\begin{figure*}
\centering
\includegraphics[width=0.75\textwidth]{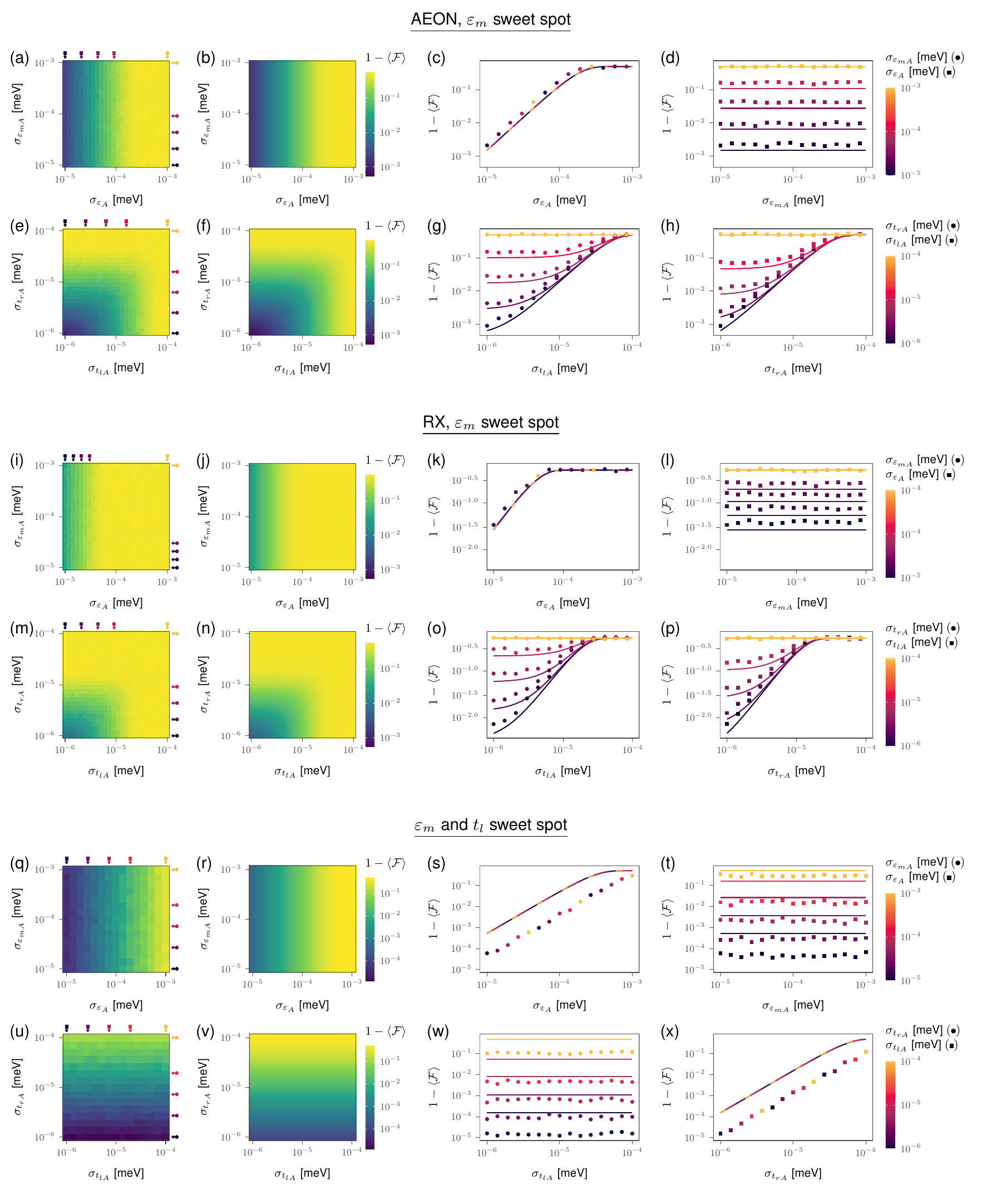}
\caption{ \linespread{1}\selectfont{}
Gate infidelities under detuning or tunneling noise and 2QSS.
(a--h) Infidelity plots of  AEON with qubit A  in the presence of only detuning noise (panels (a--d)) and only tunneling noise  (panels (e--h)), while sitting at its single-qubit double sweet spot and the $\varepsilon_{\mathrm{m}}$ 2QSS. 
(i--p) Infidelity plots of  RX ($k=7$, for a comparable gate time with AEON) with qubit A in the presence of only detuning noise (panels (i--l)) and only tunneling noise  (panels (m--p)), while sitting at its single-qubit single sweet spot and the $\varepsilon_{\mathrm{m}}$ 2QSS. 
(q--x) Infidelity plots with qubit A in the presence of only detuning noise  (panels (q--t)) and only tunneling noise  (panels (u--x)), while sitting at the $\varepsilon_{\mathrm{m}}$ and $t_{\mathrm{l}}$ 2QSS. 
Panels (a,e,i,m,q,u) are numerical simulations (Eq.~\eqref{eq:fid}) averaged over 500 (panels a,e,i,m) and 100 (panels q, u) noise realizations.
Panels (b,f,j,n,r,v)  are analytical calculations (Eq.~\eqref{eq:fidelity-formula}) and agree well with  numerical simulations, as evident from comparisons with  line cuts. 
Corresponding analytical and  numerical plots share the same color scale.
Panels (c, k, s) show  horizontal line cuts (circles), and panels (d, l, t) show vertical line cuts (squares) from the numerical result in panels (a, i, q), and agree very well with analytical calculations (lines) at the $\varepsilon_{\mathrm{m}}$ 2QSS, while analytical infidelity overestimates by an order of magnitude for the $\varepsilon_{\mathrm{m}}$ and $t_{\mathrm{l}}$ double 2QSS. 
Panels (d, l, t) show that infidelity is independent of middle detuning noise $\sigma_{\varepsilon_{\mathrm{mA}}}$ which confirms the $\varepsilon_{\mathrm{mA}}$ 2QSS.
Panel (w) shows that infidelity is also independent of the left tunneling noise $\sigma_{t_{\mathrm{lA}}}$ which confirms that it is the  double 2QSS of $\varepsilon_{\mathrm{mA}}$ and $t_{\mathrm{lA}}$. 
In contrast, infidelity increases with noise detuning $\sigma_{\varepsilon_\mathrm{A}}$ (panels c, k, s) because there is no such 2QSS, and in right tunneling $\sigma_{t_{\mathrm{rA}}}$ when not operated at the $t_{\mathrm{rA}}$ 2QSS (panels h, p, x). 
Color scales in rightmost column represent the numerical value of noise amplitude at which the line cuts are taken.
Results for noisy qubit B or both qubits noisy are qualitatively similar (see Supplementary Discussion 2).
}\label{fig:fid-results}
\end{figure*}

Numerical simulations of Eq.~\eqref{eq:fid} with noisy parameters are averaged to give infidelity $1- \left \langle \mathcal{F} \right \rangle$, and compared with that from analytics, $1- \left \langle \mathcal{F_\mathrm{an}} \right \rangle$. Results are plotted in Fig.~\ref{fig:fid-results} for noise affecting the non-local gate for qubit A only. The reason is that  results for noisy qubit B or both qubits noisy are qualitatively similar and the analytic fidelity acquires a simple expression,  $\langle \mathcal{F_\mathrm{an}} \rangle = \langle \mathcal{F_\mathrm{an}} \rangle_\mathrm{A} \times \langle \mathcal{F_\mathrm{an}} \rangle_\mathrm{B}  $ (see Supplementary Discussion 2).

Infidelity results are shown in Fig.~\ref{fig:fid-results}, which we refer to in the rest of this section. Numerical simulations are calculated by averaging over 500 noise realizations for every point in detuning space, except at double $t_{\mathrm{l}}$ and $\varepsilon_{\mathrm{m}}$ 2QSS which are averaged over 100 realizations. The reason is the much longer gate times (up to $\sim 100 \mu$s) at those points result in impractical runtimes.  Infidelities from numerical simulations are shown in the leftmost column (Fig.~\ref{fig:fid-results}(a,e,i,m,q,u)), analytical infidelity (Eq.~\eqref{eq:fidelity-formula}) in the second column (Fig.~\ref{fig:fid-results}(b,f,j,n,r,v)). Each pair of infidelity color plots have the same scale to facilitate comparison.

Horizontal linecuts (circles) from numerical infidelity results are plotted with analytical infidelity (lines) in the third column (Fig.~\ref{fig:fid-results}(c,g,k,o,s,w)). Similarly, vertical linecuts (squares) are plotted in the rightmost column (Fig.~\ref{fig:fid-results}(d,h,l,p,t,x)). Color scales for linecuts indicate the standard deviation of the noisy parameter at which the linecut was taken. All plots display good agreement between numerical and analytical calculations, except at  $\varepsilon_{\mathrm{m}}$ and $t_{\mathrm{l}}$ double 2QSS where $t_0$ increases significantly and the analytical formula over-predicts infidelity.

We analyze infidelity  due to noise in detuning vs tunneling parameters separately because the contribution to infidelity depends on noise variance as well as the first derivative of the squares of charge admixtures.  The latter yields terms that are  smaller by a factor of $\sim t/(U+\varepsilon)$ for detuning derivatives compared to tunneling derivatives (see Supplementary Eqs.~(134, 135), Supplementary Note 5). When comparing  infidelities from only detuning noise with that from only tunneling noise at the same working point, e.g. between panels (a,b) and (e,f), or between panels (i,j) and (m,n) of Fig.~\ref{fig:fid-results}, it is clear that infidelity is less sensitive to noise in detuning than tunneling.

\begin{figure*}
\centering
\includegraphics[width=0.8\textwidth]{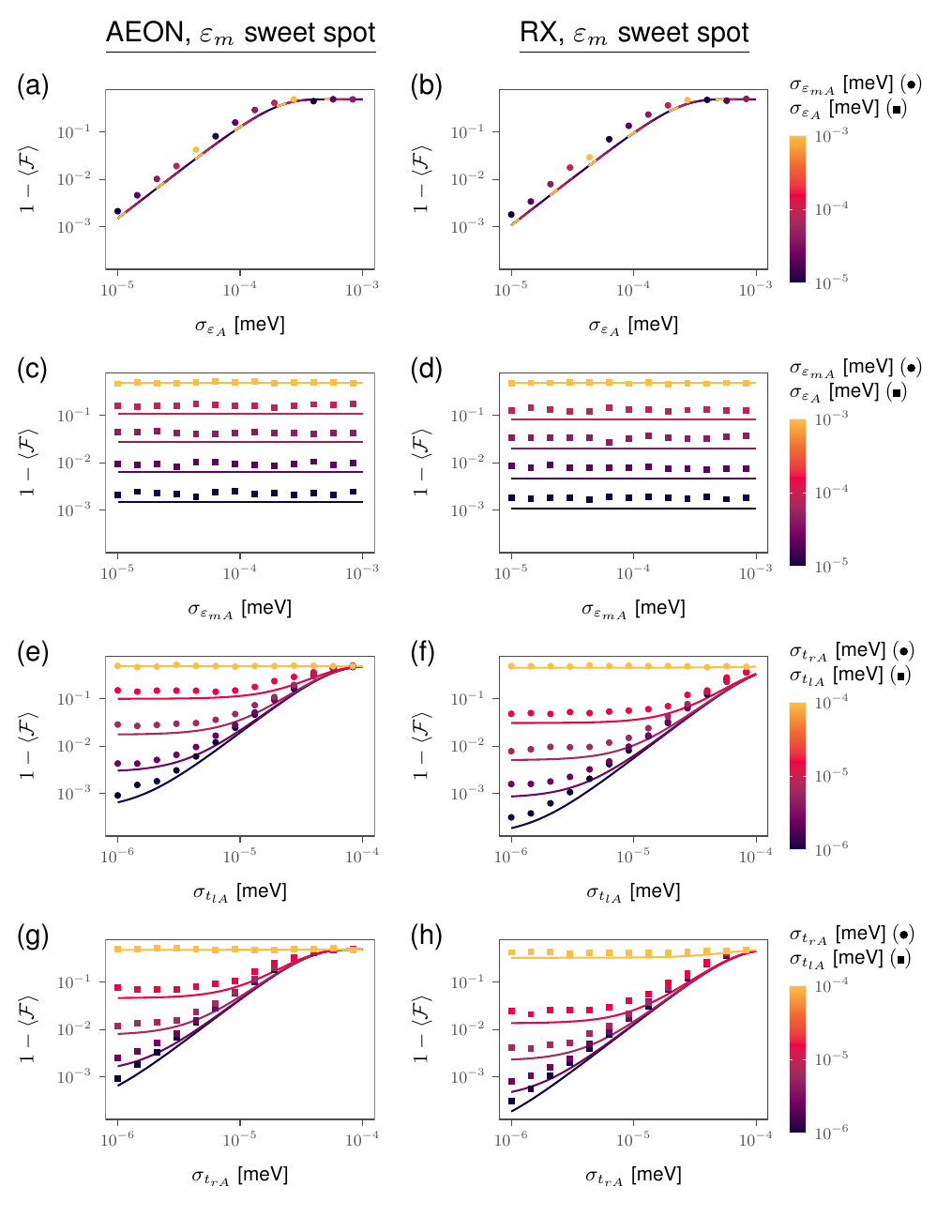}
\caption{ \linespread{1}\selectfont{}
Comparison of infidelities for the fastest gate times ($k=1$) for the AEON ($t_0 = 450$~ns) and RX ($t_0 = 64$~ns), when the 
$\varepsilon_{\mathrm{mA}}$ 2QSS overlaps with the $\varepsilon_{\mathrm{A}} = 0$ line. Left column (panels a,c,e,g) are identical with panels (c,d,g,h) of Fig.~\ref{fig:fid-results}. Right column (panels b,d,f,h) are similar to panels (k,l,o,p) of Fig.~\ref{fig:fid-results}, except that they are calculated for $k=1$. Comparing each row, it is clear that the RX performs slightly better than the AEON, as expected for a faster qubit.
}\label{fig:k=1}
\end{figure*}

Now, we examine the case when the RX and AEON qubits are operated at $\varepsilon_{\mathrm{m}}$ 2QSS and $\varepsilon$ 1QSS. When $\arctan(t_{\mathrm{lA}}/t_{\mathrm{rA}})=57^\circ$, the $\varepsilon_{\mathrm{mA}}$ 2QSS lies on $\varepsilon_\mathrm{A} =0$ 1QSS. Vertical linecuts across $\sigma_{\varepsilon_{\mathrm{mA}}}$ (panels d, l) show excellent agreement between analytical and numerical calculations. In order to fairly compare  RX and AEON infidelities at these points, we chose $k$ (Eq.~\eqref{eq:t0}) such that they have comparable gate times of 450~ns ($k=1$, AEON) and 448~ns ($k=7$, RX). For comparable gate times, it is favorable to work at the AEON 1QSS, suggesting that while the RX operates faster due to a greater dipole-dipole interaction, it is more susceptible to charge noise. However, with the fastest gate time ($k=1$) of 64~ns for RX, a slightly better fidelity (Fig.~\ref{fig:k=1}(b,d,f,h)) is achieved, demonstrating the natural relationship between fast gates and improved fidelity.

Next, when the $\varepsilon_{\mathrm{mA}}$ 2QSS line is slanted, e.g. for $\arctan(t_{\mathrm{lA}}/t_{\mathrm{rA}})=45^\circ$, it intersects the $t_{\mathrm{lA}}$ 2QSS. At this double 2QSS, where gate time is longer than when operating on RX or AEON 1QSS, vertical linecuts along $\sigma_{\varepsilon_{\mathrm{mA}}}$ (Fig.~\ref{fig:fid-results}(t)) and horizontal linecuts along $\sigma_{t_{\mathrm{lA}}}$ (Fig.~\ref{fig:fid-results}(w)) display constant infidelities, demonstrating the double 2QSS character of the working point. However, the analytical formula over-predicts infidelity compared with numerical simulations, as shown in (Fig.~\ref{fig:fid-results}(s,t,w,x)).

Finally, there is no $\varepsilon_\mathrm{A}$ 2QSS.  As expected, linecuts (Fig.~\ref{fig:fid-results}(c, g, h, k, o, p)) show rising infidelity with greater $\sigma_{\varepsilon_\mathrm{A}}$, and excellent analytical and numerical agreement.

\section{Discussion}

Here, we describe optimal choices of 2QSS working points, all of which fall on the $\varepsilon_{\mathrm{m}}$ 2QSS. Firstly, if it is desired that qubits are protected from noise during both single- and two-qubit operation with minimal experimental control, then it is best to work on the intersection of $\varepsilon_{\mathrm{m}}$ 2QSS and $\varepsilon$ 1QSS. This requires tunnel coupling ratios to be tuned to $\arctan(t_{\mathrm{lA}}/t_{\mathrm{rA}}) = 57^\circ$ during non-local operations. At this ratio, the $\varepsilon_{\mathrm{mA}}$ 2QSS is a vertical line passing through AEON and RX $\varepsilon$ 1QSS. During single-qubit gates, in order to return to the RX 1QSS, tunnel couplings must be re-equalized. The AEON does not require retuning since its 1QSS is independent of tunnelling.

\begin{figure*}
\centering
\includegraphics[width=0.8\textwidth]{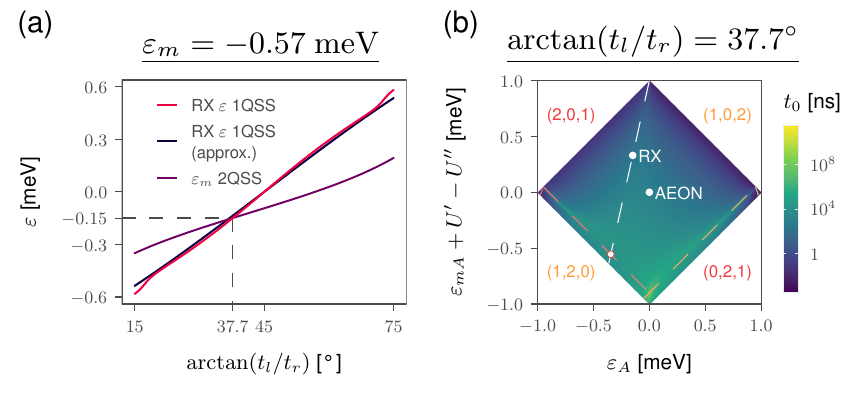}
\caption{\linespread{1}\selectfont{}
RX single-qubit sweet spot (1QSS) and two-qubit sweet spot (2QSS) dependence on tunneling ratio. (a) The red curve shows the exact dependence of RX $\varepsilon$ 1QSS on $\arctan(t_{\mathrm{l}}/t_{\mathrm{r}})$. The blue line, from Ref.~\cite{Taylor2013}, shows the approximate dependence, $\varepsilon\approx -(8\Delta/5)y$, where $\Delta = U-2U'+U''-\varepsilon_{\mathrm{m}}$ and small tunneling asymmetry $y = \sin(\pi/4-\arctan(t_{\mathrm{l}}/t_{\mathrm{r}}))$. The purple curve shows the $\varepsilon_{\mathrm{m}}$ 2QSS position as a function of $\arctan(t_{\mathrm{l}}/t_{\mathrm{r}})$. At the intersection where both 1QSS and 2QSS share the same tunneling ratio, $\varepsilon = -0.15~\mathrm{meV}$, $\arctan(t_{\mathrm{l}}/t_{\mathrm{r}}) = 37.7^\circ$, for parameters used in this study. (b) The gate times at the tunneling ratio $\arctan(t_{\mathrm{l}}/t_{\mathrm{r}}) = 37.7^\circ$. At the intersection of $\varepsilon_{\mathrm{m}}$ 2QSS (white dashed line) and RX 1QSS, non-local gate time is $t_0 = 198~\mathrm{ns}$. 
}\label{fig:RX_12QSS}
\end{figure*}

Further optimization for RX can be done. Because tunnelling asymmetry shifts its $\varepsilon$ 1QSS, this can be made to coincide with the $\varepsilon_{\mathrm{m}}$ 2QSS for one particular tunneling ratio, $\arctan(t_{\mathrm{l}}/t_{\mathrm{r}}) = 37.7^\circ$ for the parameters used. This is presented in Fig.~\ref{fig:RX_12QSS}, where panel (a) plots 1QSS and 2QSS against tunneling ratio, and panel (b) illustrates where they coincide on a color plot of non-local gate time. Left of the midline, gate time is longer, $t_0 = 198$~ns.

Next, even when these advantages cannot be exploited due to say, limited tunability of tunnel coupling, the intersection of $\varepsilon_{\mathrm{m}}$ 2QSS line with one of the tunneling 2QSS provides another avenue for fidelity improvement. This  intersection between $\varepsilon_{\mathrm{m}}$ and $t_{\mathrm{l}}$ 2QSS is shown by the white circle at the lower left of Fig.~\ref{fig:gateresults}(b, e).  Comparing infidelity plots in Fig.~\ref{fig:fid-results}, the best fidelities are found when working at the $\varepsilon_{\mathrm{m}}$ and $t_{\mathrm{l}}$ double 2QSS.  The experimental complication to working at this point is the movement to and from the 1QSS, during single-qubit gates.

The conditions for the existence of both $t_{\mathrm{l}}$ and $t_{\mathrm{r}}$ 2QSS are equivalent to the requirement for the qubit to remain in the (1,1,1) region, $-(U-U^\prime)< (\varepsilon_{\mathrm{mA}}+U^\prime - U^{\prime \prime}) \pm \varepsilon_{\mathrm{A}} < U-U^\prime$. The conditions for $\varepsilon_{\mathrm{mA}}$ and $\varepsilon_{\mathrm{mB}}$ 2QSS to exist are $\text{sgn} (C_{51} - C_{11}) = \text{sgn} (C_{21} - C_{11})$ and $\text{sgn} (C_{13} - C_{11}) = \text{sgn} (C_{14} - C_{11})$ respectively. These conditions are automatically satisfied in a linear QD array. The full derivation is given in Supplementary Note 6.

\begin{figure*}
\centering
\includegraphics[width=0.6\textwidth]{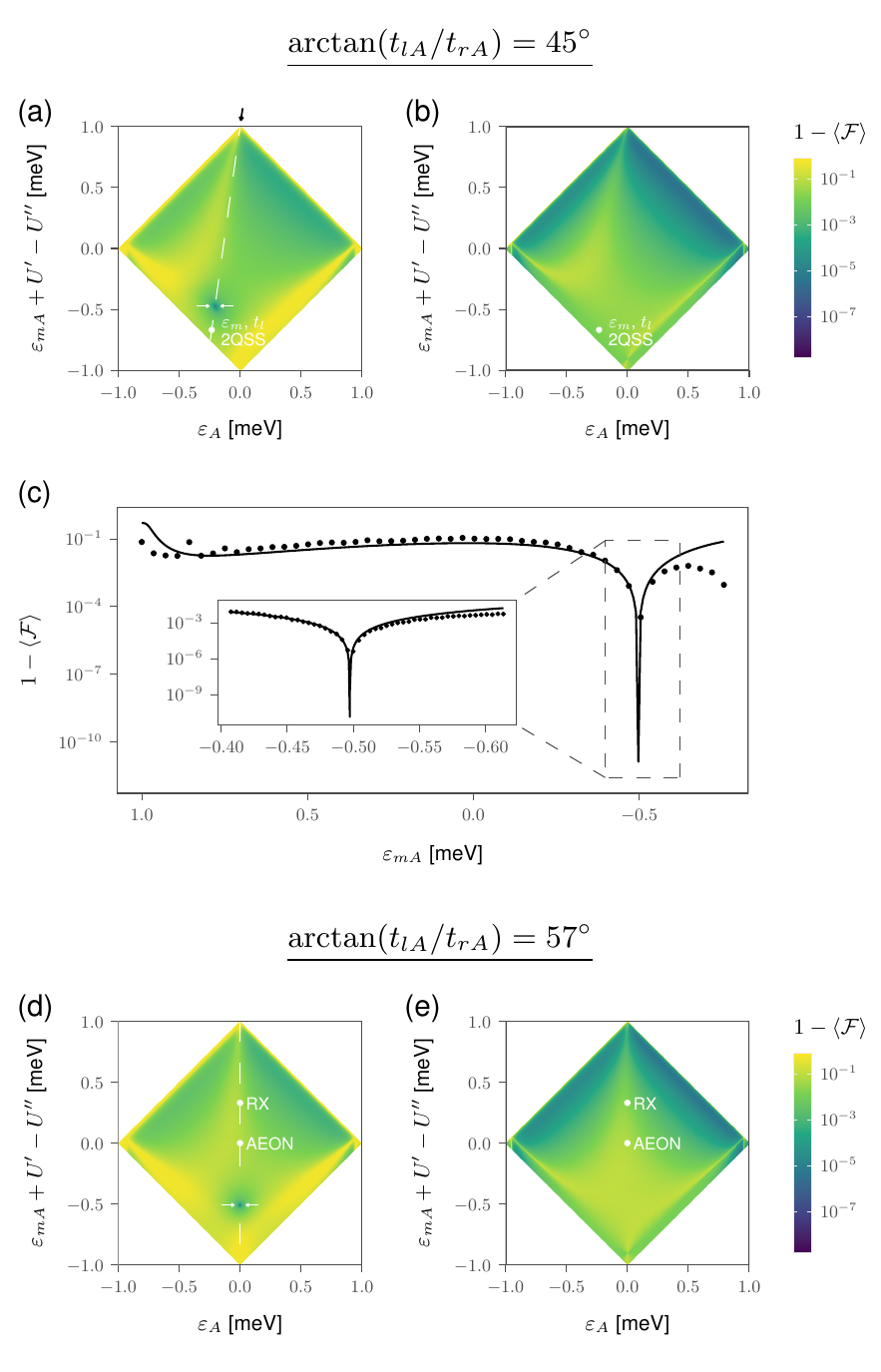}
\caption{\linespread{1}\selectfont{}
Color plots of analytical formula (Eq.\eqref{eq:fidelity-formula}) of infidelity  in detuning parameter space with fixed standard deviation of noise. There is only detuning noise, $\sigma_{\varepsilon_{\mathrm{mA}}} = \sigma_{\varepsilon_\mathrm{A}} = 10^{-4}~\textrm{meV}$ for (a,c,d), and only tunneling noise, $\sigma_{t_{\mathrm{lA}}} = \sigma_{t_{\mathrm{rA}}} = 10^{-5}~\textrm{meV}$ for (b,e).
Tunneling parameters are $t_{\mathrm{lA}} = t_{\mathrm{rA}}$ in panels (a,b,c), and $\arctan(t_{\mathrm{lA}} /t_{\mathrm{rA}})=57^\circ$ in panels (d,e). (a) The global optimum when there is only detuning noise lies along the $\varepsilon_{\mathrm{m}}$ 2QSS. This global optimum is in a region of large $\varepsilon_{\mathrm{m}}$ for tunneling noise.The global optimum lies at a point which has extremely small gate times that require timing precision of $\sim$ps or better, which may be currently out of reach experimentally.  (b) When there is only tunneling noise, the global optimum lies in the region near the upper right boundary of the (1,1,1) region. The optimal point in panel (a) is now a point with significantly larger infidelity ($1- \langle\mathcal{F_\mathrm{an}} \rangle \approx 10^{-2}$). (c) Infidelity linecut along the $\varepsilon_{\mathrm{m}}$ 2QSS. The analytical  (line) and numerically simulated (points) infidelities agree well, although they start to deviate past the global optimum. Global infidelity optimum is better than $10^{-10}$ from analytical calculations, and $10^{-6}$ from numerical simulations.
(d) With only detuning noise, and tuning tunneling parameters so that the $\varepsilon_{\mathrm{m}}$ 2QSS is along $\varepsilon = 0$, the global optimum still lies on the $\varepsilon_{\mathrm{m}}$ 2QSS. However the gate times become very large near bottom corner of the (1,1,1) region (see Fig.~\ref{fig:gateresults}) and becomes impractical to implement.
(e) With only tunneling noise and the $\varepsilon_{\mathrm{m}}$ 2QSS is along $\varepsilon = 0$, the infidelity again rises significantly at the region where it was the global optimum in panel (d) when there was only detuning noise. In reality, there should be noise in both detuning and tunneling, and infidelities are approximately additive,  demonstrating the difficulty of finding a truly global optimal working point.
}\label{fig:global-opt}
\end{figure*}

Having studied fidelities of specific working points on 2QSS, it is reasonable to ask if a global fidelity optimum might exist. In Fig.~\ref{fig:global-opt}, when there is only detuning noise in qubit A (panels a, c, d), the global fidelity optimum is a single point lying on $\varepsilon_{\mathrm{mA}}$ 2QSS, in the lower half of the (1,1,1) charge region. Analytical calculations show infidelity $\approx 10^{-10}$; numerical calculations give infidelity $\approx 10^{-6}$. Both meet fault-tolerance thresholds, $1-\mathcal{F} < 10^{-4}$~\cite{Aliferis2007, Aliferis2008} and $10^{-6}$~\cite{Aharonov2008}. This is significant because when tunneling noise is negligible, working at this global optimum will achieve fault-tolerance. 

However, when there is only tunneling noise, the global fidelity optimum is located in the upper right quadrant of the (1,1,1) region (panels b, e), and infidelity in the lower half of the (1,1,1) region increases significantly.  Because we expect both detuning and tunneling noise to affect EO qubits, and infidelity is approximately additive, the optimum point for global fidelity from detuning noise may  be limited  by infidelity from tunneling noise. 

Above, we analyzed results when qubit A is noisy. Similar results apply when qubit B or both qubits are noisy.
We also assumed noiseless single-qubit gates.  Next, we discuss fidelity optimization with noisy single-qubit gates.

CPHASE involves an additional $z$-rotation for each qubit. At simultaneous 1QSS and 2QSS, $z$-rotations times are $t_1 = t_2  \approx 1$~ns (AEON) and $0.4$~ns (RX) for $\hbar\omega_{\mathrm{A/B}}\approx 1$~meV. Both single-qubit $z$-gates are 2 orders of magnitude faster than the non-local gate. Therefore, at these sweet spot intersections, it is likely that the non-local gate limits fidelity.

As discussed, the global minimum need not lie on SS intersections, and is dependent on the dominant noise parameter. This necessitates a complete understanding of the noise power of each parameter. In addition, noise may be correlated. However, noise spectroscopy may not be trivial to implement since noise acts on multiple axes in these qubits, although theoretical progress in multiaxis noise spectroscopy had been made~\cite{Green2013, Paz-Silva2019}. Fortunately, the same parameter space governs single and two-qubit gates; perhaps a simple formula might relate single and two-qubit fidelities.

CNOT requires single-qubit $x$- and $z$-rotations. Because pulse gating for AEON rotates the qubit around the $-(x+z)$ axis, single-qubit rotations might benefit from optimal control pulses~\cite{Hanson2007a}. On the other hand, RX uses microwave control and can directly perform $x$-rotations, the speed of which depends on drive amplitude. Because AEON and RX can transform into each other, they should be able to take advantage of the features each one offers for further optimization.

In summary, we studied capacitive two-qubit CPHASE and CNOT gates for EO qubits with a focus on AEON and RX proposals. We demonstrated the existence of $\varepsilon_{\mathrm{m}}$, $t_{\mathrm{l}}$ and $t_{\mathrm{r}}$ 2QSS for each qubit, and provided conditions for their existence. We showed how the $\varepsilon_{\mathrm{m}}$ 2QSS can be tuned to intersect with $\varepsilon$ 1QSS, requiring only tuning of tunnel coupling ratios. This has the benefit of operating the qubit at both 1QSS and 2QSS. We also showed that double 2QSS also exist -- $\varepsilon_{\mathrm{mA}}$ with $t_{\mathrm{lA}}$ (qubit A) or $\varepsilon_{\mathrm{mB}}$ with $t_{\mathrm{rB}}$ (qubit B) --  providing another avenue for fidelity improvements. Importantly, the global fidelity optimum lies along the $\varepsilon_{\mathrm{m}}$ 2QSS, with a fidelity better than the fault tolerance threshold when tunneling noise is negligible.  

Our infidelity results illustrate the stringent requirement for qubits. Best fidelities are obtained when working at the double $\varepsilon_{\mathrm{mA}}$ and $t_{\mathrm{lA}}$ 2QSS. However, only with extremely low noise, e.g. $\sigma_{\varepsilon_\mathrm{A}} < 10^{-5}$ meV  or $\sigma_{t_{\mathrm{rA}}} < 10^{-5}$ meV, can the fault tolerance conditions be met. The fidelities in our study were computed for noisy non-local gate and noiseless single-qubit gates. In reality, because both qubits will be noisy and single-qubit gates will similarly be afflicted, the requirements are likely to be even stricter.

Given recent experimental progress in scaling up of QD  arrays and capacitive coupling, our results should contribute towards the realization of high fidelity two-qubit gates.

\section{Methods}

\subsection{Numerical simulations}

We numerically calculate the average fidelity (Eq.~\eqref{eq:fid}) of the non-local two-qubit gate by averaging over 500 different simulations of noise for each noisy parameter $\tilde{n}_i$, except at double $t_{\mathrm{l}}$ and $\varepsilon_{\mathrm{m}}$ 2QSS points which are averaged over 100 realizations. Each noisy time series $\delta n_i (t)$ is simulated with the desired spectrum of Eq.~\eqref{eq:S(w)} from the algorithm of Refs.~\onlinecite{Timmer1995, pycolorednoise}, which 
generates for every positive $\omega_k$ value, two Gaussian distributed random numbers to represent the real and imaginary parts of the spectrum. After scaling by $\sqrt{S(\omega_k)/2}$, an inverse FFT  produces the desired noisy time series. Our modification consists of shifting the mean of the generated time series to zero and then rescaling its variance to the desired value (see Supplementary Method 4).
At each time step, the charge admixtures $\alpha_i, \beta_i$ and  the interaction terms $\tilde{V}_i$ are computed, and the full unitary evolution with the exact gate sequence is  calculated.
All simulations in Figs.~\ref{fig:fid-results},~\ref{fig:k=1} are performed with cutoff frequencies, $\omega_\mathrm{l}/2\pi = $ 66.7~kHz, $\omega_\mathrm{h}/2\pi = $  50~GHz.

\subsection{Triple quantum dot potential and parameters}

The Hubbard model describes both intra-TQD and inter-TQD interactions. Intra-TQD interactions comprise QD detunings, tunnel couplings as well as intra and inter-dot Coulomb energies. Inter-TQD interaction comprise inter-dot Coulomb energies only, when tunnel coupling between the TQDs (QD 3 and 4) is zero. Parameters of inter-TQD interactions are calculated from a model confinement potential and intra-TQD parameters from estimates in literature. See Supplementary Note 7 for a discussion of this approach.

The TQD potential is modeled as a 2D tri-quadratic potential, $V_\text{pot} (\bm{r})= \min [v_{\text{pot},1}(\bm{r}),$ $v_{\text{pot},2}(\bm{r}), v_{\text{pot},3}(\bm{r}) ]$, where the $i$-th QD  well at position $\bm{R}_i$ is $v_{\text{pot},i} (\bm{r})= \frac{1}{2}m\omega_0^2 (|\bm{x}-\bm{R}_i|^2 + \bm{y}^2)+\varepsilon_i$, whose eigenfunctions are the 2D Fock-Darwin wavefunctions~\cite{Davies1997}. The 2D character of the potential is a good approximation to electrostatically gated QDs, given the tight confinement in the $z$-direction. The confinement $m \omega_0 = \hbar/ a_\mathrm{B}^2$, where $a_\mathrm{B}$ is the Bohr radius. Treating  neighboring potential wells as perturbations, dot-centered, normalized single-electron wavefunctions $\psi_i$ of the TQD potential are constructed from the Fock-Darwin wavefunctions using the method of L\"{o}wdin Orthogonalization~\cite{Annavarapu2013, Zhang2018}, from which the three-electron wavefunctions are  formulated.  Numerical values take reference from Refs.~\cite{DasSarma2011, Neyens2019a}: $a = 50$~nm, $a_\mathrm{B} = 25$~nm;  $R = 160$~nm; $U_i = U = 2.8$~meV; $U_{12} = U_{23} = U_{45} = U_{56} = U^\prime = 1.8$ meV; $U_{13} = U_{46} = U^{\prime \prime} = 0.9$~meV.
For both AEON and RX, 
$t_{\mathrm{lA}} = t_{\mathrm{lB}} = 0.12 $~meV, $\varepsilon_{\mathrm{A}} = \varepsilon_{\mathrm{B}} = 0$.
For AEON,
$\varepsilon_{\mathrm{mA}} = \varepsilon_{\mathrm{mB}} = -0.9$~meV.
For RX,
$\varepsilon_{\mathrm{mA}} = \varepsilon_{\mathrm{mB}} = -0.57$~meV.
Direct Coulomb integrals in the capacitive interaction are $\mathcal{V}_{ij} = (q^2 / 4\pi \epsilon_\mathrm{r} \epsilon_0) \int \int |\psi_i (\bm{r}_1)|^2 \frac{1}{|\bm{r}_1 - \bm{r}_2 |} |\psi_j (\bm{r}_2)|^2 d\bm{r}_1 d\bm{r}_2 $,
where we take silicon relative permittivity, $\epsilon_\mathrm{r} = 11.68$. Even though the exact form of QD confinement potential depends on the device, an advantage of modeling the TQD potential as tri-quadratic is that each integral is analytically tractable.
With these parameters, a check shows that  direct Coulomb integrals are a factor of $10^4$ greater than the spin-dependent exchange Coulomb integrals (see Supplementary Discussion 3), validating our assumption of capacitive non-local gating.

\section{Data Availability}

The data that support the findings of this study are available at https://doi.org/10.21979/N9/TYUUVS

\section{Code Availability}

The computer code used in generating the data are available from the corresponding author on reasonable request.

\section{Acknowledgements}

MKF was supported by a Singapore Ministry of Education AcRF Tier 1 grant (RG177/16), and acknowledges useful discussions with Jun Yoneda. LHZ was supported by the SGUnited program (CP0002392). We thank the Nanyang Technological University (NTU) High Performance Computing Centre for computing support, and the Division of Physics and Applied Physics, School of Physical and Mathematical Sciences, NTU, for financial support.

\section{Author contributions}
TSK and MKF contributed to the idea of capacitive two-qubit gates. MKF and LHZ performed the numerical simulations and analytical calculations. All authors analyzed the data and results. TSK wrote the main manuscript with input from all authors. MKF and LHZ wrote the supplementary material. The project was carried out under the supervision of TSK.

\section{Competing Interests}

The authors declare no competing interests.



\newpage

\begin{thebibliography}{99}

\bibitem[1]{Loss1998} Loss, D. \& DiVincenzo, D. P. Quantum computation with quantum dots. Phys. Rev. A \textbf{57}, 120–126 (1998).

\bibitem[2]{Zwanenburg2013}Zwanenburg, F. A. et al. Silicon quantum electronics. Rev. Mod. Phys. \textbf{85}, 961–1019 (2013). 

\bibitem[3]{Yoneda2018}Yoneda, J. et al. A quantum-dot spin qubit with coherence limited by charge noise and fidelity higher than 99.9\%. Nat. Nanotechnol. \textbf{13}, 102–106 (2018).

\bibitem[4]{Zajac2018}Zajac, D. M. et al. Resonantly driven CNOT gate for electron spins. Science \textbf{359}, 439–442 (2018). 

\bibitem[5]{Foletti2009}Foletti, S., Bluhm, H., Mahalu, D., Umansky, V. \& Yacoby, A. Universal quantum control of two- electron spin quantum bits using dynamic nuclear polarization. Nat. Phys. \textbf{5}, 903–908 (2009). 

\bibitem[6]{Bluhm2011a}Bluhm, H. et al. Dephasing time of GaAs electron-spin qubits coupled to a nuclear bath exceeding 200 $\mu$s. Nat. Phys. \textbf{7}, 109–113 (2011).

\bibitem[7]{Kim2014}Kim, D. et al. Quantum control and process tomography of a semiconductor quantum dot hybrid qubit. Nature \textbf{511}, 70–74 (2014).

\bibitem[8]{Kim2015}Kim, D. et al. High-fidelity resonant gating of a silicon-based quantum dot hybrid qubit. npj Quantum Inf. \textbf{1}, 15004 (2015).

\bibitem[9]{Thorgrimsson2017}Thorgrimsson, B. et al. Extending the coherence of a quantum dot hybrid qubit. npj Quantum Inf. \textbf{3}, 32 (2017).

\bibitem[10]{Cerfontaine2020}Cerfontaine, P. et al. Closed-loop control of a GaAs-based singlet-triplet spin qubit with 99.5\% gate fidelity and low leakage. Nat. Commun. \textbf{11}, 4144 (2020).

\bibitem[11]{Schroer2007}Schr\"{o}er, D. et al. Electrostatically defined serial triple quantum dot charged with few electrons. Phys. Rev. B \textbf{76}, 075306 (2007).

\bibitem[12]{Laird2010}Laird, E. A. et al. Coherent spin manipulation in an exchange-only qubit. Phys. Rev. B \textbf{82}, 075403 (2010).

\bibitem[13]{Gaudreau2012}Gaudreau, L. et al. Coherent control of three-spin states in a triple quantum dot. Nat. Phys. \textbf{8}, 54–58 (2012). 

\bibitem[14]{DiVincenzo2000}DiVincenzo, D. P., Bacon, D., Kempe, J., Burkard, G. \& Whaley, K.B. Universal quantum computation with the exchange interaction. Nature \textbf{408}, 339–342 (2000).

\bibitem[15]{Shi2012}Shi, Z. et al. Fast hybrid silicon double-quantum-dot qubit. Phys. Rev. Lett. \textbf{108}, 140503 (2012). 

\bibitem[16]{Koh2012}Koh, T. S., Gamble, J.K., Friesen, M., Eriksson, M. A. \& Coppersmith, S. N. Pulse-gated quantum-dot hybrid qubit. Phys. Rev. Lett. \textbf{109}, 250503 (2012).

\bibitem[17]{Russ2017}Russ, M. \& Burkard, G. Three-electron spin qubits. J. Phys. Condens. Matter \textbf{29}, 393001 (2017). 

\bibitem[18]{Srinivasa2015}Srinivasa, V. \& Taylor, J. M. Capacitively coupled singlet-triplet qubits in the double charge resonant regime. Phys. Rev. B \textbf{92}, 235301 (2015).

\bibitem[19]{Calderon-Vargas2015}Calderon-Vargas, F. A. \& Kestner, J. P. Directly accessible entangling gates for capacitively coupled singlet-triplet qubits. Phys. Rev. B \textbf{91}, 035301 (2015).

\bibitem[20]{Neyens2019a}Neyens, S. F. et al. Measurements of capacitive coupling within a quadruple-quantum-dot array. Phys. Rev. Appl. \textbf{12}, 064049 (2019).

\bibitem[21]{MacQuarrie2020}MacQuarrie, E. R. et al. Progress toward a capacitively mediated CNOT between two charge qubits in Si/SiGe. npj Quantum Inf. \textbf{6}, 81 (2020).

\bibitem[22]{Fujita2017}Fujita, T., Baart, T. A., Reichl, C., Wegscheider, W. \& Vandersypen, L. M. Coherent shuttle of electron-spin states. npj Quantum Inf. \textbf{3}, 22 (2017).

\bibitem[23]{Feng2018}Feng, M., Kwong, C. J., Koh, T. S. \& Kwek, L. C. Coherent transfer of singlet-triplet qubit states in an architecture of triple quantum dots. Phys. Rev. B \textbf{97}, 245428 (2018).

\bibitem[24]{Mills2019}Mills, A. R. et al. Shuttling a single charge across a one-dimensional array of silicon quantum dots. Nat. Commun. \textbf{10}, 1063 (2019).

\bibitem[25]{Yoneda2020}Yoneda, J., Huang, W., Feng, M. et al. Coherent spin qubit transport in silicon. Nat. Commun. \textbf{12}, 4114 (2021).

\bibitem[26]{Ginzel2020}Ginzel, F., Mills, A. R., Petta, J. R. \& Burkard, G. Spin shuttling in a silicon double quantum dot. Phys. Rev. B \textbf{102}, 195418 (2020).

\bibitem[27]{Mi2017}Mi, X. et al. Circuit quantum electrodynamics architecture for gate-defined quantum dots in silicon. App. Phys. Lett. \textbf{110}, 43502 (2017). 

\bibitem[28]{Stockklauser2017}Stockklauser, A. et al. Strong coupling cavity QED with gate-defined double quantum dots enabled by a high impedance resonator. Phys. Rev. X \textbf{7}, 11030 (2017).

\bibitem[29]{Fong2011}Fong, B. H. \& Wandzura, S. M. Universal quantum computation and leakage reduction in the 3-qubit decoherence free subsystem. Quantum Info. Comput. \textbf{11}, 1003–1018 (2011).

\bibitem[30]{Doherty2013}Doherty, A. C. \& Wardrop, M. P. Two-qubit gates for resonant exchange qubits. Phys. Rev. Lett. \textbf{111}, 050503 (2013).

\bibitem[31]{Shim2016}Shim, Y. P. \& Tahan, C. Charge-noise-insensitive gate operations for always-on, exchange-only qubits. Phys. Rev. B \textbf{93}, 121410(R) (2016).

\bibitem[32]{Paladino2014}Paladino, E., Galperin, Y., Falci, G. \& Altshuler, B. L. 1/ f noise: Implications for solid-state quantum information. Rev. Mod. Phys. \textbf{86}, 361–418 (2014).

\bibitem[33]{Fei2015}Fei, J. et al. Characterizing gate operations near the sweet spot of an exchange-only qubit. Phys. Rev. B \textbf{91}, 205434 (2015).

\bibitem[34]{Frees2019}Frees, A., Mehl, S., Gamble, J. K., Friesen, M. \& Coppersmith, S. N. Adiabatic two-qubit gates in capacitively coupled quantum dot hybrid qubits. npj Quantum Inf. \textbf{5}, 73 (2019).

\bibitem[35]{Medford2013}Medford, J. et al. Quantum-dot-based resonant exchange qubit. Phys. Rev. Lett. \textbf{111}, 050501 (2013).

\bibitem[36]{Taylor2013}Taylor, J. M., Srinivasa, V. \& Medford, J. Electrically protected resonant exchange qubits in triple quantum dots. Phys. Rev. Lett. \textbf{111}, 050502 (2013).

\bibitem[37]{Russ2015}Russ, M. \& Burkard, G. Asymmetric resonant exchange qubit under the influence of electrical noise. Phys. Rev. B \textbf{91}, 235411 (2015).

\bibitem[38]{Wardrop2016}Wardrop, M. P. \& Doherty, A. C. Characterization of an exchange-based two-qubit gate for resonant exchange qubits. Phys. Rev. B \textbf{93}, 075436 (2016).

\bibitem[39]{DasSarma2011}Das Sarma, S., Wang, X. \& Yang, S. Hubbard model description of silicon spin qubits: Charge stability diagram and tunnel coupling in Si double quantum dots. Phys. Rev. B \textbf{83}, 235314 (2011). 

\bibitem[40]{Schrieffer1966}Schrieffer, J. R. \& Wolff, P. A. Relation between the Anderson and Kondo Hamiltonians. Phys. Rev. \textbf{149}, 491–492 (1966).

\bibitem[41]{Gros1987}Gros, C., Joynt, R. \& Rice, T. M. Antiferromagnetic correlations in almost-localized Fermi liquids. Phys. Rev. B \textbf{36}, 381–393 (1987).

\bibitem[42]{MacDonald1988}MacDonald, A. H., Girvin, S. M. \& Yoshioka, D. t/U expansion for the Hubbard model. Phys. Rev. B \textbf{37}, 9753–9756 (1988).

\bibitem[43]{Chan2018}Chan, K. W. et al. Assessment of a silicon quantum dot spin qubit environment via noise spectroscopy. Phys. Rev. Appl. \textbf{10}, 44017 (2018).

\bibitem[44]{Struck2020a}Struck, T. et al. Low-frequency spin qubit energy splitting noise in highly purified $^{28}$Si/SiGe. npj Quantum Inf. \textbf{6}, 1–7 (2020).

\bibitem[45]{Russ2016}Russ, M., Ginzel, F. \& Burkard, G. Coupling of three-spin qubits to their electric environment. Phys. Rev. B \textbf{94}, 165411 (2016).

\bibitem[46]{Huang2018a}Huang, P., Zimmerman, N. M. \& Bryant, G. W. Spin decoherence in a two-qubit CPHASE gate: the critical role of tunneling noise. npj Quantum Inf. \textbf{4}, 62 (2018).

\bibitem[47]{Timmer1995}Timmer, J. \& Koenig, M. On generating power law noise. Astron. Astrophys. \textbf{300}, 707–707 (1995).

\bibitem[48]{pycolorednoise}Patzelt, F. Python package to generate gaussian (1/f)**beta noise (2017–). https://github.com/felixpatzelt/colorednoise. Online; accessed 15 Aug 2019.

\bibitem[49]{Makhlin2004a}Makhlin, Y. Nonlocal properties of two-qubit gates and mixed states and optimization of quantum computations. Quant. Inf. Proc. \textbf{1}, 243–252 (2004). 

\bibitem[50]{Pal2015}Pal, A., Rashba, E. I. \& Halperin, B. I. Exact CNOT gates with a single nonlocal rotation for quantum-dot qubits. Phys. Rev. B \textbf{92}, 125409 (2015).

\bibitem[51]{Luczak2018}{\L}uczak, J. \& Bu{\l}ka, B. R. Two-qubit logical operations in three quantum dots system. J. Phys. Condens. Matter \textbf{30}, 225601 (2018).

\bibitem[52]{Wang2008}Wang, X., Yu, C. S. \& Yi, X. X. An alternative quantum fidelity for mixed states of qudits. Phys. Lett. A \textbf{373}, 58–60 (2008).

\bibitem[53]{Green2013}Green, T. J., Sastrawan, J., Uys, H. \& Biercuk, M. J. Arbitrary quantum control of qubits in the presence of universal noise. New J. Phys. \textbf{15}, 095004 (2013).

\bibitem[54]{Nielsen2010}Nielsen, M. A. \& Chuang, I. L. Quantum Computation and Quantum Information (Cambridge University Press, 2010).

\bibitem[55]{Kubo1962}Kubo, R. Generalised Cumulant Expansion Method. J. Phys. Soc. Japan \textbf{17}, 1100–1120 (1962).

\bibitem[56]{Kubo1963}Kubo, R. Stochastic Liouville equations. J. Math. Phys. \textbf{4}, 174–183 (1963). 

\bibitem[57]{Aliferis2007}Aliferis, P. \& Cross, A. W. Subsystem fault tolerance with the Bacon-Shor code. Phys. Rev. Lett. \textbf{98}, 220502 (2007).

\bibitem[58]{Aliferis2008}Aliferis, P., Gottesman, D. \& Preskill, J. Accuracy threshold for postselected quantum computation. Quantum Info. Comput. \textbf{8}, 181–244 (2008).

\bibitem[59]{Aharonov2008}Aharonov, D. \& Ben-Or, M. Fault-tolerant quantum computation with constant error rate. SIAM J. Comput. \textbf{38}, 1207–1282 (2008).

\bibitem[60]{Paz-Silva2019}Paz-Silva, G. A., Norris, L. M., Beaudoin, F. \& Viola, L. Extending comb-based spectral estimation to multiaxis quantum noise. Phys. Rev. A \textbf{100}, 042334 (2019). 

\bibitem[61]{Hanson2007a}Hanson, R. \& Burkard, G. Universal set of quantum gates for double-dot spin qubits with fixed interdot coupling. Phys. Rev. Lett. \textbf{98}, 050502 (2007).

\bibitem[62]{Davies1997}Davies, J. The Physics of Low-dimensional Semiconductors (Cambridge University Press, 1997). 

\bibitem[63]{Annavarapu2013}Annavarapu, R. N. Singular Value Decomposition and the Centrality of L\"{o}wdin Orthogonalizations. Am. J. Comput. Appl. Math. \textbf{3}, 33–35 (2013).

\bibitem[64]{Zhang2018}Zhang, C., Yang, X. C. \& Wang, X. Leakage and sweet spots in triple-quantum-dot spin qubits: A molecular-orbital study. Phys. Rev. A \textbf{97},  042326 (2018).


\end{thebibliography}
\end{document}


\title{Two-qubit sweet spots for capacitively coupled exchange-only spin qubits: Supplementary Information}
\author{MengKe Feng}\thanks{Present address: School of Electrical Engineering and Telecommunications,
	University of New South Wales, Sydney, New South Wales 2052, Australia.}
\affiliation{Division of Physics and Applied Physics, School of Physical and Mathematical Sciences, 21 Nanyang Link, Singapore 637371, Singapore.}
\author{Lin Htoo Zaw} \thanks{Present address: Centre for Quantum Technologies, National University of Singapore, 3 Science Drive 2, Singapore 117543, Singapore.}
\affiliation{Division of Physics and Applied Physics, School of Physical and Mathematical Sciences, 21 Nanyang Link, Singapore 637371, Singapore.}
\author{Teck Seng Koh} \thanks{Corresponding author:~\href{mailto:kohteckseng@ntu.edu.sg}{kohteckseng@ntu.edu.sg} }
\affiliation{Division of Physics and Applied Physics, School of Physical and Mathematical Sciences, 21 Nanyang Link, Singapore 637371, Singapore.}

\maketitle

\tableofcontents

\section{Supplementary Methods}

\subsection{\label{supp:singlequbit}Single Qubit Hamiltonian}
We describe a single qubit system consisting of  three electrons in a triple quantum dot (TQD) with the Hubbard model~\cite{Yang:2011p161301,DasSarma:2011p235314}:
\begin{align}
\mathcal{\hat{H}} = \hat{H}_\varepsilon + \hat{H}_U +  \hat{H}_t
\end{align}
where $\hat{H}_\varepsilon$ describes the detuning of the individual quantum dots (QD), $\hat{H}_t$ describes tunneling between the dots, $\hat{H}_U$ describes the intra-dot and inter-dot Coulomb energies:
\begin{align}
\hat{H}_\varepsilon &= \sum_{i,\sigma} \varepsilon_i \hat{n}_{i\sigma} \\
\hat{H}_t &= \sum_{\langle i,j \rangle, \sigma} t_{ij} \hat{c}^\dagger_{i,\sigma} \hat{c}_{j,\sigma} \\
\hat{H}_U &= \sum_{i} U_{i} \hat{n}_{i \uparrow} \hat{n}_{i \downarrow} +  \frac{1}{2}\sum_{i}\sum_{j\neq i} U_{ij}   \hat{n}_{i }\hat{n}_{j}
\end{align}
Here, $\hat{n}_{i,\sigma}=\hat{c}^\dagger_{i,\sigma}\hat{c}_{i,\sigma}$ is the number operator and $\hat{n}_i = \hat{n}_{i\uparrow}+\hat{n}_{i \downarrow}$,  and $\varepsilon_i$ is the detuning of the $i$th-dot, $t_{ij}$ is the tunneling coupling between neighboring dots, $U_{i}$ is the intra-dot Coulomb energy and $U_{ij}$ are inter-dot Coulomb energy between all dots in the TQD. We assume identical QDs: $U_i \equiv U $ for all $i$, $U_{12}=U_{23}=U_{45}=U_{56}\equiv U^\prime$ and $U_{13}=U_{46} \equiv U^{\prime \prime}$. We consider two identical qubits (A and B) in a linear array. The basis states for qubit A are given by
\begin{align}
	\label{eqn:singlequbitbasisi}
	\ket{0_A} \equiv &  \left( 2\ket{\downarrow_1 \uparrow_2 \downarrow_3} - \ket{\downarrow_1 \downarrow_2 \uparrow_3} - \ket{\uparrow_1 \downarrow_2 \downarrow_3} \right)/ \sqrt{6}  \\
	\ket{1_A} \equiv & \left(\ket{\downarrow_1 \downarrow_2 \uparrow_3} - \ket{ \uparrow_1 \downarrow_2 \downarrow_3} \right)/ \sqrt{2} \\
	\ket{A_1} \equiv & \ket{\downarrow_1 \uparrow_3 \downarrow_3} \\
	\ket{A_2} \equiv & \ket{\uparrow_1 \downarrow_1 \downarrow_3} \\
	\ket{A_3} \equiv & \ket{\downarrow_1 \uparrow_2 \downarrow_2} \\
	\ket{A_4} \equiv & \ket{\uparrow_2 \downarrow_2 \downarrow_3}.
	\label{eqn:singlequbitbasisf}
\end{align}
Those for qubit B follow accordingly. The doubly occupied states $|A_{i}\rangle$ correspond to $(1,0,2), (2,0,1), (1,2,0)$ and $(0,2,1)$ charge states for  $i=1~\text{to}~4$. Note that these basis states are in the $S_z=-1/2$ subspace, and the singlet and triplet states are symmetric about the middle dot. Qubit A/B is the control/target qubit in CNOT and CPHASE gates. The $(0,1,2)$ and $(2,1,0)$ charge configurations can be neglected because they involve two tunneling processes and lead to higher order admixtures. Dot detunings are defined by
\begin{align}
\varepsilon &= \frac{1}{2}(\varepsilon_1-\varepsilon_3) \\
\varepsilon_m &= \varepsilon_2 - \frac{1}{2}(\varepsilon_1+\varepsilon_3)
\end{align}
By shifting the energy of the system by the sum of the individual detunings, the Hamiltonian can be expressed in terms of $\varepsilon$ and $\varepsilon_m$. Performing a Schrieffer-Wolff transformation, we obtain the effective Hamiltonian in the computational basis,
\begin{align}
\label{eqn:heff}
\hat{H}_\text{eff} = -\frac{\hbar\omega}{2}\hat{\sigma}_z - \frac{\hbar g}{2}\hat{\sigma}_x
\end{align}
where $\hbar\omega=2\left(t_l^2b_-+t_r^2b_+\right)$ and $\hbar g = 2\sqrt{3}\left(t_l^2b_- -  t_r^2b_+\right)$, with $b_\pm=\frac{1}{\left(U-U^{\prime\prime}\pm\epsilon+\epsilon_m\right)}+\frac{1}{\left(U-2U^\prime+U^{\prime\prime}\mp\epsilon-\epsilon_m\right)}$. The basis states for each qubit are now
\begin{align}
	\label{eqn:singlequbithybridi}
	|{1}^\prime_A\rangle =& \frac{1}{\sqrt{N_A}} \left(|1_A\rangle + \sum_{i=1}^4 \alpha_i |A_{i}\rangle \right),~~\text{and} ~~ |0_A\rangle, \\
	|{1}^\prime_B\rangle =& \frac{1}{\sqrt{N_B}}\left(|1_B\rangle +  \sum_{i=1}^4 \beta_i |B_{i}\rangle \right),~~\text{and} ~~ |0_B\rangle .
	\label{eqn:singlequbithybridf}
\end{align}
where $\sqrt{N_{A/B}}$ are the normalization constants ($N_A=1+\alpha_1^2+\alpha_2^2+\alpha_3^2+\alpha_4^2, N_B=1+\beta_1^2+\beta_2^2+\beta_3^2+\beta_4^2$) and each of the $\alpha_i, \beta_i$ terms are coefficients describing the admixtures of the doubly occupied singlet states, given by
\begin{align}
\alpha_1 &= \frac{t_{rA}}{\sqrt{2}(-U+U^{\prime\prime}+2\varepsilon_A)}, \\
\alpha_2 &= \frac{t_{lA}}{\sqrt{2}(-U+U^{\prime\prime}-2\varepsilon_A)}, \\
\alpha_3 &= \frac{t_{rA}}{\sqrt{2}(-U+2U^\prime-U^{\prime\prime}+\varepsilon_A+\varepsilon_{mA})}, \\
\alpha_4 &= \frac{t_{lA}}{\sqrt{2}(-U+2U^\prime-U^{\prime\prime}-\varepsilon_A+\varepsilon_{mA})}, \\
\beta_1 &= \frac{t_{rB}}{\sqrt{2}(-U+U^{\prime\prime}+2\varepsilon_B)}, \\
\beta_2 &= \frac{t_{lB}}{\sqrt{2}(-U+U^{\prime\prime}-2\varepsilon_B)}, \\
\beta_3 &= \frac{t_{rB}}{\sqrt{2}(-U+2U^\prime-U^{\prime\prime}+\varepsilon_B+\varepsilon_{mB})}, \\
\beta_4 &=  \frac{t_{lB}}{\sqrt{2}(-U+2U^\prime-U^{\prime\prime}-\varepsilon_B+\varepsilon_{mB}).}
\end{align}

With the effective Hamiltonian (Eq.~\eqref{eqn:heff}) we are able to find the single-qubit sweet spots, defined as flat points in the energy landscape satisfying~\cite{russ2017three}
\begin{align}
\frac{\partial \omega'_{A/B}}{\partial \varepsilon_{A/B}} =& 0,\\
\frac{\partial \omega'_{A/B}}{\partial \varepsilon_{mA/B}} =& 0,
\end{align}
where $\omega'_{A/B} = \sqrt{\omega_{A/B}^2 + g_{A/B}^2}$ is the eigenenergy difference between the ground and excited states for qubits $A$ and $B$. The double sweet spot where the AEON qubit is operated satisfies both conditions. This exists at $\varepsilon_{A/B}=0$~meV and $\varepsilon_{mA/B}=U^{\prime\prime}-U^\prime=-0.9~\mathrm{meV}$. On the other hand, the RX qubit is generally defined to be operating in a region of increased $\varepsilon_{mA/B}$~\cite{russ2017three}. In our study, the RX qubit  operates at the same $\varepsilon_{A/B}=0~\mathrm{meV}$ single-qubit sweet spot, but at a different middle detuning value, $\varepsilon_{mA/B}=-0.57~\mathrm{meV}$. These choices of $\varepsilon_{mA/B}$ are so that we are able to obtain comparable gate times for both the AEON and RX qubits, but generally, the RX qubit can be operated at any value of $\varepsilon_m$ such that there is strong admixture between the singly and doubly occupied states.


\subsection{\label{app:micromodel}Triple Quantum Dot Potential}
The confinement potential for a triple quantum dot system is modeled as a tri-quadratic potential,
\begin{align}
V_{\mathrm{pot}}(\vec{r})=\min{[v_{\mathrm{pot},1}(\vec{r}),v_{\mathrm{pot},2}(\vec{r}),v_{\mathrm{pot},3}(\vec{r})]},
\end{align}
where $v_{\mathrm{pot},i}(\vec{r})$ refers to the potential well of each dot centered at $\vec{R}_i$, given by
\begin{align}
v_{\mathrm{pot},i}(\vec{r})=\frac{m\omega_0^2}{2}\left(|\vec{x}-\vec{R}_i|^2 + \vec{y}^2\right) + \varepsilon_i.
\end{align}
Here, $\omega_0$ is the confinement energy of the dot which depends on its dimensions, and $\varepsilon_i$ is the detuning applied on the $i$-th dot. The 2-dimensional character of this potential is a good approximation given the tight confinement in the $z$-direction that is typical of electrostatically gated quantum dots. The ground orbital of an electron in such a potential is given by the Fock-Darwin wavefunction,
\begin{align}
\label{eqn:phifunc}
\phi_i(\vec{r}) = \left(\frac{1}{a_B \sqrt{\pi}}\right) \exp(-\frac{1}{2a_B^2}\left(|\vec{x}-\vec{R}_i|^2 + \vec{y}^2\right)).
\end{align}
This expression is normalized and $a_B=\sqrt{\hbar/m\omega_0}$ is the radius of the quantum dot. We can now formulate dot-centered, orthonomalized single-electron wavefunctions $\psi_i$ of the triple quantum dot potential using the method of L\"{o}wdin orthogonalization \cite{Zhang2018,Annavarapu2013}. These are given by
\begin{align}
\Psi = \Phi M^{-1/2}
\label{eq:Lowdin}
\end{align}
where $\Psi=\{\psi_1,\psi_2,\psi_3\}$ is the set of dot-centered, orthonomalized, single-electron wavefunctions, $\Phi=\{\phi_1,\phi_2,\phi_3\}$ is the set of Fock-Darwin wavefunctions, and $M$ is the overlap matrix given by
\begin{align}
\label{eqn:overlap}
M = \begin{pmatrix}
1	&	s	&	s^4	\\
s	&	1	&	s	\\
s^4	&	s	&	1
\end{pmatrix},
\end{align}
where $s=\exp(-a^2/a_B^2)$ and $2a$ is the distance between two neighboring dots.

The single-electron wavefunctions are crucial in the calculation of the terms in the interaction Hamiltonian, which is given by all pairwise direct Coulomb interactions between electrons in each qubit.
For three electrons in a triple quantum dot, there are eighty-one pairwise direct Coulomb integrals. With the quadratic potentials, each integral is analytically tractable and given generally by\\
\begin{align}
	\label{eqn:directcoulomb}
	& \int \int \phi_i^\star(\vec{r}_1)\phi_j^\star(\vec{r}_2) \frac{\kappa}{ |\vec{r}_1-\vec{r}_2|} \phi_k(\vec{r}_1) \phi_l(\vec{r}_2) ~d\vec{r}_1 d\vec{r}_2 \nonumber\\
	&= \kappa \sqrt{\frac{\pi}{2}} \frac{1}{a_B} \exp{-\frac{\left(R_i - R_k\right)^2-\left(R_j-R_l\right)^2}{4 a_B^2}}  \exp{-\frac{\left(R_i+R_k-R_j-R_l\right)^2}{16 a_B^2}} I_0\left(\frac{\left(R_i+R_k-R_j-R_l\right)^2}{16 a_B^2}\right)
\end{align}
where $\kappa=q^2/4\pi\epsilon_0\epsilon_r$ and $I_0$ is the zeroth order modified Bessel function of the first kind. The Bessel function appears as the solution to a standard integral of the form $\int_0^1 \exp{-at^2}/\sqrt{1-t^2}~dt$. This result is similar to that derived in Ref.~\cite{Calderon-Vargas2015}. In our calculations, we take relative permittivity $\epsilon_r = 11.68$ to be that for silicon.


\subsection{Interaction Hamiltonian}
The two-qubit capacitive interaction arises from the inter-dot Coulomb interaction between the  TQDs, given by
\begin{align}\label{eq:Hint}
	\hat{H}_\text{int} =& \sum_{i=1}^{3} \sum_{j=4}^{6} \mathcal{V}_{ij} \hat{n}_i \hat{n}_j,
\end{align}
where $\mathcal{V}_{ij}$ is given by the direct Coulomb integral between the orthonormalized electron densities of $i$-th and $j$-th QD,
\begin{align}
\mathcal{V}_{ij} = \int \int |\psi_i(\vec{r}_1)|^2 \frac{\kappa}{ |\vec{r}_1-\vec{r}_2|} |\psi_j(\vec{r}_2)|^2 d\vec{r}_1 d\vec{r}_2.
\end{align}
Here, $i$-th dot-centered, normalized, single-electron wavefunctions $\psi_i(\vec{r})$ are constructed using L\"{o}wdin's orthogonalization method from single QD wavefunctions as shown in Eq.~\eqref{eq:Lowdin}.  The three-electron wavefunctions for qubits A and B follow from Eqs.~\eqref{eqn:singlequbitbasisi}--\eqref{eqn:singlequbitbasisf}, and the corresponding hybridized basis states are as outlined in Eqs.~\eqref{eqn:singlequbithybridi}--\eqref{eqn:singlequbithybridf}. Hence, the computational basis for the interaction Hamiltonian (Eq.~\eqref{eq:Hint}) comprises
\begin{align}
	|0_A 0_B\rangle & ,  \\
	|0_A 1_B^\prime\rangle & =   \frac{1}{\sqrt{N_B}} |0_A\rangle \left(|1_B\rangle + \sum_{n=1}^{4} \beta_n |B_n\rangle \right), \\
	|1_A^\prime 0_B\rangle & = \frac{1}{\sqrt{N_A}}\left(|1_A\rangle + \sum_{m=1}^{4} \alpha_m |A_m\rangle \right) |0_B\rangle, \\
	|1_A^\prime 1_B^\prime \rangle & = \frac{1}{\sqrt{N_A N_B}}\left(|1_A\rangle + \sum_{m=1}^{4} \alpha_m |A_m\rangle \right) \left(|1_B\rangle + \sum_{n=1}^{4} \beta_n |B_n\rangle \right).
\end{align}
Because the basis states are eigenstates of the number operators,  Eq.~\eqref{eq:Hint} must be a diagonal matrix in the computational basis $\{\ket{0_A 0_B}, \ket{0_A 1_B^\prime}, \ket{1_A^\prime 0_B}, \ket{1_A^\prime 1_B^\prime} \}$, i.e.
\begin{align}
	\hat{H}_\text{int} =  \hbar \times \text{diag}\{V_1, V_2, V_3, V_4 \}.
\end{align}
The diagonal elements are explicitly calculated as follows.
\begin{align}
	\hbar V_1 \equiv &  \langle 0_A 0_B | \left ( \sum_{i=1}^{3} \sum_{j=4}^{6} \mathcal{V}_{ij} \hat{n}_i \hat{n}_j \right) |0_A 0_B \rangle \nonumber \\
	= & \mathcal{V}_{14} + \mathcal{V}_{24} + \mathcal{V}_{34} + \mathcal{V}_{15} +  \mathcal{V}_{25} +  \mathcal{V}_{35} + \mathcal{V}_{16} +  \mathcal{V}_{26} + \mathcal{V}_{36} \equiv C_{11},
\end{align}
where we have defined $C_{11}$ as the sum of all the possible inter-qubit capacitive interaction terms. The $V_2$  term is given as follows.
\begin{align}
	\hbar V_2 \equiv &  \langle 0_A 1_B^\prime | \left ( \sum_{i=1}^{3} \sum_{j=4}^{6} \mathcal{V}_{ij} \hat{n}_i \hat{n}_j \right) |  0_A 1_B^\prime \rangle \nonumber \\
	= & \frac{1}{N_B} \left( \langle 0_A | \langle 1_B | + \sum_{m=1}^4 \beta_m^\star \langle 0_A | \langle B_{m}|  \right)	\left ( \sum_{i=1}^{3} \sum_{j=4}^{6} \mathcal{V}_{ij} \hat{n}_i \hat{n}_j \right)  \left( | 0_A  \rangle  | 1_B \rangle + \sum_{n=1}^4 \beta_n | 0_A \rangle | B_{n} \rangle \right)	 \nonumber \\
	=&  \frac{1}{N_B} \left(\mathcal{V}_{14} + \mathcal{V}_{24} + \mathcal{V}_{34} + \mathcal{V}_{15} +  \mathcal{V}_{25} +  \mathcal{V}_{35} + \mathcal{V}_{16} +  \mathcal{V}_{26} + \mathcal{V}_{36}  \right)  \nonumber \\
	& + \frac{1}{N_B}\left (\beta_1^2 \left(\mathcal{V}_{14} + \mathcal{V}_{24} + \mathcal{V}_{34}  + 2 \mathcal{V}_{16} + 2 \mathcal{V}_{26} + 2 \mathcal{V}_{36}  \right) \right) \nonumber \\
	& + \frac{1}{N_B}\left (\beta_2^2 \left(2\mathcal{V}_{14} + 2\mathcal{V}_{24} + 2\mathcal{V}_{34}  +  \mathcal{V}_{16} +  \mathcal{V}_{26} +  \mathcal{V}_{36}   \right) \right) \nonumber \\
	& + \frac{1}{N_B}\left (\beta_3^2 \left(\mathcal{V}_{14} + \mathcal{V}_{24} + \mathcal{V}_{34}  + 2 \mathcal{V}_{15} + 2 \mathcal{V}_{25} + 2 \mathcal{V}_{35}  \right) \right) \nonumber \\
	& + \frac{1}{N_B}\left (\beta_4^2 \left( 2 \mathcal{V}_{15} + 2 \mathcal{V}_{25} + 2 \mathcal{V}_{35} +\mathcal{V}_{16} + \mathcal{V}_{26} + \mathcal{V}_{36}   \right) \right) \nonumber \\
	\equiv & \frac{1}{N_B} \left ( C_{11} + \beta_1^2 C_{12} + \beta_2^2 C_{13} + \beta_3^2 C_{14} + \beta_4^2 C_{15}  \right) \nonumber\\
	\equiv & \frac{1}{N_B} \left ( C_{11} + \vec{\nu} \cdot \vec{F}_B \right),
\end{align}
where the exact terms can be derived from the basis states as defined in Eq.~\eqref{eqn:singlequbitbasisi}--\eqref{eqn:singlequbitbasisf} and we have defined
\begin{align}
	C_{12} \equiv & \mathcal{V}_{14} + \mathcal{V}_{24} + \mathcal{V}_{34}  + 2 \mathcal{V}_{16} + 2 \mathcal{V}_{26} + 2 \mathcal{V}_{36}  ,\\
	C_{13} \equiv & 2\mathcal{V}_{14} + 2\mathcal{V}_{24} + 2\mathcal{V}_{34}  +  \mathcal{V}_{16} +  \mathcal{V}_{26} +  \mathcal{V}_{36} ,\\
	C_{14} \equiv & \mathcal{V}_{14} + \mathcal{V}_{24} + \mathcal{V}_{34}  + 2 \mathcal{V}_{15} + 2 \mathcal{V}_{25} + 2 \mathcal{V}_{35},\\
	C_{15} \equiv & 2 \mathcal{V}_{15} + 2 \mathcal{V}_{25} + 2 \mathcal{V}_{35} +\mathcal{V}_{16} + \mathcal{V}_{26} + \mathcal{V}_{36} ,\\
	\vec{F}_B \equiv & \{C_{12}, C_{13}, C_{14}, C_{15} \}^\text{T}, \\
	\vec{\nu} \equiv & \{\beta_1^2,\beta_2^2,\beta_3^2,\beta_4^2 \}^\text{T}.
\end{align}
Next,
\begin{align}
	\hbar V_3 \equiv &  \langle 1_A^\prime 0_B | \left ( \sum_{i=1}^{3} \sum_{j=4}^{6} \mathcal{V}_{ij} \hat{n}_i \hat{n}_j \right) |1_A^\prime 0_B \rangle \nonumber \\
	= & \frac{1}{N_A}  \left( \langle 1_A | \langle 0_B | + \sum_{m=1}^{4} \alpha_m^\star  \langle A_m | \langle 0_B | \right)	\left ( \sum_{i=1}^{3} \sum_{j=4}^{6} \mathcal{V}_{ij} \hat{n}_i \hat{n}_j \right)  \left(|1_A\rangle | 0_B\rangle + \sum_{n=1}^{4} \alpha_n |A_m\rangle |0_B\rangle \right)		 \nonumber \\
	=&  \frac{1}{N_A} \left(\mathcal{V}_{14} + \mathcal{V}_{24} + \mathcal{V}_{34} + \mathcal{V}_{15} +  \mathcal{V}_{25} +  \mathcal{V}_{35} + \mathcal{V}_{16} +  \mathcal{V}_{26} + \mathcal{V}_{36}  \right)  \nonumber \\
	& + \frac{1}{N_A}\left (\alpha_1^2 \left(\mathcal{V}_{14} + \mathcal{V}_{15} + \mathcal{V}_{16}  + 2 \mathcal{V}_{34} + 2 \mathcal{V}_{35} + 2 \mathcal{V}_{36}  \right) \right) \nonumber \\
	& + \frac{1}{N_A}\left (\alpha_2^2 \left(2\mathcal{V}_{14} + 2\mathcal{V}_{15} + 2\mathcal{V}_{16}  +  \mathcal{V}_{34} +  \mathcal{V}_{35} +  \mathcal{V}_{36}   \right) \right) \nonumber \\
	& + \frac{1}{N_A}\left (\alpha_3^2 \left(\mathcal{V}_{14} + \mathcal{V}_{15} + \mathcal{V}_{16}  + 2 \mathcal{V}_{24} + 2 \mathcal{V}_{25} + 2 \mathcal{V}_{26}   \right) \right) \nonumber \\
	& + \frac{1}{N_A}\left (\alpha_4^2 \left( 2\mathcal{V}_{24} + 2\mathcal{V}_{25} + 2\mathcal{V}_{26}  +  \mathcal{V}_{34} +  \mathcal{V}_{35} + \mathcal{V}_{36}     \right) \right) \nonumber \\
	\equiv & \frac{1}{N_A} \left ( C_{11} + \alpha_1^2 C_{21} + \alpha_2^2 C_{31} + \alpha_3^2 C_{41} + \alpha_4^2 C_{51}  \right) \nonumber \\
	\equiv & \frac{1}{N_A} \left ( C_{11} + \vec{\xi} \cdot \vec{F}_A \right).
\end{align}
where we have defined
\begin{align}
	C_{21} \equiv & \mathcal{V}_{14} + \mathcal{V}_{15} + \mathcal{V}_{16}  + 2 \mathcal{V}_{34} + 2 \mathcal{V}_{35} + 2 \mathcal{V}_{36}  ,\\
	C_{31} \equiv & 2\mathcal{V}_{14} + 2\mathcal{V}_{15} + 2\mathcal{V}_{16}  +  \mathcal{V}_{34} +  \mathcal{V}_{35} +  \mathcal{V}_{36},\\
	C_{41} \equiv & \mathcal{V}_{14} + \mathcal{V}_{15} + \mathcal{V}_{16}  + 2 \mathcal{V}_{24} + 2 \mathcal{V}_{25} + 2 \mathcal{V}_{26} ,\\
	C_{51} \equiv & 2\mathcal{V}_{24} + 2\mathcal{V}_{25} + 2\mathcal{V}_{26}  +  \mathcal{V}_{34} +  \mathcal{V}_{35} +  \mathcal{V}_{36} , \\
	\vec{F}_{A}  \equiv & \{C_{21}, C_{31}, C_{41}, C_{51} \}^\text{T}, \\
	\vec{\xi} \equiv & \{\alpha_1^2, \alpha_2^2, \alpha_3^2, \alpha_4^2 \}^\text{T}
\end{align}

The $V_4$ term is
\begin{align}\label{eq:V4}
	\hbar V_4 \equiv & \langle 1^\prime_A 1^\prime_B | \left ( \sum_{i=1}^{3} \sum_{j=4}^{6} \mathcal{V}_{ij} \hat{n}_i \hat{n}_j \right) |  1^\prime_A 1^\prime_B \rangle \nonumber \\
	= & \frac{1}{N_A N_B} \left [ \left( \langle 1_A | + \sum_{m=1}^4 \alpha_m^\star \langle A_{m} | \right) \left( \langle 1_B | + \sum_{n=1}^4 \beta_n^\star \langle B_{n} | \right)  \right ] 	\left ( \sum_{i=1}^{3} \sum_{j=4}^{6} \mathcal{V}_{ij} \hat{n}_i \hat{n}_j \right)  \times \nonumber \\
	&
	\left [ \left(|1_A\rangle + \sum_{p=1}^4 \alpha_p |A_{p}\rangle \right) \left(|1_B\rangle + \sum_{q=1}^4 \beta_q |B_{q}\rangle \right)  \right ].
\end{align}

Expanding the kets, we group terms according to the pre-factors,
\begin{align}
	\left(|1_A\rangle + \sum_{p=1}^4 \alpha_p |A_{p}\rangle \right) \left(|1_B\rangle + \sum_{q=1}^4 \beta_q |B_{q}\rangle \right)
	& = |1_A \rangle | 1_B\rangle + \sum_{p=1}^4 \alpha_p |A_{p}\rangle |1_B\rangle + \sum_{q=1}^4 \beta_q |1_A\rangle |B_{q}\rangle + \sum_{p=1}^4 \sum_{q=1}^4 \alpha_p \beta_q |A_{p}\rangle  |B_{q}\rangle .
\end{align}
There are no cross terms since all states are orthogonal, so
the terms in $V_4$ can be grouped into those without admixture factors, and terms with admixture factors $\alpha_p^2$,  $\beta_q^2$ and  $\alpha_p^2 \beta_q^2$.

The first group of terms without admixtures are
\begin{align}
	\langle 1_A | \langle 1_B | \left ( \sum_{i=1}^{3} \sum_{j=4}^{6} \mathcal{V}_{ij} \hat{n}_i \hat{n}_j \right) |1_A \rangle |1_B \rangle
	= & \mathcal{V}_{14} + \mathcal{V}_{24} + \mathcal{V}_{34} + \mathcal{V}_{15} +  \mathcal{V}_{25} +  \mathcal{V}_{35} + \mathcal{V}_{16} +  \mathcal{V}_{26} + \mathcal{V}_{36} \equiv C_{11}.
\end{align}

The second group of $\alpha_p^2$ terms are
\begin{align}
	& \left( \sum_{m=1}^4 \alpha_m \langle A_{p} | \langle 1_B | \right) \left ( \sum_{i=1}^{3} \sum_{j=4}^{6} \mathcal{V}_{ij} \hat{n}_i \hat{n}_j \right) \left(\sum_{p=1}^4 \alpha_p |A_{p}\rangle |1_B\rangle  \right)   \nonumber \\
	= & \alpha_1^2 \left(\mathcal{V}_{14} + \mathcal{V}_{15} + \mathcal{V}_{16} + 2\mathcal{V}_{34} +  2\mathcal{V}_{35} +  2\mathcal{V}_{36}  \right) \nonumber \\
	& + \alpha_2^2 \left(2\mathcal{V}_{14} +2 \mathcal{V}_{15} + 2\mathcal{V}_{16} + \mathcal{V}_{34} +  \mathcal{V}_{35} +  \mathcal{V}_{36} \right) \nonumber \\
	& + \alpha_3^2 \left( \mathcal{V}_{14} + \mathcal{V}_{15} + \mathcal{V}_{16} + 2\mathcal{V}_{24} +  2\mathcal{V}_{25} +  2\mathcal{V}_{26} \right) \nonumber \\
	& + \alpha_4^2 \left(2\mathcal{V}_{24} + 2\mathcal{V}_{25} + 2\mathcal{V}_{26} + \mathcal{V}_{34} +  \mathcal{V}_{35} +  \mathcal{V}_{36} \right) \nonumber \\
	= & \alpha_1^2 C_{21} +\alpha_2^2 C_{31} +\alpha_3^2 C_{41} +\alpha_4^2 C_{51} = \vec{\xi}\cdot \vec{F}_A.
\end{align}

The third group of $\beta_q^2$ terms are
\begin{align}
	& \left( \sum_{n=1}^4 \beta_n \langle 1_A | \langle B_{n} |  \right) \left ( \sum_{i=1}^{3} \sum_{j=4}^{6} \mathcal{V}_{ij} \hat{n}_i \hat{n}_j \right) \left(\sum_{q=1}^4 \beta_q | 1_A\rangle   |B_{q}\rangle \right)   \nonumber \\
	= & \beta_1^2 \left(\mathcal{V}_{14} + \mathcal{V}_{24} + \mathcal{V}_{34} + 2\mathcal{V}_{16} +  2\mathcal{V}_{26} +  2\mathcal{V}_{36}  \right) \nonumber \\
	& + \beta_2^2 \left(2\mathcal{V}_{14} +2 \mathcal{V}_{24} + 2\mathcal{V}_{34} + \mathcal{V}_{16} +  \mathcal{V}_{26} +  \mathcal{V}_{36} \right) \nonumber \\
	& + \beta_3^2 \left( \mathcal{V}_{14} + \mathcal{V}_{24} + \mathcal{V}_{34} + 2\mathcal{V}_{15} +  2\mathcal{V}_{25} +  2\mathcal{V}_{35} \right) \nonumber \\
	& + \beta_4^2 \left(2\mathcal{V}_{15} + 2\mathcal{V}_{25} + 2\mathcal{V}_{35} + \mathcal{V}_{16} +  \mathcal{V}_{26} +  \mathcal{V}_{36} \right) \nonumber \\
	= & \beta_1^2 C_{12} +\beta_2^2 C_{13} +\beta_3^2 C_{14} +\beta_4^2 C_{15} = \vec{\nu} \cdot \vec{F}_B.
\end{align}

The fourth group of $\alpha_p^2 \beta_q^2$ terms are
\begin{align}
	&\left( \sum_{m=1}^4 \sum_{n=1}^4 \alpha_m^\star \beta_n^\star \langle A_{m} | \langle  B_{n}| \right) \left ( \sum_{i=1}^{3} \sum_{j=4}^{6} \mathcal{V}_{ij} \hat{n}_i \hat{n}_j \right) \left( \sum_{p=1}^4 \sum_{q=1}^4 \alpha_p \beta_q |A_{p}\rangle  |B_{q}\rangle \right)\nonumber \\
	= & \alpha_1^2 \left( \beta_1^2 C_{22} + \beta_2^2  C_{23}  + \beta_3^2  C_{24}  + \beta_4^2  C_{25}  \right ) \nonumber \\
	& + \alpha_2^2  \left ( \beta_1^2 C_{32} + \beta_2^2  C_{33}  + \beta_3^2  C_{34}  + \beta_4^2  C_{35}  \right ) \nonumber \\
	& + \alpha_3^2  \left ( \beta_1^2 C_{42} + \beta_2^2  C_{43}  + \beta_3^2  C_{44}  + \beta_4^2  C_{45}  \right ) \nonumber \\
	& + \alpha_4^2  \left ( \beta_1^2 C_{52} + \beta_2^2  C_{53}  + \beta_3^2  C_{54}  + \beta_4^2  C_{55}  \right ) \nonumber \\
	= & \vec{\xi}~^\text{T} \overleftrightarrow{F} \vec{\nu}.
\end{align}
In the last line, we have defined the matrix $\overleftrightarrow{F} \equiv C_{ij}$ for $i , j =2, 3, 4,5$, and column vectors $\vec{\xi}$, $\vec{\nu}$. The definitions of $C_{ij}$ above are
\begin{align}
	C_{22} \equiv &  \mathcal{V}_{14} + 2\mathcal{V}_{16} + 2\mathcal{V}_{34} + 4\mathcal{V}_{36}, \\
	C_{23} \equiv & 2\mathcal{V}_{14} + \mathcal{V}_{16} + 4\mathcal{V}_{34} + 2\mathcal{V}_{36}, \\
	C_{24} \equiv &  \mathcal{V}_{14} +2\mathcal{V}_{15} + 2\mathcal{V}_{34} + 4\mathcal{V}_{35}, \\
	C_{25} \equiv & 2\mathcal{V}_{15} + \mathcal{V}_{16} + 4\mathcal{V}_{35} + 2\mathcal{V}_{36}, \\
	C_{32} \equiv & 2\mathcal{V}_{14} + 4\mathcal{V}_{16} + \mathcal{V}_{34} + 2\mathcal{V}_{36}, \\
	C_{33} \equiv & 4\mathcal{V}_{14} + 2\mathcal{V}_{16} + 2\mathcal{V}_{34} + \mathcal{V}_{36}, \\
	C_{34} \equiv & 2\mathcal{V}_{14} + 4\mathcal{V}_{15} + \mathcal{V}_{34} + 2\mathcal{V}_{35}, \\
	C_{35} \equiv & 4\mathcal{V}_{15} + 2\mathcal{V}_{16} + 2\mathcal{V}_{35} + \mathcal{V}_{36}, \\
	C_{42} \equiv & \mathcal{V}_{14} + 2\mathcal{V}_{16} + 2\mathcal{V}_{24} + 4\mathcal{V}_{26}, \\
	C_{43} \equiv & 2\mathcal{V}_{14} + \mathcal{V}_{16} + 4\mathcal{V}_{24} + 2\mathcal{V}_{26}, \\
	C_{44} \equiv & \mathcal{V}_{14} + 2\mathcal{V}_{15} + 2\mathcal{V}_{24} + 4\mathcal{V}_{25}, \\
	C_{45} \equiv & 2\mathcal{V}_{15} + \mathcal{V}_{16} + 4\mathcal{V}_{25} + 2\mathcal{V}_{26}, \\
	C_{52} \equiv & 2\mathcal{V}_{24} + 4\mathcal{V}_{26} + \mathcal{V}_{34} + 2\mathcal{V}_{36}, \\
	C_{53} \equiv & 4\mathcal{V}_{24} + 2\mathcal{V}_{26} + 2\mathcal{V}_{34} + \mathcal{V}_{36}, \\
	C_{54} \equiv & 2\mathcal{V}_{24} + 4\mathcal{V}_{25} + \mathcal{V}_{34} + 2\mathcal{V}_{35}, \\
	C_{55} \equiv & 4\mathcal{V}_{25} + 2\mathcal{V}_{26} + 2\mathcal{V}_{35} + \mathcal{V}_{36}.
\end{align}

Therefore,
\begin{align}
	\hbar V_4 = & \frac{1}{N_A N_B}   \left ( C_{11} +\vec{\xi} \cdot \vec{F}_A + \vec{\nu} \cdot \vec{F}_B  + \vec{\xi}~^\text{T} ~ \overleftrightarrow{F} \vec{\nu} \right ) .
\end{align}

In summary, the interaction Hamiltonian in the computational basis is
\begin{align}
	\hat{H}_\text{int} =  \hbar \times \text{diag}\{V_1, V_2, V_3, V_4 \},
	\label{eqn:hint2}
\end{align}
where the matrix elements are
\begin{align}
	\hbar	V_1 = & C_{11} , \\
	\hbar	V_2 = & \frac{1}{N_B} \left ( C_{11} + \vec{\nu} \cdot \vec{F}_B \right) ,\\
	\hbar	V_3 = & \frac{1}{N_A} \left ( C_{11} + \vec{\xi} \cdot \vec{F}_A \right), \\
	\hbar	V_4 = & \frac{1}{N_A N_B}  \left ( C_{11} +\vec{\xi} \cdot \vec{F}_A + \vec{\nu} \cdot \vec{F}_B  + \vec{\xi}~^\text{T} ~ \overleftrightarrow{F} \vec{\nu} \right ).
\end{align}


\subsection{\label{sec:1fnoise}Noise Simulations and Fidelity Calculations}
\begin{figure}[ht!]
	\includegraphics[width=0.6\linewidth]{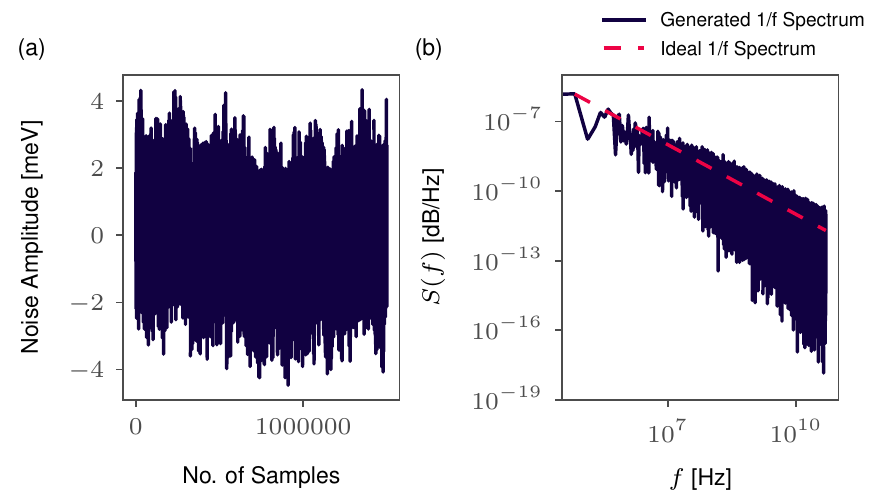}
	\caption{$1/f$ noise spectrum. (a) Typical numerically generated $1/f$ noise spectrum ($\delta\varepsilon$, $\delta\varepsilon_m$ or $\delta t_{l(r)}$) with $15\times10^5$ samples.  (b) Power density spectrum of a typical noise spectrum with red dashed line showing the $1/f$ dependence. The spectrum corresponds to that which is described in Eq.~\eqref{eqn:noise}, with  higher and lower frequency cutoffs of 50 GHz and 66.7 kHz respectively.
	}
	\label{fig:noisespec}
\end{figure}

In this section, we will explain how $1/f$ noise is generated and incorporated into the Hamiltonian, and therefore forming the noisy gate, of which we can then calculate the fidelity. The typical noise spectrum, Fig.~\ref{fig:noisespec}(a) is generated in Python, modified from code written by Felix Patzelt \cite{pycolorednoise,AA1995} using an algorithm in which we generate Gaussian distributed noise that is inversely proportional to the number of samples. The power spectrum density, Fig.~\ref{fig:noisespec}(b), is also plotted to demonstrate that the noise has $1/f$ characteristics.

The noise is generated in the following way:
\begin{enumerate}
	\item First, based on the number of samples (i.e. number of data points) that is to be generated, we calculate the frequency spectrum. In our simulations, we use a total of $15\times10^5$ samples for a time interval of $[0, \tau_\mathrm{expt}]$, where $\tau_\mathrm{expt}$ is the total length of time of the experiment. In terms of frequency, the sampling frequency will be defined as $f_{\mathrm{samp}}=1/\Delta t$, where $\Delta t$ is the size of the time steps as defined by the time evolution of the Hamiltonian. This means that $\tau_\mathrm{expt}=N\Delta t$ can change depending on the size of the time evolution time step. The higher frequency cutoff will consequentially be $f_h=f_{\mathrm{samp}}/2$ and the lower frequency cutoff will be defined as $f_l=f_{\mathrm{samp}}/N$ for all simulations except for results where the value of $f_l$ is independently adjusted to examine its effect on fidelity.

	\item Second, calculate the scaling factors corresponding to each frequency. The lower frequency cutoff is applied here with all lower values of frequency set to the cutoff value. Following which, we construct the scaling factors, $s$, in this way:
	\begin{align}
		s = f^{-\alpha/2}
	\end{align}
	such that we account for the frequency characteristic of the spectrum as given by $\alpha$ in the exponent of the frequency. Note that the exponent is divided by $1/2$ because the noise spectrum is proportional to the square root of the power spectrum density, $\sqrt{S(2\pi f)}$.
	\item Then, we calculate the ideal variance of the spectrum assuming a Gaussian distribution. This determines the amplitude of the output noise spectrum and will be proportional to the scaling factors calculated earlier.
	\item Next, based on the scaling, we generate random normal distributed values for both the power and the phase of the noise spectrum.
	\item Finally, a Fourier transformation is performed on these randomly generated values to obtain a real time series based on the previously generated noise spectrum, which is then scaled to unit variance. The use of the scaling factor ensures that the generated spectrum follows the intended frequency power law, which is also shown in Fig.~\ref{fig:noisespec}(b). The final noise spectrum can be described as follows:
	\begin{align}
		S(2\pi f) = \begin{cases}
			\Delta^2_{n_i}/f_l & f \leq f_l \\
			\Delta^2_{n_i}/f & f_l \leq f \leq f_h \\
			0 & \mathrm{,otherwise}
		\end{cases}
	\label{eqn:noise}
	\end{align}
\end{enumerate}

In the simulations of time evolution, time steps of 0.01 ns is used, indicating a sampling frequency, $f_\mathrm{samp} = 100$ GHz. Correspondingly, the higher frequency cutoff, $f_h$ is 50 GHz, and the lower frequency cutoff, $f_l$ is 66.7 kHz.

These noise spectra will be generated over many iterations of noisy time evolution, and we will now describe here how noise enters the Hamiltonian. In each iteration of noisy time evolution, we generate two separate noise spectra for four different noisy parameters:
\begin{align}
	\{\delta\varepsilon, \delta\varepsilon_m, \delta t_l, \delta t_r\}
	\label{eqn:deltanoise}
\end{align}
We need only two separate noise spectra for four parameters because we are considering uncorrelated noise with independent spectra, so we are only effectively comparing the effects of only two noisy parameters at the same time which would then mean that only two independent noise spectra are needed at the same time.

Depending on the parameter being considered, we consider a fixed magnitude of fluctuations in the parameter, with differing magnitudes for the detuning and tunneling parameters since they have different magnitudes. The output spectrum generated from the procedure is scaled by a value of variance that is calculated from the scaling factors which may not be of the amplitude that we want. In addition, the mean of the generated spectrum is not exactly at zero. Therefore, we want to first scale the noise spectra to unit variance and zero mean:
\begin{align}
	y_i = \frac{x_i - \bar{x}}{\bar{\sigma_x}}
\end{align}
where $x$ is the generated noise spectrum and $y$ is the normalized noise spectrum. After this normalization, we can now adjust freely the amplitude and standard deviation of the noise spectrum. The spectra is then multiplied by the desired noise amplitude (which changes effectively the standard deviation of the noise) as shown in Eq.~\eqref{eqn:deltanoise}. We define the noise levels in these parameters to vary from $10^{-5}$ to $10^{-3}$ for detuning parameters and $10^{-6}$ to $10^{-4}$ for tunneling parameters. These noisy parameters are now added to the corresponding parameters such that
\begin{align}
	\{
		\varepsilon \rightarrow \varepsilon + \delta\varepsilon,~
		\varepsilon_m \rightarrow \varepsilon_m + \delta\varepsilon_m,~
		t_l \rightarrow t_l+\delta t_l,~
		t_r\rightarrow t_r+\delta t_r
	\}
\end{align}
Fluctuations in the detuning and tunnel coupling parameters will directly result in fluctuations in the admixtures ($\alpha_i$ and $\beta_i$) of the doubly occupied singlet states in the dressed singlet excited state ($\ket{1^\prime_A}$ and $\ket{1^\prime_B}$).
\begin{align}
	\alpha^\prime_1 &= \frac{t_{rA}+\delta t_{rA}}{\sqrt{2}(-U+U^{\prime\prime}+2(\varepsilon_A+\delta \varepsilon_A))} \\
	\alpha^\prime_2 &= \frac{t_{lA}+\delta t_{lA}}{\sqrt{2}(-U+U^{\prime\prime}-2(\varepsilon_A+\delta\varepsilon_A))} \\
	\alpha^\prime_3 &= \frac{t_{rA}+\delta t_{rA}}{\sqrt{2}(-U+2U^\prime-U^{\prime\prime}+\varepsilon_A+\delta \varepsilon_A+\varepsilon_{mA}+\delta \varepsilon_{mA})} \\
	\alpha^\prime_4 &= \frac{t_{lA}+\delta t_{lA}}{\sqrt{2}(-U+2U^\prime-U^{\prime\prime}-\varepsilon_A-\delta \varepsilon_A+\varepsilon_{mA}+\delta\varepsilon_{mA})} \\
	\beta^\prime_1 &= \frac{t_{rB}+\delta t_{rB}}{\sqrt{2}(-U+U^{\prime\prime}+2(\varepsilon_B+\delta \varepsilon_B))} \\
	\beta^\prime_2 &= \frac{t_{lB}+\delta t_{lB}}{\sqrt{2}(-U+U^{\prime\prime}-2(\varepsilon_B+\delta\varepsilon_B))} \\
	\beta^\prime_3 &= \frac{t_{rB}+\delta t_{rB}}{\sqrt{2}(-U+2U^\prime-U^{\prime\prime}+\varepsilon_B+\delta \varepsilon_B+\varepsilon_{mB}+\delta \varepsilon_{mB})} \\
	\beta^\prime_4 &= \frac{t_{lB}+\delta t_{lB}}{\sqrt{2}(-U+2U^\prime-U^{\prime\prime}-\varepsilon_B-\delta \varepsilon_B+\varepsilon_{mB}+\delta\varepsilon_{mB})}
\end{align}
This will lead to fluctuations in the Hamiltonian terms ($V_1$, $V_2$, $V_3$, and $V_4$), such that the interaction Hamiltonian as outlined in Eq.~\eqref{eqn:hint2} is now $\mathrm{diag}\{V_1,V_2+\delta V_2,V_3+\delta V_3,V_4+\delta V_4\}$.

\section{Supplementary Notes}
\subsection{Comparison to the Ising Coupling}
Since the interaction Hamiltonian  is diagonal, it includes only the $\Bqty{
	\hat{\mathbb{1}}_A,\hat{\sigma}_{z,A}
} \otimes \Bqty{
	\hat{\mathbb{1}}_B,\hat{\sigma}_{z,B}
}$ terms. We can recast the Hamiltonian in the Pauli basis by first writing the Hamiltonian in terms of the two-qubit basis states, then rewriting the state projectors with the Pauli operators
\begin{equation}
	\begin{aligned}
		\hat{H}_{\text{int}}
		&= \ket{0_A 0_B} \hbar V_1 \bra{0_A 0_B} +
		\ket{0_A 1^\prime_B} \hbar V_2 \bra{0_A 1^\prime_B} +
		\ket{1^\prime_A 0_B} \hbar V_3 \bra{1^\prime_A 0_B} +
		\ket{1^\prime_A 1^\prime_B} \hbar V_4 \bra{1^\prime_A 1^\prime_B} \\
		&= \hbar V_1\pqty\Big{
			\tfrac{1}{2}\pqty{\hat{\mathbb{1}}_A + \hat{\sigma}_{z,A}} \otimes
			\tfrac{1}{2}\pqty{\hat{\mathbb{1}}_B + \hat{\sigma}_{z,B}}
		} +
		\hbar V_2\pqty\Big{
			\tfrac{1}{2}\pqty{\hat{\mathbb{1}}_A + \hat{\sigma}_{z,A}} \otimes
			\tfrac{1}{2}\pqty{\hat{\mathbb{1}}_B - \hat{\sigma}_{z,B}}
		} \\
		&\qquad+
		\hbar V_3\pqty\Big{
			\tfrac{1}{2}\pqty{\hat{\mathbb{1}}_A - \hat{\sigma}_{z,A}} \otimes
			\tfrac{1}{2}\pqty{\hat{\mathbb{1}}_B + \hat{\sigma}_{z,B}}
		} +
		\hbar V_4\pqty\Big{
			\tfrac{1}{2}\pqty{\hat{\mathbb{1}}_A - \hat{\sigma}_{z,A}} \otimes
			\tfrac{1}{2}\pqty{\hat{\mathbb{1}}_B - \hat{\sigma}_{z,B}}
		} \\
		&= \tfrac{1}{4}\pqty{\hbar V_1 + \hbar V_2 + \hbar V_3 + \hbar V_4}
		~ \hat{\mathbb{1}}_A \otimes \hat{\mathbb{1}}_B  +
		\tfrac{1}{4}\pqty{\hbar V_1 - \hbar V_2 + \hbar V_3 - \hbar V_4}
		~ \hat{\sigma}_{z,A} \otimes \hat{\mathbb{1}}_B  \\
		&\qquad+
		\tfrac{1}{4}\pqty{\hbar V_1 + \hbar V_2 - \hbar V_3 - \hbar V_4}
		~ \hat{\mathbb{1}}_A\otimes\hat{\sigma}_{z,B}  +
		\tfrac{1}{4}\pqty{\hbar V_1 - \hbar V_2 - \hbar V_3 + \hbar V_4}
		~ \hat{\sigma}_{z,A}\otimes\hat{\sigma}_{z,B}
	\end{aligned}
\end{equation}
Upon the recognition that the $\hat{\sigma}_{z,A} \otimes \hat{\mathbb{1}}_B$ and $\hat{\mathbb{1}}_A\otimes\hat{\sigma}_{z,B}$ terms are single-qubit operations that can be grouped into the single-qubit Hamiltonians as adjustments to the single-qubit energies, and ignoring the global phase from $\hat{\mathbb{1}}_A \otimes \hat{\mathbb{1}}_B$, it leaves the $Z$-$Z$ coupling term
\begin{equation}
	\begin{aligned}
		&\tfrac{1}{4}\pqty{\hbar V_1 - \hbar V_2 - \hbar V_3 + \hbar V_4}
		~ \hat{\sigma}_{z,A}\otimes\hat{\sigma}_{z,B} \\
		&\qquad=
		\frac{1}{4}\pqty{
			\pqty{1-\tfrac{1}{N_A}}\pqty{1-\tfrac{1}{N_B}}C_{11}
			- \pqty{1-\tfrac{1}{N_A}}\pqty{\tfrac{\vec{\nu}}{N_B}}\cdot\vec{F}_B
			- \pqty{1-\tfrac{1}{N_B}}\pqty{\tfrac{\vec{\xi}}{N_A}}\cdot\vec{F}_A
			+ \pqty{\tfrac{\vec{\xi}}{N_A}}^T \overleftrightarrow{F}\pqty{\tfrac{\vec{\nu}}{N_B}}
		}~\hat{\sigma}_{z,A}\otimes\hat{\sigma}_{z,B}.
	\end{aligned}
\end{equation}
This agrees with previous work done on capacitive coupling that approximates a capacitively-coupled two-qubit interaction as an Ising Hamiltonian of the form $\hat{H}^{(\text{Ising})} =  -J\pqty{\hat{\sigma}_{z}\otimes\hat{\sigma}_{z}}$\cite{Watson2018,Frees2019}.


\subsection{Noise Autocorrelation Function}
In this paper, the control parameters are taken to have $1/f$ noise. This has the spectral density
\begin{equation}
	S_{n_i}(\omega) = \begin{cases}
		\Delta_{n_i}^2\frac{2\pi}{\omega_l}, & \text{for } \abs{\omega} \leq \omega_l\\
		\Delta_{n_i}^2\frac{2\pi}{\abs{\omega}}, & \text{for } \omega_l \leq \abs{\omega} \leq \omega_h \\
		0, & \text{otherwise}
	\end{cases},
\end{equation}
where $\Delta_{n_i}^2$ is the noise amplitude. The exact form of the auto-correlation function $C_{n_i}(t)$ can be worked out using the Wiener--Khinchin theorem \cite{Chatfield2003} to give
\begin{equation}
	\begin{aligned}
		C_{n_i}(t)
		&= \int_{-\infty}^{\infty}\frac{\dd{\omega}}{2\pi} e^{i\omega t} S_{n_i}(\omega) \\
		&= \Delta_{n_i}^2\pqty{
			\int_{-\omega_l}^{\omega_l}\dd{\omega}\frac{e^{i\omega t}}{\omega_l} +
			\int_{\omega_l}^{\omega_h}\dd{\omega}\frac{e^{i\omega t}}{\omega} +
			\int_{-\omega_h}^{-\omega_l}\dd{\omega}\frac{e^{i\omega t}}{-\omega}
		} \\
		&= \Delta_{n_i}^2\pqty{
			\int_{0}^{\omega_l}\dd{\omega}\frac{e^{i\omega t}+e^{-i\omega t}}{\omega_l} +
			\int_{\omega_l}^{\omega_h}\dd{\omega}\frac{e^{i\omega t}+e^{-i\omega t}}{\omega}
		} \\
		&= 2\Delta_{n_i}^2\pqty{
			\int_{0}^{\omega_l}\dd{\omega}\frac{\cos(\omega t)}{\omega_l} +
			\int_{\omega_l}^{\omega_h}\dd{\omega}\frac{\cos(\omega t)}{\omega}
		} \\
		&= 2\Delta_{n_i}^2\pqty{
			\frac{\sin(\omega_l t)}{\omega_l t} +
			\operatorname{Ci}\pqty{\omega_h t} -
			\operatorname{Ci}\pqty{\omega_l t}
		}.
	\end{aligned}
\end{equation}
Meanwhile, the noise amplitude can be found by relating the auto-correlation to the variance, $\lim_{t\to0}C_{n_i}(t) = \sigma_{n_i}^2$, given by
\begin{equation}
	\begin{aligned}
		\sigma_{n_i}^2 &= \lim_{t\to0}C_{n_i}(t) \\
		&= 2\Delta_{n_i}^2\lim_{t\to0}\pqty{
			\frac{\sin(\omega_l t)}{\omega_l t} +
			\operatorname{Ci}\pqty{\omega_h t} -
			\operatorname{Ci}\pqty{\omega_l t}
		} \\
		&= 2\Delta_{n_i}^2 \lim_{t\to0}\pqty{
			1 + \sum_{k=1}^{\infty}\pqty{-1}^k\frac{\pqty{\omega_l t}^{2k}}{\pqty{2k+1}!} +
			\pqty{\gamma + \ln(\omega_h t) + \sum_{k=1}^{\infty}\frac{\pqty{-\omega_h^2t^2}^k}{2k\pqty{2k}!}} -
			\pqty{\gamma + \ln(\omega_l t) + \sum_{k=1}^{\infty}\frac{\pqty{-\omega_l^2t^2}^k}{2k\pqty{2k}!}}
		} \\
		&= 2\Delta_{n_i}^2\Bigg(
		1 + \ln(\omega_h/\omega_l) +
		\lim_{t\to0}\sum_{k=1}^{\infty}\underbrace{\pqty{
				\pqty{-1}^k\frac{\pqty{\omega_l t}^{2k}}{\pqty{2k+1}!} +
				\frac{\pqty{-\omega_h^2t^2}^k}{2k\pqty{2k}!} -
				\frac{\pqty{-\omega_l^2t^2}^k}{2k\pqty{2k}!}
		}}_{\text{goes to } 0 \text{ when } t \rightarrow 0}
		\Bigg) \\
		&= 2\Delta_{n_i}^2\pqty{1 + \ln(\omega_h/\omega_l)}.
	\end{aligned}
\end{equation}
Hence, the auto-correlation function is
\begin{equation}
	C_{n_i}(t) = \sigma_{n_i}^2\times\overline{C}(t) = \sigma_{n_i}^2\times\frac{1}{1 + \ln(\omega_h/\omega_l)}\pqty{
		\frac{\sin(\omega_l t)}{\omega_l t} +
		\operatorname{Ci}\pqty{\omega_h t} -
		\operatorname{Ci}\pqty{\omega_l t}
	},
\end{equation}
where $\overline{C}(t)$ is the normalized auto-correlation function for $1/f$ noise.

A common appearance in the analytical formula for fidelity is the double integral $\varsigma(t) = \int_0^t\dd{t'}\int_0^{t'}\dd{t''} \overline{C}(t)$, which is
\begin{equation}
	\begin{aligned}
		\varsigma(t) &= \frac{1}{2\pqty{1+\ln(\omega_h/\omega_l)}}\bigg(
		\operatorname{Ci}(\omega_ht)t^2
		- \operatorname{Ci}(\omega_lt)t^2
		- \frac{\sin(\omega_ht)t}{\omega_h}
		+ \frac{\sin(\omega_lt)t}{\omega_l}
		+ \frac{2\operatorname{Si}(\omega_lt)t}{\omega_l}
		+ \frac{\cos(\omega_ht)-1}{\omega_h^2}
		+ \frac{\cos(\omega_lt)-1}{\omega_l^2}
		\bigg)
	\end{aligned}
\end{equation}
for $1/f$ noise. This can be easily differentiated twice to verify the result.


\subsection{\label{app:makhlin-invariants}Analytic Expression for Makhlin Invariants}
We use Makhlin invariants to determine if the interaction Hamiltonian is able to form a two-qubit CPHASE or CNOT gate \cite{Makhlin:2002p243}. Given the interaction Hamiltonian in Eq.~\eqref{eq:Hint}, we find the two Makhlin invariants to be:
\begin{align*}
G_1 &= \frac{1}{4}\exp(-i(V_1+V_2+V_3+V_4) t_0)\times\left[\exp(i(V_2+V_3) t_0)+\exp(i(V_1+V_4) t_0)\right]^2 \\
G_2 &= 2 + \frac{1}{2}\exp(-i(V_1+V_2+V_3+V_4) t_0)\times\left[\exp(2i(V_2+V_3) t_0)+\exp(2i(V_1+V_4) t_0)\right]
\end{align*}
and there exists a value of gate time $t_0$ where $G_1 = 0$ and $G_2 = 1$, which indicates that the two-qubit gate of interest is indeed a CNOT or CPHASE gate up to a sequence of local gates. Expressing $G_1$ and $G_2$ in their trigonometric identities, we find that the two-qubit gate time $t_0$ must fulfill:
\begin{align}
t_0 = \frac{k\pi}{(V_1-V_2-V_3+V_4)}
\label{eqn:makhlintime}
\end{align}
where $k$ is any odd integer value.


\subsection{\label{sec:fidelityEquivalence}Equivalence of Fidelity Definitions}
In this paper, we have used the operator fidelity for both the numerical simulations and the analytical calculations. However, as unitary operators cannot be directly measured in experiments, this figure of merit might not be very appropriate for use with experimental data. To enable a more direct comparison, and to ensure that our analysis is still applicable with experimental considerations, we show in this section that the operator fidelity is equivalent to a figure of merit more commonly used and directly measureable in experiments.

\subsubsection*{The Pauli Basis}
Every $n$-qubit operator $\hat{A}$ can be expanded into the Pauli basis
\begin{equation}
	\hat{A} = \sum_{i=0}^{d^2-1} a_i \hat{\sigma}_i^{(n)},
\end{equation}
where $d = 2^n$ is the dimension of the $n$-qubit Hilbert space. The Pauli basis operators for $n=1$ are the standard Pauli operators
\begin{align}
	\hat{\sigma}_0^{(1)} &= \hat{\mathbb{1}} &
	\hat{\sigma}_1^{(1)} &= \hat{\sigma}_x &
	\hat{\sigma}_2^{(1)} &= \hat{\sigma}_y &
	\hat{\sigma}_3^{(1)} &= \hat{\sigma}_z,
\end{align}
while the basis operators for $n$ qubits are formed via the tensor product of the single-qubit Pauli operators
\begin{equation}
	\hat{\sigma}_i^{(n)} \equiv \hat{\sigma}_{i_1}^{(1)} \otimes \hat{\sigma}_{i_2}^{(1)} \otimes \dots \otimes \hat{\sigma}_{i_n}^{(1)}.
\end{equation}
These Pauli operators form an orthogonal basis
\begin{equation}
	\tr(\hat{\sigma}_i^{(n)}\hat{\sigma}_j^{(n)}) = d\delta_{ij},
\end{equation}
which, together with the linearity of the trace, can be used to find the coefficients $a_i$ from the operator $\hat{A}$
\begin{equation}
	\begin{aligned}
		\frac{1}{d}\tr(\hat{\sigma}_i^{(n)}\hat{A}) &= \frac{1}{d}\tr(\hat{\sigma}_i^{(n)} \sum_{j} a_j\hat{\sigma}_j^{(n)})\\
		&= \sum_{j} \frac{a_j}{d}\underbrace{\tr(\hat{\sigma}_i^{(n)}\hat{\sigma}_j^{(n)})}_{d\delta_{ij}}\\
		&= a_i.
	\end{aligned}
\end{equation}

For the rest of the section, the superscripts of $\hat{\sigma}_{\square}^{(n)}$ will be dropped with the understanding that the appropriate Pauli operators will be used based on the number of qubits of the system, the ideal unitary operator will be written as $\hat{U} = \sum_iu_i\hat{\sigma}_i$, and the noisy unitary operator as $\widetilde{U} = \sum_i\tilde{u}_i\hat{\sigma}_i$.

\subsubsection*{Operator Fidelity}

The operator fidelity is defined~\cite{Green_2013} as
\begin{equation}
	\mathcal{F} \equiv \frac{1}{d^2}\abs{\tr(\hat{U}^\dag\widetilde{U})}^2
	\label{eq:operatorFidelity},
\end{equation}
where $d$ is the dimension of the Hilbert space, $\hat{U}$ is the ideal operator, and $\widetilde{U}$ is its noisy implementation. $\mathcal{F} = 1$ only in the noiseless case where $\widetilde{U} = \hat{U} \implies \hat{U}^\dag\widetilde{U} = \hat{\mathbb{1}}$. In the presence of noise, $\mathcal{F} < 1$, and the fidelity decreases with an increase in noise.

Following from the definition, the operator fidelity in terms of the Pauli basis coefficients is
\begin{align}
	\mathcal{F}
	&= \frac{1}{d^2}\abs{\tr(\hat{U}^\dagger \widetilde{U})}^2 \nonumber\\
	&= \frac{1}{d^2}\abs{\tr(\sum_{i} u_i^*\hat{\sigma}_i \sum_j \tilde{u}_j \hat{\sigma}_j)}^2 \nonumber\\
	&= \frac{1}{d^2}\abs{\sum_{ij}u_i^*\tilde{u}_j\operatorname{tr}(\hat{\sigma}_i \hat{\sigma}_j)}^2\\
	&= \frac{1}{d^2}\abs{\sum_{ij}u_i^*\tilde{u}_jd\delta_{ij}}^2 \nonumber\\
	&= \abs{\sum_{i}u_i^*\tilde{u}_i}^2 \nonumber
\end{align}

\subsubsection*{Process Fidelity}
In \cite{Yang2019}, the process fidelity is defined as
\begin{equation}
	\mathcal{F}^{(p)} = \tr(\chi\widetilde{\chi})\label{eq:processFidelity},
\end{equation}
where the $\chi$ is the process matrix, whose elements $\chi_{ij}$ are the coefficients that appear when expanding a quantum process in the Pauli basis
\begin{equation}
	\mathcal{E}\pqty{\hat{\rho}} = \sum_{ij} \chi_{ij}\, \hat{\sigma}_i\hat{\rho}\hat{\sigma}_j.
\end{equation}
For an ideal unitary map, this is
\begin{equation}
	\begin{aligned}
		\mathcal{U}\pqty{\hat{\rho}}
		&= \hat{U}\hat{\rho}\hat{U}^\dagger \\
		&= \pqty{\sum_iu_i\hat{\sigma}_i}\hat{\rho}\pqty{\sum_ju_j^*\hat{\sigma}_j} \\
		&= \sum_{ij} \underbrace{u_iu_j^*}_{\chi_{ij}} ~ \hat{\sigma}_i\hat{\rho}\hat{\sigma}_j.
	\end{aligned}
\end{equation}
Similarly, the noisy unitary map has the process matrix elements $\tilde{\chi}_{ij} = \tilde{u}_i\tilde{u}_j^*$. Placing this back into Eq.~\eqref{eq:processFidelity},
\begin{align}
	\mathcal{F}^{(p)} &= \tr(\chi\widetilde{\chi}) \nonumber\\
	&= \sum_{ij} \chi_{ij}\widetilde{\chi}_{ji} \nonumber\\
	&= \sum_{ij} u_iu_j^* \tilde{u}_j\tilde{u}_i^* \\
	&= \pqty{\sum_{j} u_j^*\tilde{u}_j}
	\pqty{\sum_{i} u_i\tilde{u}_i^*} \nonumber\\
	&= \abs{ \sum_{i}u_i^*\tilde{u}_i}^2 \nonumber
\end{align}
Therefore, we show that the operator fidelity and process fidelity are equivalent.

\subsubsection*{Experimentally Measured Fidelity and the Ensemble Average}
In actual setups, the fidelity is obtained by first measuring the process matrix elements through an experimental procedure~\cite{James2001,Obrien2004,Yamamoto2010}.~As the experimentally-determined process matrix requires the averaging of repeated measurements of the matrix elements to obtain an expected value of $\ev{\widetilde{\chi}_{ij}}$, what can actually be measured in an experiment is in fact
\begin{equation}
	\mathcal{F}^{(\text{exp})} =
	\tr(\chi\!\ev{\widetilde{\chi}}) =
	\ev{\tr(\chi\widetilde{\chi})} =
	\ev{\mathcal{F}},
\end{equation}
which is the average fidelity. The cumulant average can be brought out of the trace as the trace is a linear operation.

Since the two definitions of fidelity are equivalent, we report the operator fidelity Eq.~\eqref{eq:operatorFidelity}, averaged over noise realizations, to emulate the result of experimentally-determined fidelities.


\subsection{Analytical Derivation of Fidelity}
\subsubsection*{Time Evolution in the Presence of Noise}
Consider the control parameters $\mathbf{n}_{A/B} \equiv(
t_{lA/B},~t_{rA/B},~\varepsilon_{A/B},~\varepsilon_{mA/B}
)^T$. Note that this notation is slightly different from the main text, where the parameters are instead combined into a single vector $\vec{n} = (\mathbf{n}_A,~\mathbf{n}_B)^T$. In the presence of noise, the noisy parameters $\tilde{\mathbf{n}}_{A/B}(t) = \mathbf{n}_{A/B} + \delta\mathbf{n}_{A/B}(t)$ introduce fluctuations into the interaction Hamiltonian. We take these fluctuations to have zero mean as we are assuming that upon many repetitions of the experiment, the noisy parameters average out to the target (ideal) parameters $\ev{\tilde{\mathbf{n}}_{A/B}(t)} = \mathbf{n}_{A/B}$.

The resulting interaction due to the noise can be written as the ideal Hamiltonian with a noise term
\begin{equation}
	\widetilde{H}_{\text{int}} =
	\hat{H}_{\text{int}} + \delta\hat{H}_\text{int}.
\end{equation}
As we do not expect the magnitude of the noise to be very large in comparison to the control parameters, $\delta\hat{H}_\text{int}$ can be found by taking the first term of the series expansion of $\hat{H}$. This means that
\begin{equation}
	\delta\hat{H}_\text{int} \approx \grad_{\!A}\hat{H}_\text{int} ~ \delta\mathbf{n}_A
	+ \grad_{\!B}\hat{H}_\text{int} ~ \delta\mathbf{n}_B
\end{equation}
where $\grad_{A/B} = (
	\partial_{t_{lA/B}},~
	\partial_{t_{rA/B}},~
	\partial_{\varepsilon_{A/B}},~
	\partial_{\varepsilon_{mA/B}}
)$. In addition, only the noise terms with coefficients to the linear order of $t_{l/r,A/B}/(U+\dots)$ are kept. This is because the tunneling parameters are much smaller in magnitude than the inter/intra-dot energies when operating a triple quantum dot qubit.

Writing the normalization factor as $N_{A} = 1+\vec{\xi}\cdot\vec{1}$ and $N_{B} = 1+\vec{\nu}\cdot\vec{1}$ where $\vec{1} =
\{1,1,1,1\}^T$, the last diagonal term in $\delta\hat{H}$ is
\begin{align}
	\hbar\delta V_4
	&\approx  \grad_{\!A}\bqty{\frac{
			C_{11} +
			\vec{\xi}\cdot\vec{F}_{A} +
			\vec{\nu}\cdot\vec{F}_{B} +
			{\vec{\xi}}^T\overleftrightarrow{F}\vec{\nu}
		}{N_AN_B}}\delta\mathbf{n}_A +
	\grad_{\!B}\bqty{\frac{C_{11} +
			\vec{\xi}\cdot\vec{F}_A +
			\vec{\nu}\cdot\vec{F}_B +
			{\vec{\xi}}^T\overleftrightarrow{F}\vec{\nu}}{N_A N_B}
	}\delta\mathbf{n}_B \nonumber\\
	&= \frac{1}{N_A N_B}\Big(
	{\vec{F}_A}^T \grad_{\!A} \vec{\xi} ~ \delta\mathbf{n}_A +
	{\vec{F}_B}^T \grad_{\!B} \vec{\nu} ~ \delta\mathbf{n}_B +
	{\vec{\xi}}^T\overleftrightarrow{F}\grad_{\!B}\vec{\nu} ~ \delta\mathbf{n}_B +
	{\vec{\nu}}^T\overleftrightarrow{F}^T\grad_{\!A}\vec{\xi} ~ \delta\mathbf{n}_A \nonumber\\
	&\qquad-
	\pqty{
		N_B\grad_{\!A}N_A~\delta\mathbf{n}_A +
		N_A\grad_{\!B}N_B~\delta\mathbf{n}_B
	}~\hbar V_4
	\Big) \nonumber\\
	&=
	\frac{1}{N_A N_B}\bigg(
	{\vec{F}_A}^T \grad_{\!A} \vec{\xi} ~ \delta\mathbf{n}_A +
	{\vec{F}_B}^T \grad_{\!B} \vec{\nu} ~ \delta\mathbf{n}_B +
	{\vec{\xi}}^T\overleftrightarrow{F}\grad_{\!B}\vec{\nu} ~ \delta\mathbf{n}_B +
	{\vec{\nu}}^T\overleftrightarrow{F}^T\grad_{\!A}\vec{\xi} ~ \delta\mathbf{n}_A \nonumber\\
	&\qquad-
	\pqty{
		\pqty{1+\vec{\nu}\cdot\vec{1}}
		\pqty{\vec{1}^T\grad_{\!A}\vec{\xi}} ~ \delta\mathbf{n}_A +
		\pqty{1+\vec{\xi}\cdot\vec{1}}
		\pqty{\vec{1}^T\grad_{\!B}\vec{\nu}} ~ \delta\mathbf{n}_B
	}~\hbar V_4\bigg)\nonumber\\
	&=
	\frac{1}{N_A N_B}\bigg(
	{\vec{F}_A}^T \grad_{\!A} \vec{\xi} ~ \delta\mathbf{n}_A +
	{\vec{F}_B}^T \grad_{\!B} \vec{\nu} ~ \delta\mathbf{n}_B -
	\pqty{
		{\vec{1}^T\grad_{\!A}\vec{\xi}}  ~ \delta\mathbf{n}_A +
		{\vec{1}^T\grad_{\!B}\vec{\nu}} ~ \delta\mathbf{n}_B
	}~\hbar V_4\nonumber\\
	&\qquad+
	{\vec{\xi}}^T\overleftrightarrow{F}\grad_{\!B}\vec{\nu} \;\delta\mathbf{n}_B +
	{\vec{\nu}}^T\overleftrightarrow{F}^T\grad_{\!A}\vec{\xi} \;\delta\mathbf{n}_A -
	\pqty{
		\pqty{\vec{\nu}\cdot\vec{1}}
		\pqty{{\vec{1}}^T\grad_{\!A}\vec{\xi}} \delta\mathbf{n}_A +
		\pqty{\vec{\xi}\cdot\vec{1}}
		\pqty{{\vec{1}}^T\grad_{\!B}\vec{\nu}} \delta\mathbf{n}_B
	}~\hbar V_4
	\bigg).
\end{align}
The second row of the final expression clearly has terms larger than $\mathcal{O}(\xi,\nu)=\mathcal{O}(\alpha^2,\beta^2)$, so they will be dropped. Meanwhile, since the first row already includes linear-ordered terms, the normalisation also has to be expanded to $1/N_{A/B} = 1 - \mathcal{O}(\alpha^2,\beta^2)$. Finally, the only term that survives from $\hbar V_4$ on the right-hand side of the equation is the constant $C_{11}$, as it is multiplied by a factor that is already of the order $\mathcal{O}(\alpha,\beta)$. Therefore, $\hbar \delta V_4$ is
\begin{equation}
	\hbar \delta V_4 = \pqty{\vec{F}_A-C_{11}\vec{1}}^T \grad_{\!A} \vec{\xi} ~ \delta\mathbf{n}_A +
	\pqty{\vec{F}_B-C_{11}\vec{1}}^T \grad_{\!B} \vec{\nu} ~ \delta\mathbf{n}_B + \mathcal{O}(\alpha^2,\beta^2)
\end{equation}
Applying a similar process to the remaining matrix elements,
\begin{align}
	\label{eq:noise-diagonal-elements0}
	\hbar \delta V_1 &\approx 0 \\
	\hbar \delta V_2 &\approx \pqty{\vec{F}_B-C_{11}\vec{1}}^T \grad_{\!B} \vec{\nu} ~ \delta\mathbf{n}_B\\
	\hbar \delta V_3 &\approx \pqty{\vec{F}_A-C_{11}\vec{1}}^T \grad_{\!A} \vec{\xi} ~ \delta\mathbf{n}_A \\
	\hbar \delta V_4 &\approx  \pqty{\vec{F}_A-C_{11}\vec{1}}^T \grad_{\!A} \vec{\xi} ~ \delta\mathbf{n}_A +
	\pqty{\vec{F}_B-C_{11}\vec{1}}^T \grad_{\!B} \vec{\nu} ~ \delta\mathbf{n}_B.\label{eq:noise-diagonal-elements}
\end{align}
The time evolution of the system in the presence of noise can be then found from
the noisy Hamiltonian.
\begin{equation}
	\begin{aligned}
		\widetilde{U}(t)
		&=
		\mathcal{T}_+ \exp\Bqty{-\frac{i}{\hbar}\int_0^{t}\dd{t'}\widetilde{H}_\text{int}(t')}\\
		&=
		\underbrace{\exp\Bqty{-\frac{it}{\hbar}\hat{H}_\text{int}}}_{\equiv\hat{U}(t)}
		\underbrace{\exp\Bqty{-\frac{i}{\hbar}\int_0^{t}\dd{t'}\delta\hat{H}_\text{int}(t')}}_{\equiv\hat{U}_\delta(t)}\label{eq:noise-time-operator}
	\end{aligned}
\end{equation}
Here, $\hat{U}(t)$ is the ideal time evolution of the system in the absence of noise, while $\hat{U}_\delta(t)$ consists of just the contributions due to noise. Note that the second step can be arrived from the first step only because
\begin{enumerate}
	\item
	The interaction Hamiltonian, its ideal term, and its noise term are all diagonal. So, the time-ordering operator is not required as they commute among themselves, and for the same reason, the sum in the exponent can be seperated into a product of exponentials
	\item
	The control parameters are constant over the operation of the two-qubit gate, so the only time-dependent terms are their fluctuations
\end{enumerate}

\subsubsection*{Cumulant Expansion of Fidelity}
Considering the operator fidelity, Eq.~\eqref{eq:operatorFidelity} in the context of Eq.~\eqref{eq:noise-time-operator}, the fidelity of the
noisy two-qubit interaction is
\begin{equation}
	\begin{aligned}
		\mathcal{F} &= \frac{1}{4^2}\abs{\tr(\hat{U}^\dag\widetilde{U})}^2\\
		&= \frac{1}{16}\abs{\tr(\hat{U}^\dag\pqty{\hat{U}\hat{U}_\delta})}^2 \\
		&= \frac{1}{16}\abs{\tr(\hat{U}_\delta)}^2.
	\end{aligned}
\end{equation}

In an actual setup, this fidelity will be obtained through the repetition of many experiments --- or in other words, over many noise realizations. This means that the final value obtained will be a cumulant average
\begin{equation}
	\begin{aligned}
		\ev{\mathcal{F}}
		&= \frac{1}{16}\ev{\abs{
				\tr(\exp\Bqty{-\frac{i}{\hbar}\int_0^{t}\dd{t'}\delta\hat{H}_\text{int}(t')}
				)
			}^2} \\
		&= \frac{1}{16}\ev{\abs{
				\sum_i\exp\Bqty{-i\int_0^{t}\dd{t'}\delta V_i(t')}
			}^2} \\
		&= \frac{1}{16}
		\sum_{i,j}\ev{\exp\Bqty{
				-i\int_0^{t}\dd{t'}
				\pqty{\delta V_i(t')-\delta V_j(t')}
		}}.
	\end{aligned}
\end{equation}
This averaged operator fidelity is the same as the experimentally-measurable fidelity --- this equivalence is laid out in Sec. \ref{sec:fidelityEquivalence}.

From the cumulant expansion~\cite{Kubo1962}, to the first order,
\begin{equation}
	\begin{aligned}
		\ev{\mathcal{F}}
		&\approx \frac{1}{16}
		\sum_{i,j}\exp\Bqty{
			-i\int_0^{t}\dd{t'}
			\underbrace{\ev{\delta V_i(t')-\delta V_j(t')}}_{0}
			-\frac{1}{2}\int_0^{t}\dd{t'}\int_0^{t'}\dd{t''}
			\ev{\pqty{\delta V_i(0)-\delta V_j(0)}\pqty{\delta V_i(t'')-\delta V_j(t'')}}
		} \\
		&= \frac{1}{16}
		\sum_{i\neq j}\exp\Bqty{
			-\frac{1}{2}\int_0^{t}\dd{t'}\int_0^{t'}\dd{t''}
			\ev{\delta V_i(0)\delta V_i(t'')}
			- 2\ev{\delta V_i(0)\delta V_j(t'')}
			+ \ev{\delta V_j(0)\delta V_j(t'')}
		},
	\end{aligned}
\end{equation}
where the first term vanishes as the noise has zero mean. In addition, it is assumed that the correlation is only defined by the time difference between the two factors, so $\ev{\delta V_i(0)\delta V_j(t'')} = \ev{\delta V_i(t'')\delta V_j(0)}$.

This expression can be further simplified by substituting $\delta V_1=0$ and $\delta V_4 = \delta V_2 + \delta V_3$ from Eq.~\eqref{eq:noise-diagonal-elements0} to Eq.~\eqref{eq:noise-diagonal-elements}
\begin{equation}
	\begin{aligned}
		\ev{\mathcal{F}}
		&= \frac{1}{4} +
		\frac{1}{4}\exp\Bqty{
			-\frac{1}{2}\int_0^{t}\dd{t'}\int_0^{t'}\dd{t''}
			\ev{\delta V_2(0)\delta V_2(t'')}
		} +
		\frac{1}{4}\exp\Bqty{
			-\frac{1}{2}\int_0^{t}\dd{t'}\int_0^{t'}\dd{t''}
			\ev{\delta V_3(0)\delta V_3(t'')}
		} \\
		&\qquad~ +
		\frac{1}{8}\exp\Bqty{
			-\frac{1}{2}\int_0^{t}\dd{t'}\int_0^{t'}\dd{t''}
			\ev{\delta V_2(0)\delta V_2(t'')}
			- 2\ev{\delta V_2(0)\delta V_3(t'')}
			+ \ev{\delta V_3(0)\delta V_3(t'')}
		} \\
		&\qquad~ +
		\frac{1}{8}\exp\Bqty{
			-\frac{1}{2}\int_0^{t}\dd{t'}\int_0^{t'}\dd{t''}
			\ev{\delta V_2(0)\delta V_2(t'')}
			+ 2\ev{\delta V_2(0)\delta V_3(t'')}
			+ \ev{\delta V_3(0)\delta V_3(t'')}
		}.
	\end{aligned}
\end{equation}

Finally, $\ev{\delta V_2(0)\delta V_3(t'')}$ can be taken to be zero if there is no correlation between the noise in the two qubits,
as $\delta V_2$ only includes terms in qubit B and $\delta V_3$
only includes terms in qubit A
\begin{equation}
	\begin{aligned}
		\ev{\mathcal{F}}
		&= \frac{1}{4} +
		\frac{1}{4}\exp\Bqty{
			-\frac{1}{2}\int_0^{t}\dd{t'}\int_0^{t'}\dd{t''}
			\ev{\delta V_2(0)\delta V_2(t'')}
		} +
		\frac{1}{4}\exp\Bqty{
			-\frac{1}{2}\int_0^{t}\dd{t'}\int_0^{t'}\dd{t''}
			\ev{\delta V_3(0)\delta V_3(t'')}
		} \\
		&\qquad~ +
		\frac{1}{4}\exp\Bqty{
			-\frac{1}{2}\int_0^{t}\dd{t'}\int_0^{t'}\dd{t''}
			\ev{\delta V_2(0)\delta V_2(t'')} + \ev{\delta V_3(0)\delta V_3(t'')}
		} \\
		&= \frac{1}{4}\pqty{1+\exp\Bqty{
				-\frac{1}{2}\int_0^{t}\dd{t'}\int_0^{t'}\dd{t''}
				\ev{\delta V_2(0)\delta V_2(t'')}
		}}\pqty{1+\exp\Bqty{
				-\frac{1}{2}\int_0^{t}\dd{t'}\int_0^{t'}\dd{t''}
				\ev{\delta V_3(0)\delta V_3(t'')}
		}}
	\end{aligned}
\end{equation}

As the expression for $\delta V_{2}$ and $\delta V_{3}$ are identical,
it suffices to work out just one of the expressions and apply it to the other by switching $A \leftrightarrow B$ and $\vec{\xi}\leftrightarrow \vec{\nu}$. However, note that this similarity is purely symbolic, in the sense that $\vec{F}_{A}$ and $\vec{F}_{B}$ do not share the same numerical values in general, so care must be taken to ensure that $\vec{F}_A$ always appears alongside $\vec{\xi}$ in $\delta V_3$ and $\vec{F}_B$ alongside $\vec{\nu}$ in $\delta V_2$. With that in mind, the qubit subscripts will be neglected in the intermediate steps.
\begin{equation}
	\begin{aligned}
		\ev{\delta V(0)\delta V(t'')}
		&=
		\ev{
			\pqty{\vec{F}-C_{11}\vec{1}}^T \grad \vec{\xi} ~ \delta\mathbf{n}(0)
			\pqty{\vec{F}-C_{11}\vec{1}}^T \grad \vec{\xi} ~ \delta\mathbf{n}(t'')
		} \\
		&=
		\sum_{i,j}
		\ev{
			\pqty{\vec{F}-C_{11}\vec{1}}^T \partial_i \vec{\xi} ~ \delta n_i(0) \;
			\pqty{\vec{F}-C_{11}\vec{1}}^T \partial_j \vec{\xi} ~ \delta n_j(t'')
		}\\
		&=
		\sum_{i,j}
		\pqty{\pqty{\vec{F}-C_{11}\vec{1}}^T \partial_i \vec{\xi}}
		\pqty{\pqty{\vec{F}-C_{11}\vec{1}}^T \partial_j \vec{\xi}}
		\ev{\delta n_i(0)\delta n_j(t'')}
	\end{aligned}
\end{equation}

Here, two further simplifying assumptions are made:
\begin{enumerate}
	\item The noise in the control parameters are independent of each other. Hence,
	\begin{equation}
		\ev{\delta n_i(0)\delta n_j(t)}
		= \delta_{ij}\ev{\delta n_i(0)\delta n_i(t)}
	\end{equation}
	where $\delta_{ij}$ is the Kronecker delta
	\item
	Every noise statistics follow the same behaviour, in the sense that apart from the scale, the noise distributions are identical. So,
	\begin{equation}
		\ev{\delta n_i(0)\delta n_i(t)}
		= C_{n_i}(t) = \sigma_{n_i}^2 \overline{C}(t)
	\end{equation}
	where $\sigma_{n_i}$ is the standard deviation of the noise in
	control parameter $n_i$,
	and $\overline{C}(t)$ is the autocorrelation function normalized to $\overline{C}(0) = 1$, so that $C_{n_i}(0) = \sigma_{n_i}^2$
\end{enumerate}
If necessary, these assumptions can be relaxed by assigning an auto-correlation function $\overline{C}_{n_i,n_j}(t)$ to the (auto-)correlation of parameter(s). For the purposes of our analysis, however, the simplifications above suffices to investigate the effects of noise on the fidelity of our system.

With that,
\begin{equation}
	\begin{aligned}
		\ev{\delta V(0)\delta V(t'')}
		&=
		\sum_{i,j}
		\pqty{\pqty{\vec{F}-C_{11}\vec{1}}^T \partial_i \vec{\xi}}
		\pqty{\pqty{\vec{F}-C_{11}\vec{1}}^T \partial_j \vec{\xi}}
		\delta_{ij}\sigma_{n_i}^2 C(t) \\
		&=
		\sum_{i}\pqty{\pqty{\vec{F}-C_{11}\vec{1}}^T\partial_i \vec{\xi}}^2\sigma_{n_i}^2 C(t)
	\end{aligned}
\end{equation}

Therefore, the fidelity is
\begin{equation}
	\begin{aligned}
		\ev{\mathcal{F}}
		&=
		\frac{1}{4}\pqty{1+\exp\Bqty{
				-\frac{1}{2}\sum_{i}\pqty{\pqty{\vec{F}_A-C_{11}\vec{1}}^T \partial_i \vec{\xi}}^2\sigma_{n_{iA}}^2
				\varsigma(t)
		}}\pqty{1+\exp\Bqty{
				-\frac{1}{2}\sum_{i}\pqty{\pqty{\vec{F}_B-C_{11}\vec{1}}^T \partial_i \vec{\nu}}^2\sigma_{n_{iB}}^2
				\varsigma(t)
		}}
	\end{aligned}\label{eq:fidelity-formula}
\end{equation}
which leaves the time dependence of the fidelity purely in $\varsigma(t)\equiv\int_0^{t}\dd{t'}\int_0^{t'}\dd{t''} C(t'')$,
the double integral of the autocorrelation function.

This final form is instructive in approaching the optimization of the
control parameters to maximize the fidelity of the two-qubit gate.
As $\sigma_{n_i}$ and $C(t)$ arise from the noise and cannot be controlled,
its coefficients ${\vec{F}_A}^T \partial_i \vec{\xi}$ and ${\vec{F}_B}^T \partial_i \vec{\nu}$ are the parameters to be minimized instead.

These coefficients for qubit A are, explicitly,
\begin{subequations}
	\begin{align}
		\pqty{\vec{F}_A-C_{11}\vec{1}}^T \partial_{t_{lA}} \vec{\xi}
		&= \frac{t_{lA}\pqty{C_{31}-C_{11}}}{\pqty{U-U''+2\varepsilon_{A}}^2}
		+ \frac{t_{lA}\pqty{C_{51}-C_{11}}}{\pqty{U-2U'+U''-\varepsilon_{mA}+\varepsilon_A}^2}\label{eq:coef-tlA}\\
		\pqty{\vec{F}_A-C_{11}\vec{1}}^T \partial_{t_{rA}} \vec{\xi}
		&= \frac{t_{rA}\pqty{C_{21}-C_{11}}}{\pqty{U-U''-2\varepsilon_{A}}^2}
		+ \frac{t_{rA}\pqty{C_{41}-C_{11}}}{\pqty{U-2U'+U''-\varepsilon_{mA}-\varepsilon_A}^2}\label{eq:coef-trA}\\
		\pqty{\vec{F}_A-C_{11}\vec{1}}^T \partial_{\varepsilon_{A}} \vec{\xi}
		&= \frac{2t_{rA}^2\pqty{C_{21}-C_{11}}}{\pqty{U-U''-2\varepsilon_{A}}^3}
		- \frac{2t_{lA}^2\pqty{C_{31}-C_{11}}}{\pqty{U-U''+2\varepsilon_{A}}^3} \nonumber\\
		&\qquad\quad
		+ \frac{t_{rA}^2\pqty{C_{41}-C_{11}}}{\pqty{U-2U'+U''-\varepsilon_{mA}-\varepsilon_A}^3}
		- \frac{t_{lA}^2\pqty{C_{51}-C_{11}}}{\pqty{U-2U'+U''-\varepsilon_{mA}+\varepsilon_A}^3} \label{eq:coef-epA}\\
		\pqty{\vec{F}_A-C_{11}\vec{1}}^T \partial_{\varepsilon_{mA}} \vec{\xi}
		&= \frac{t_{rA}^2\pqty{C_{41}-C_{11}}}{\pqty{U-2U'+U''-\varepsilon_{mA}-\varepsilon_A}^3}
		+ \frac{t_{lA}^2\pqty{C_{51}-C_{11}}}{\pqty{U-2U'+U''-\varepsilon_{mA}+\varepsilon_A}^3}\label{eq:coef-epmA}
	\end{align}  \label{coef-A-all}
\end{subequations}
and for qubit B,
\begin{subequations}
	\begin{align}
		\pqty{\vec{F}_B-C_{11}\vec{1}}^T \partial_{t_{lB}} \vec{\xi}
		&= \frac{t_{lB}\pqty{C_{13}-C_{11}}}{\pqty{U-U''+2\varepsilon_{B}}^2}
		+ \frac{t_{lB}\pqty{C_{15}-C_{11}}}{\pqty{U-2U'+U''-\varepsilon_{mB}+\varepsilon_B}^2}\\
		\pqty{\vec{F}_B-C_{11}\vec{1}}^T \partial_{t_{rB}} \vec{\xi}
		&= \frac{t_{rB}\pqty{C_{12}-C_{11}}}{\pqty{U-U''-2\varepsilon_{B}}^2}
		+ \frac{t_{rB}\pqty{C_{14}-C_{11}}}{\pqty{U-2U'+U''-\varepsilon_{mB}-\varepsilon_B}^2}\\
		\pqty{\vec{F}_B-C_{11}\vec{1}}^T \partial_{\varepsilon_{B}} \vec{\xi}
		&= \frac{2t_{rB}^2\pqty{C_{12}-C_{11}}}{\pqty{U-U''-2\varepsilon_{B}}^3}
		- \frac{2t_{lB}^2\pqty{C_{13}-C_{11}}}{\pqty{U-U''+2\varepsilon_{B}}^3} \nonumber\\
		&\qquad\quad
		+ \frac{t_{rB}^2\pqty{C_{14}-C_{11}}}{\pqty{U-2U'+U''-\varepsilon_{mB}-\varepsilon_B}^3}
		- \frac{t_{lB}^2\pqty{C_{15}-C_{11}}}{\pqty{U-2U'+U''-\varepsilon_{mB}+\varepsilon_B}^3}\\
		\pqty{\vec{F}_B-C_{11}\vec{1}}^T \partial_{\varepsilon_{mB}} \vec{\xi}
		&= \frac{t_{rB}^2\pqty{C_{14}-C_{11}}}{\pqty{U-2U'+U''-\varepsilon_{mB}-\varepsilon_B}^3}
		+ \frac{t_{lB}^2\pqty{C_{15}-C_{11}}}{\pqty{U-2U'+U''-\varepsilon_{mB}+\varepsilon_B}^3}
	\end{align}  \label{coef-B-all}
\end{subequations}


\subsection{Two-Qubit Sweet Spots}
As defined in the main text, two-qubit sweet spots (2QSS) are points in the parameter space where the first-order derivatives (Eqs.~\eqref{eq:noise-diagonal-elements0}--\eqref{eq:noise-diagonal-elements}) vanish. While the 2QSS do not arise from the maximization of the fidelity, the same first-order derivatives also appear as coefficients in the exponent of the fidelity in Eqs.~\eqref{coef-A-all} \& \eqref{coef-B-all}. Hence, when operating at the sweet spot, the fidelity is also maximized as a consequence.

\subsubsection*{Tunneling Sweet Spots}
Before attempting to find the sweet spots, we shall take a closer look at the coefficients $\vec{F}_{A/B}- C_{11}\vec{1}$
\begin{equation}
	\begin{aligned}
		\vec{F}_{A} - C_{11}\vec{1}
		&= \pmqty{
			C_{21} - C_{11} \\
			C_{31} - C_{11} \\
			C_{41} - C_{11} \\
			C_{51} - C_{11}
		} \\
		&= \sum_{j=4}^6\pmqty{
			\pqty{\mathcal{V}_{1j} + 2\mathcal{V}_{3j}} -  \pqty{\mathcal{V}_{1j} + \mathcal{V}_{2j} + \mathcal{V}_{3j}} \\
			\pqty{2\mathcal{V}_{1j} + \mathcal{V}_{3j}} -  \pqty{\mathcal{V}_{1j} + \mathcal{V}_{2j} + \mathcal{V}_{3j}} \\
			\pqty{\mathcal{V}_{1j} + 2\mathcal{V}_{2j}} -  \pqty{\mathcal{V}_{1j} + \mathcal{V}_{2j} + \mathcal{V}_{3j}} \\
			\pqty{2\mathcal{V}_{2j} + \mathcal{V}_{3j}} -  \pqty{\mathcal{V}_{1j} + \mathcal{V}_{2j} + \mathcal{V}_{3j}} \\
		} \\
		&= \sum_{j=4}^6 \pmqty{
			\pqty{\mathcal{V}_{1j} + 2\mathcal{V}_{3j}} -  \pqty{\mathcal{V}_{1j} + \mathcal{V}_{2j} + \mathcal{V}_{3j}} \\
			\pqty{\mathcal{V}_{1j} + \mathcal{V}_{2j} + \mathcal{V}_{3j}} -  \pqty{2\mathcal{V}_{2j} + \mathcal{V}_{3j}} \\
			\pqty{\mathcal{V}_{1j} + \mathcal{V}_{2j} + \mathcal{V}_{3j}} -  \pqty{\mathcal{V}_{1j} + 2\mathcal{V}_{3j}} \\
			\pqty{2\mathcal{V}_{2j} + \mathcal{V}_{3j}} -  \pqty{\mathcal{V}_{1j} + \mathcal{V}_{2j} + \mathcal{V}_{3j}} \\
		}
	\end{aligned}
\end{equation}
\begin{equation}
	\implies
	\begin{aligned}
		C_{31} - C_{11} &= -\pqty{C_{51} - C_{11}}, \\
		C_{41} - C_{11} &= -\pqty{C_{21} - C_{11}}.\label{eq:col-coef-negative-A}
	\end{aligned}
\end{equation}
Following the same steps with $\vec{F}_{B}$ leads us to
\begin{equation}
	\begin{aligned}
		C_{12} - C_{11} &= -\pqty{C_{14} - C_{11}}, \\
		C_{15} - C_{11} &= -\pqty{C_{13} - C_{11}}.\label{eq:col-coef-negative-B}
	\end{aligned}
\end{equation}
With this in hand, Eqs.~\eqref{eq:coef-tlA} \& \eqref{eq:coef-trA} give
\begin{align}
	\pqty{\vec{F}_A-C_{11}\vec{1}}^T \partial_{t_{lA}} \vec{\xi}
	&= -t_{lA}\pqty{C_{51}-C_{11}}\pqty{\frac{1}{\pqty{U-U''+2\varepsilon_{A}}^2}
		- \frac{1}{\pqty{U-2U'+U''-\varepsilon_{mA}+\varepsilon_A}^2}}\\
	\pqty{\vec{F}_A-C_{11}\vec{1}}^T \partial_{t_{rA}} \vec{\xi}
	&= t_{rA}\pqty{C_{21}-C_{11}}\pqty{\frac{1}{\pqty{U-U''-2\varepsilon_{A}}^2}
		- \frac{1}{\pqty{U-2U'+U''-\varepsilon_{mA}-\varepsilon_A}^2}}.
\end{align}
When the terms in the larger brackets go to zero, we get the sweet spots for the tunneling parameters. This occurs when the denominators are equal

\begin{subequations}
	\begin{minipage}[t]{.49\columnwidth}
		\begin{equation}
			\begin{aligned}
				U-U''\pm2\varepsilon_{A} &=U-2U'+U''-\varepsilon_{mA}\pm\varepsilon_A\\
				\varepsilon_{mA} &= \mp \varepsilon_A -2\pqty{U' - U''},\label{eq:tltrSS}
			\end{aligned}
		\end{equation}\\
	\end{minipage}%
	\begin{minipage}[t]{.02\columnwidth}
		\vspace{3mm}\hspace{3mm}
	\end{minipage}%
	\begin{minipage}[t]{.49\columnwidth}
		\begin{equation}
			\begin{aligned}
				-(U-U'')\mp2\varepsilon_{A} &=U-2U'+U''-\varepsilon_{mA}\pm\varepsilon_A \\
				\varepsilon_{mA} &= \pm 3\varepsilon_A +2\pqty{U-U'}.\label{eq:tltrSS2}
			\end{aligned}
		\end{equation}
	\end{minipage}\\
\end{subequations}
The equations with $-\varepsilon_A$ and $+3\varepsilon_A$ terms are the $t_{lA}$ two-qubit sweet spots ($t_{lA}$ SS), while the $+\varepsilon_A$ and $-3\varepsilon_A$ terms are the $t_{rA}$ SS. The corresponding equations for qubit B can be found by swapping $t_{lA}\to t_{rB}$, $t_{rA}\to t_{lB}$, and $\varepsilon_{mA} \to -\varepsilon_{mB}$.

\begin{subequations}
	\begin{minipage}[t]{.49\columnwidth}
		\begin{equation}
			\begin{aligned}
				\varepsilon_{mB} &= \mp \varepsilon_B -2\pqty{U' - U''},\label{eq:tltrbSS}
			\end{aligned}
		\end{equation}\\
	\end{minipage}%
	\begin{minipage}[t]{.02\columnwidth}
		\vspace{3mm}\hspace{3mm}
	\end{minipage}%
	\begin{minipage}[t]{.49\columnwidth}
		\begin{equation}
			\begin{aligned}
				\varepsilon_{mB} &= \pm 3\varepsilon_B +2\pqty{U-U'}.\label{eq:tltrbSS2}
			\end{aligned}
		\end{equation}
	\end{minipage}\\
\end{subequations}
The sweet spot given by Eqs.~\eqref{eq:tltrSS2} \& \eqref{eq:tltrbSS2} rarely appears (see Fig.~{\ref{fig:gatetime}}), while the condition for Eq.~\eqref{eq:tltrSS} to exist is constrained by the requirement for the qubit to remain in the (1,1,1) region
\begin{equation}
	\begin{gathered}
		-(U-U') < \pqty{\varepsilon_{mA}+U'-U''} \pm \varepsilon_A < U-U'\\
		-(U-U') < \pqty{\mp \varepsilon_A -2\pqty{U' - U''}} + \pqty{U'-U''} \pm \varepsilon_A < U-U'\\
		\implies U'-U'' < U-U',
	\end{gathered}
\end{equation}
where only one side of the inequality was needed as $U>U'>U''$.

It seems that we can find a double sweet spot for both tunneling noises by equating both forms of Eq.~\eqref{eq:tltrSS} to find $(\varepsilon_{A},\varepsilon_{mA}) = (0,-2(U' - U''))$. Unfortunately, the Makhlin invariants, as laid out in Sec.~\ref{app:makhlin-invariants}, are not satisfied with these operating parameters. This causes the two-qubit gate times to blow up to infinity as discussed in the main text, and hence cannot be actually implemented.

\subsubsection*{Detuning Sweet Spots}
To find the middle detuning sweet spot, we bring together Eqs.~\eqref{eq:coef-epmA} \& \eqref{eq:col-coef-negative-A}, and require that ${[{(\vec{F}_A-C_{11}\vec{1})}^T \partial_{\varepsilon_{mA}} \vec{\xi}]}_{\varepsilon_{mA}\text{SS}} = 0$
\begin{equation}
	\begin{gathered}
		0 = -\frac{t_{rA}^2\pqty{C_{21}-C_{11}}}{\pqty{U-2U'+U''-\varepsilon_{mA}-\varepsilon_A}^3}
		+ \frac{t_{lA}^2\pqty{C_{51}-C_{11}}}{\pqty{U-2U'+U''-\varepsilon_{mA}+\varepsilon_A}^3} \\
		\therefore \frac{t_{lA}}{t_{rA}}
		= \sqrt{\frac{C_{51}-C_{11}}{C_{21}-C_{11}}}\pqty{\frac{U-2U'+U''-\varepsilon_{mA}-\varepsilon_A}{U-2U'+U''-\varepsilon_{mA}+\varepsilon_A}}^{\frac{3}{2}}.\label{eq:epmSS}
	\end{gathered}
\end{equation}
Hence, for the $\varepsilon_{mA}$ sweet spot, we need to adjust the ratios of the tunneling parameters to the one given in Eq.~\eqref{eq:epmSS}. This ratio has to be positive for the sweet spot to exist, and the term in $(\cdots)^{\frac{3}{2}}$ is always positive in the (1,1,1) region, so the requirement is
\begin{equation}
	\operatorname{sgn}\pqty{C_{51}-C_{11}} = \operatorname{sgn}\pqty{C_{21}-C_{11}},
\end{equation}
which is purely determined by the geometry of the setup. In the linear array, as we have used for our numerical investigations, $C_{51}-C_{11} = \sum_j \mathcal{V}_{2j} - \mathcal{V}_{1j}$ and $C_{21}-C_{11}= \sum_j \mathcal{V}_{3j} - \mathcal{V}_{2j}$ are both positive, since the first (second) dot is farther from qubit B than the second (third) dot. However, this might not be true in a different geometry, in which case an $\varepsilon_{mA}$ sweet spot will not exist.

For $\varepsilon_{mB}$, we find
\begin{equation}
	\frac{t_{lB}}{t_{rB}}
	=
	\sqrt{\frac{C_{13}-C_{11}}{C_{14}-C_{11}}}\pqty{
		\frac{U-2U'+U''-\varepsilon_{mB}-\varepsilon_B}{U-2U'+U''-\varepsilon_{mB}+\varepsilon_B}
	}^{\frac{3}{2}}
\end{equation}
with the condition
\begin{equation}
	\operatorname{sgn}\pqty{C_{13}-C_{11}} = \operatorname{sgn}\pqty{C_{14}-C_{11}},
\end{equation}
for $C_{13}-C_{11} = \sum_j \mathcal{V}_{j4} - \mathcal{V}_{j5}$ and $C_{14}-C_{11}= \sum_j \mathcal{V}_{j5} - \mathcal{V}_{j6}$, which are again positive for the linear array because the fifth (sixth) dot is farther from qubit A than the fourth (fifth) dot.

The strategy for finding the $\varepsilon_{A/B}$ sweet spots are identical, and we find
\begin{subequations}
	\begin{align}
		\frac{t_{lA}}{t_{rA}} &= \sqrt{\pqty{-\frac{C_{51}-C_{11}}{C_{21}-C_{11}}}
			\pqty{\frac{\frac{2}{\pqty{U-U''+2\varepsilon_{A}}^3}
					- \frac{1}{\pqty{U-2U'+U''-\varepsilon_{mA}+\varepsilon_A}^3}}{\frac{2}{\pqty{U-U''-2\varepsilon_{A}}^3}
					- \frac{1}{\pqty{U-2U'+U''-\varepsilon_{mA}-\varepsilon_A}^3}}}} \\
		\frac{t_{lB}}{t_{rB}} &= \sqrt{
			\pqty{-\frac{C_{13}-C_{11}}{C_{14}-C_{11}}}\pqty{
				\frac{\frac{2}{\pqty{U-U''+2\varepsilon_{B}}^3}
					- \frac{1}{\pqty{U-2U'+U''-\varepsilon_{mB}+\varepsilon_B}^3}}{
					\frac{2}{\pqty{U-U''-2\varepsilon_{B}}^3}
					- \frac{1}{\pqty{U-2U'+U''-\varepsilon_{mB}-\varepsilon_B}^3}
				}
			}
		}.
	\end{align}  \label{eq:epSSboth}
\end{subequations}
Unlike the previous scenario, there are no simple conditions for the existence of the $\varepsilon_{A/B}$ sweet spots apart from ensuring that the terms inside the square root in Eqs.~\eqref{eq:epSSboth} are not negative. For the linear array chosen for our paper, no $\varepsilon_{A/B}$ sweet spots exist.


\subsection{Parameters of the Hubbard Model}
In this study, we used the Hubbard model to describe both intra-TQD and inter-TQD interactions. Intra-TQD interactions comprise QD detunings, tunnel couplings as well as intra- and inter-dot Coulomb energies. Inter-TQD interaction comprise inter-dot Coulomb energies only, when tunnel coupling between the TQDs (QD 3 and 4) is zero.

The quadratic QD confinement potential was used to calculate inter-TQD Coulomb interactions. This approach enables the analytical derivation of the functional dependence of capacitive coupling on a minimal set of QD geometrical parameters, Bohr radius $a_B$ and interdot distance $a$. Importantly, these dependencies helped to identify the existence and conditions of 2QSS on simple experimental parameters. A key feature of the quadratic model is that simple analytical formulae connect geometrical parameters with experimental parameters of tunnel coupling and Coulomb energies, quantities utilized in the Hubbard model. This points to the compatibility between both models.

Ideally, intra-qubit tunnel couplings and Coulomb energies should also be derived with the same quadratic model, for consistency. Unfortunately, the quadratic model does not correspond perfectly with experiment -- tunnel couplings are tuned in experiment by application of gate voltages that modulate the barrier potential whereas in the quadratic model, tunnel coupling can only be varied by changing interdot distance. In addition, changing interdot distances modifies Coulomb energies in the quadratic model, which is undesirable. Even though terms corresponding to tunnel barriers could be added, or alternative Gaussian confinement models used, the drawback is that simple analytical formula relating the inputs of these models with experimental parameters do not exist. Because the goal was to utilize theoretical models that would yield desired insights into 2QSS, we chose to use the Hubbard model for its compatibility with the quadratic confinement model and its excellent theory-experiment agreement~\cite{Yang:2011p161301,DasSarma:2011p235314}:, in order to describe intra-qubit interactions. Parameters for the Hubbard model were estimated with reference to experiment~\cite{Neyens2019a} and theory~\cite{DasSarma:2011p235314} for silicon QDs, and checked against those derived from the quadratic model to be within the same order of magnitude.

\section{Supplementary Discussion}
\subsection{Effect of Lower Cutoff Frequency}
\begin{figure}[ht!]
\centering
\includegraphics[width=0.6\textwidth]{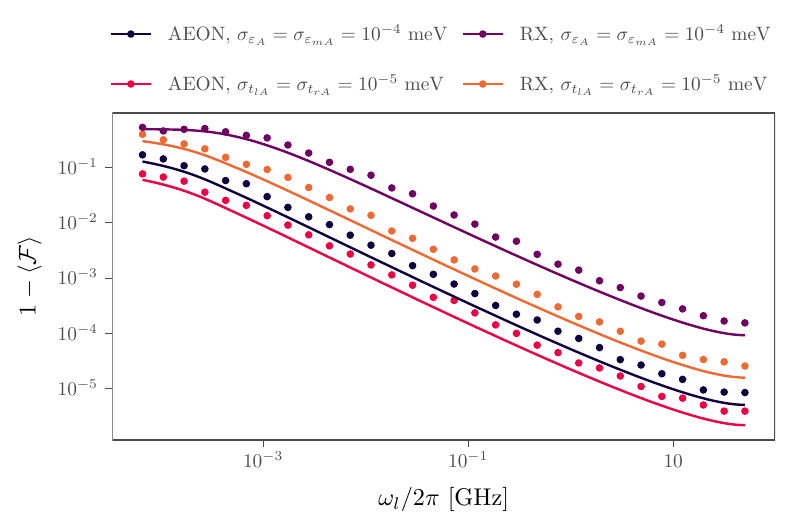}
\caption{\linespread{1}\selectfont{}
Effect of lower cutoff frequency $\omega_l/2\pi$ on infidelity $1-\left\langle \mathcal{F} \right \rangle$. The standard deviation of noise in detunings was fixed at $\sigma_{\varepsilon_{A}} = \sigma_{\varepsilon_{mA}} = 10^{-4}~\textrm{meV}$ for AEON  (blue) and RX (red), and  the standard deviation of noise in tunnel couplings was fixed at $\sigma_{t_{lA}} = \sigma_{t_{rA}} = 10^{-5}~\textrm{meV}$ for AEON (magenta) and RX (yellow), with both qubits sitting at the $\varepsilon_{mA}$ 2QSS. Circles are numerical simulations (Eq.~(13), main text), averaged over 500 noise realizations. Lines are analytical calculations (Eq.~(14), main text). As expected, infidelity decreases as the lower cutoff frequency increases, while keeping  $\omega_h$ constant.
}\label{fig:cutoff}
\end{figure}

For $1/f$ noise to have finite power, the spectrum has to be limited within a higher and lower cutoff frequency. Increasing the lower cutoff shifts the frequency range of the noise spectrum to the higher frequencies, and as the power spectral density is inversely proportional to the frequency, this should result in an improved fidelity,  as can be seen in Fig.~\ref{fig:cutoff}. The effect on fidelity arises from the term $\varsigma(t)$ in the exponent of the fidelity formula (Eq.~\eqref{eq:fidelity-formula}),
\begin{equation}
	\begin{aligned}
		\varsigma(t) =& \frac{1/2}{1+\ln(\omega_h/\omega_l)}\bigg(
		\operatorname{Ci}(\omega_ht)t^2
		- \operatorname{Ci}(\omega_lt)t^2
		- \frac{\sin(\omega_ht)t}{\omega_h}
		 + \frac{\sin(\omega_lt)t}{\omega_l}
		+ \frac{2\operatorname{Si}(\omega_lt)t}{\omega_l}
		+ \frac{\cos(\omega_ht)-1}{\omega_h^2}
		+ \frac{\cos(\omega_lt)-1}{\omega_l^2}
		\bigg)
	\end{aligned}\label{eq:cutoff}
\end{equation}
where $\text{Ci}(x)$ and $\text{Si}(x)$ are the cosine and sine integrals respectively. When $\omega_l t_0 \ll 1$, increasing lower cutoff frequency $\omega_l$ reduces the magnitude of $\varsigma(t_0)$ and increases the fidelity.

\subsection{Noise in Individual Qubits}
In the main text, we only reported the cases where noise is present in the control qubit. This is as there is not much more to be learnt from the cases where noise is present in the target qubit, or where noise is present in both qubits.

From Eq.~\eqref{eq:fidelity-formula}, we can write the total fidelity as a product of two fidelities $\ev{\mathcal{F}} = \ev{\mathcal{F}}_A \times \ev{\mathcal{F}}_B$ where $\ev{\mathcal{F}}_{A}$ ($\ev{\mathcal{F}}_{B}$) is the fidelity of the system when there is only noise in qubit A (B). Explicitly,
\begin{equation}
	\begin{aligned}
		\ev{\mathcal{F}}_A &= \left.\ev{\mathcal{F}}\right\rvert_{\sigma_{\mathbf{n}_B} = 0} \\
		&= \frac{1}{4}\pqty{1+\exp\Bqty{
				-\frac{1}{2}\sum_{i}\pqty{\pqty{\vec{F}_A - C_{11}\vec{1}}^T \partial_i \vec{\xi}}^2\sigma_{n_{iA}}^2
				\varsigma(t_0)
		}}\pqty{1+\exp\Bqty{0}}\\
		&= \frac{1}{2}\pqty{1+\exp\Bqty{
				-\frac{1}{2}\sum_{i}\pqty{\pqty{\vec{F}_A - C_{11}\vec{1}}^T \partial_i \vec{\xi}}^2\sigma_{n_{iA}}^2
				\varsigma(t_0)
		}},
	\end{aligned}
\end{equation}
and $\ev{\mathcal{F}}_B$ is obtained by swapping all appearances of $\xi \to \nu$ and $A\to B$.

\begin{figure}
	{\centering
		\includegraphics[width=\textwidth]{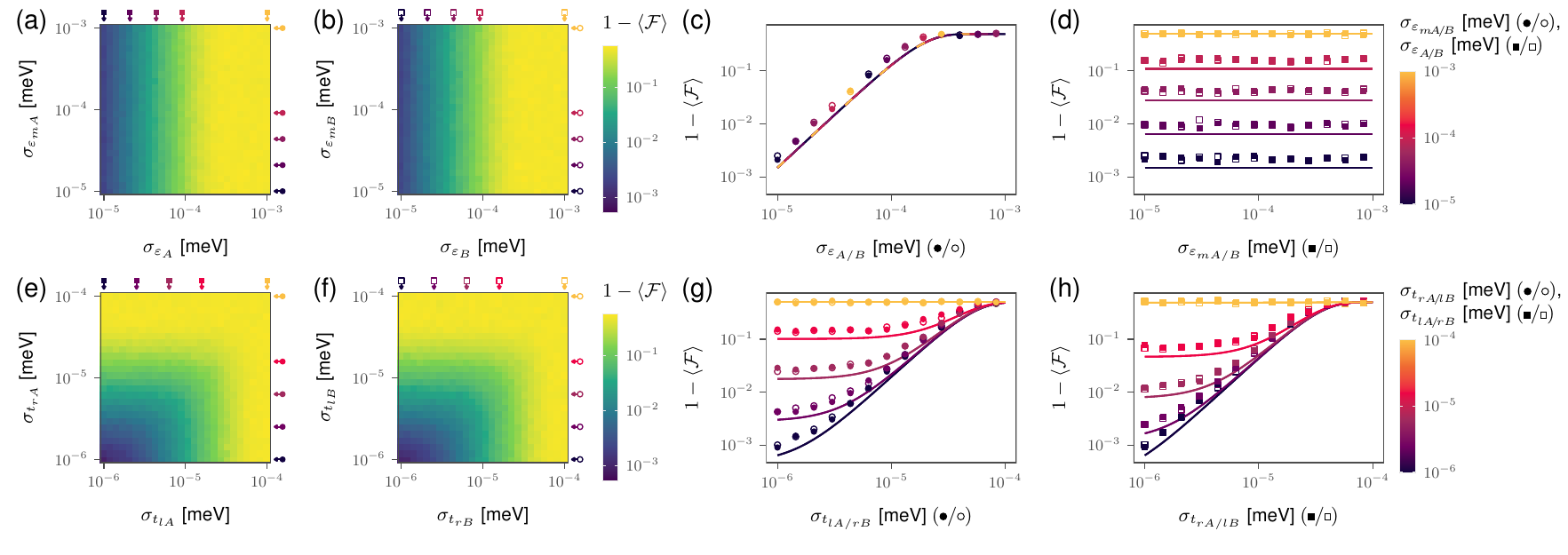}}
	\caption{Noise in control qubit versus noise in target qubit. The introduction of noise in either qubit results in the same effect on the fidelity (note that $t_{lA} \leftrightarrow t_{rB}$ mirrors $t_{rA} \leftrightarrow t_{lB}$ )}\label{fig:noiseInOneQubit}
\end{figure}

If both qubits are operated symmetrically --- $\varepsilon_A = - \varepsilon_B$, $\varepsilon_{mA} = \varepsilon_{mB}$, $t_{l/rA} = t_{r/lB}$ --- the noise coefficients ${\vec{F}_A}^T \partial_i \vec{\xi} = {\vec{F}_B}^T \partial_i \vec{\nu}$ are equal for both qubits. Since $\varsigma(t_0)$ is the same for both qubits, the behaviour of the individual fidelities for both the control and target qubits will be affected by noise in the same manner, as evident in Fig.~\ref{fig:noiseInOneQubit}.

\begin{figure}
	\centering
	\includegraphics[width=\textwidth]{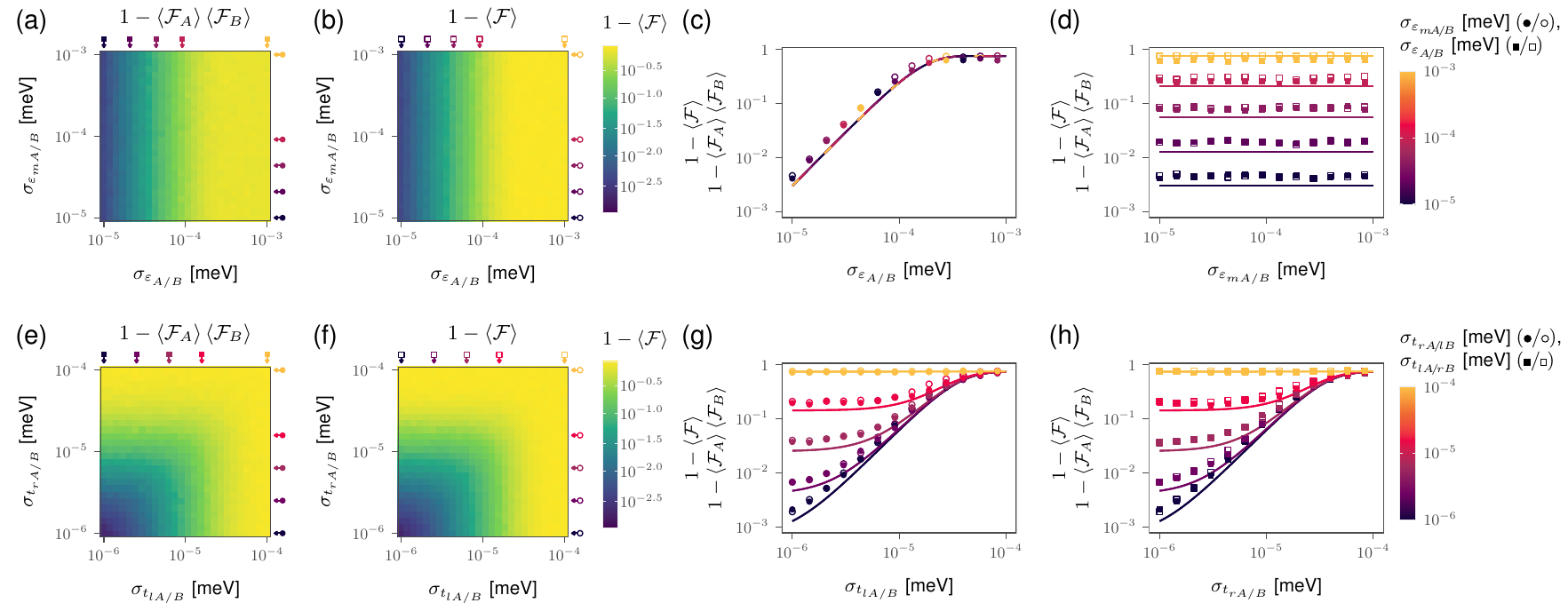}
	\caption{Comparison of noise individually affecting either qubits, versus the concurrent presence of noise in both qubits. We see that the individual fidelities multiply to give the concurrent case.}\label{fig:noiseInBothQubit}
\end{figure}

In addition, the individual fidelities combine in a predictable way. Comparing the case where noise is present in either qubit individually, against that where noise is present in both qubits concurrently, we will find that the multiplication of the two individual fidelities in the former case gives the same fidelity in the latter case, as shown through our numerical simulations in Fig.~\ref{fig:noiseInBothQubit}. The result that $\ev{\mathcal{F}}_A \times \ev{\mathcal{F}}_B = \ev{\mathcal{F}}$ holds true even when the qubit operation is not symmetric.

\subsection{\label{sec:direct}Direct vs Exchange Coulomb Energies}
\begin{figure}[ht!]
	\includegraphics[width=0.6\linewidth]{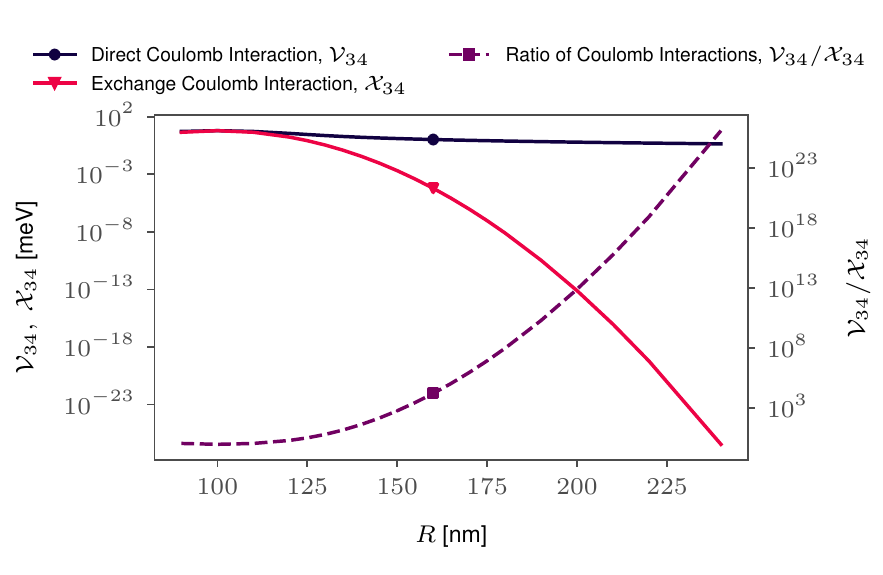}
	\caption{Dependence of direct (blue) and exchange (red) Coulomb interactions on the distance between qubits, $R$. The ratio of direct to exchange Coulomb interactions is plotted as a dashed purple line. The different markers are used to represent the values for these parameters at the value of $R$ chosen ($R=160$ nm), with the circle and triangle markers used for the direct and exchange Coulomb interactions respectively and the square marker fused for the ratio between them. At the $R$ value of choice, we achieve an exchange Coulomb interaction ($\simeq6\times10^{-5}$ meV) about four orders smaller than the direct Coulomb interaction ($\simeq1$ meV).
	}
	\label{fig:directvex}
\end{figure}

\begin{figure}[ht!]
	\includegraphics[width=0.6\linewidth]{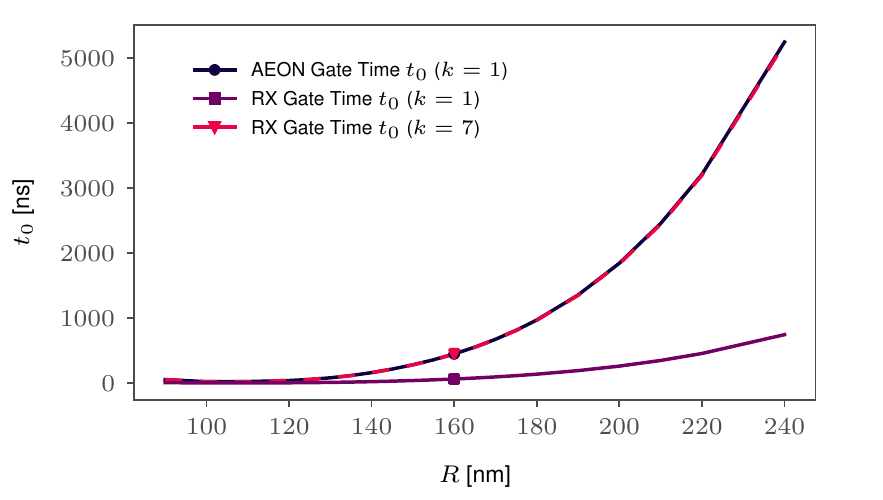}
	\caption{Plot of gate time $t_0$ with distance between qubits $R$, at the single qubit sweet spot ($\varepsilon_{A/B}=0$). The shortest possible gate times correspond to $k=1$. When $k=7$ for the RX qubit, its gate time (448~ns) is comparable with the fastest AEON qubit gate time (450~ns). At the value of $R$ chosen for the simulations ($R=160$nm) as shown by the markers, we observe that $t_0$ remains reasonable at less than 0.5 $\mu$s.
	}
	\label{fig:gatetime}
\end{figure}

One of the main motivations for this study was the advantage of using only a single two-qubit gate to achieve a CPHASE or CNOT gate. This is in stark contrast to complicated gate sequences that are required with exchange interaction \cite{Fong:2011p1003,Shi:2012p140503,DiVincenzo:2000p339}. A key assumptions in our two-qubit system is that the exchange interaction is weak enough such that it can be neglected, allowing us to describe the two-qubit interaction as a purely diagonal Hamiltonian. To achieve this, we could simply move the qubits away from each other and at large enough distances, the direct Coulomb interaction dominates as compared to the exchange Coulomb interaction, as shown in Fig.~\ref{fig:directvex}. The exchange energies are calculated similarly to the direct Coulomb energies,
\begin{align}
	\mathcal{X}_{ij}
	&= \int \int \psi_i^\star(\vec{r}_1)\psi_i^\star(\vec{r}_2) \frac{\kappa}{ |\vec{r}_1-\vec{r}_2|} \psi_j(\vec{r}_1)\psi_j(\vec{r}_2) d\vec{r}_1 d\vec{r}_2,
\end{align}
where $\kappa = q^2/4\pi\epsilon_0\epsilon_r$, as before.
In our system, we considered the quantum dot radius, $a_B = 25$ nm, the inter-dot distance, $2a = 100$ nm, and the inter-qubit distance, $2R = 320$ nm as measured from the center dot of each qubit. This ensures dots 3 and 4 are well separated with a distance of $120$ nm between them. At this distance, we compare the magnitudes of the exchange and direct Coulomb interaction between the two closest dots in the two-qubit system, namely dots 3 and 4 as labeled in Fig.~1(a) in the main text. The interaction between these two dots are expected to dominate as compared to the other inter-qubit interactions since they are the two closest dots between the two qubits. As evidenced in Fig.~\ref{fig:directvex}, the exchange energy ($\mathcal{X}_{34}$ in red) is about four orders of magnitude smaller than the corresponding direct Coulomb energy ($\mathcal{V}_{34}$ in blue) at this distance. As seen in Fig.~\ref{fig:gatetime} the trade-off is longer gate times as inter-qubit distance is increased.

